\documentclass[%
 reprint,
superscriptaddress,
 amsmath,amssymb,
 aps,
pra,
]{revtex4-2}

\usepackage{xcolor}
\usepackage{graphicx}
\usepackage{dcolumn}
\usepackage{bm}
\usepackage[caption=false]{subfig}
\usepackage[unicode=true,pdfusetitle,
 bookmarks=true,bookmarksnumbered=false,bookmarksopen=false,
 breaklinks=false,pdfborder={0 0 1},backref=false,colorlinks=true]{hyperref}
\DeclareMathOperator{\sech}{sech}

\begin{document}

\preprint{APS/123-QED}

\title{Static and dynamic properties of self-bound droplets of light in hot vapours}

\author{Heitor da Silva}
 \affiliation{Departamento de Física Teórica e Experimental, Universidade Federal do Rio Grande do Norte,
59072-970 Natal, Rio Grande do Norte, Brazil}
\affiliation{Université Côte d’Azur, CNRS, Institut de Physique de Nice, 06560 Valbonne, France}

\author{Robin Kaiser}
\affiliation{Université Côte d’Azur, CNRS, Institut de Physique de Nice, 06560 Valbonne, France}
\author{Tommaso Macrì}
\affiliation{Departamento de Física Teórica e Experimental, Universidade Federal do Rio Grande do Norte,
59072-970 Natal, Rio Grande do Norte, Brazil}
\affiliation{ITAMP, Harvard-Smithsonian Center for Astrophysics, Cambridge, Massachusetts 02138, USA}%

\date{\today}

\begin{abstract}
The propagation of light in nonlinear media is well described by a $2$D nonlinear Schrödinger equation (NLSE) within the paraxial approximation, which is equivalent to the Gross-Pitaesvskii equation (GPE), the mean-field description for the dynamics of Bose-Einstein condensates (BECs). Due to this similarity, many theoretical and experimental investigations of phenomena which have already been studied and realized in BECs have been recently analysed in alternative experimental platforms such as hot atomic vapours. In this work, we study the formation of droplets of light in these media, attempting to establish a mapping between the experimental parameters normally used in BEC experiments and those needed to observe the analogous phenomenon in hot atomic vapours. 
We obtain the energy functional for the susceptibility of the medium in the $\chi^{(3)}$ , $\chi^{(3)}+\chi^{(5)}$ and saturating regimes for a two-level atomic configuration considering the focusing (attractive) regime. We apply a Gaussian variational approach and check its predictions through numerical simulations of the NLSE for each regime. Finally, we study the real-time dynamics of the system for both the $\chi^{(3)}+\chi^{(5)}$ and saturating nonlinearities, focusing our attention on the behaviour of the breathing mode and on the analysis of droplet formation for realistic experimental conditions.
\end{abstract}

\maketitle


\section{Introduction}
The field of atomic physics has achieved great advances in the past decades, especially due to the progress seen in the experimental side. The advances obtained with experiments using ultracold atoms have made it possible to investigate many phenomena in various branches of physics. However, such experiments may bring with them some obstacles depending on the physical phenomenon one wants to study as well as their costs, which can be much higher when compared to other experimental platforms. For instance, there are many phenomena observed in Bose-Einstein condensates (BECs) which have been recently investigated in alternative platforms such as hot atomic vapours~\cite{CarusottoRevModPhys.85.299,ReviewQG}. 
A hot vapour is an extremely versatile experimental platform and it has been the workhorse in atomic physics over the years. What makes this transition possible is the existing analogy between the mean-field description for BECs, which is given by the Gross-Pitaesvskii equation (GPE), and the equation for propagation of light in a nonlinear medium in the paraxial approximation, an example of a nonlinear Schrödinger equation (NLSE). 
With that, one can then attempt to establish a mapping between the experimental parameters of the condensates with those of the thermal vapours. In fact, this analogy has already been exploited in several theoretical and experimental works (using thermal vapours) ranging from condensation of classical waves~\cite{PicoziConaughtonPRL2005, RobinPRL2018,BaudinPRL2020}, superfluidity of a paraxial fluid of light~\cite{carusotto2014superfluid,QuentinFontainePRL2018,QuentinFontainePRR2020}, vortex generation and control of their interactions~\cite{azam2022vortex}, the generation and the dynamics of dispersive shock and blast waves~\cite{KamchatnovPRA2019,KamchatnovPRE2020,Queensland2PRL020,PhysRevA.104.013515,TomBienaimePRL2021,Abuzarli_2021}, spin-orbit-coupled mixtures~\cite{MartonePRA2021} and even the investigation of analogue models in gravity, for instance, the analogue of cosmological particle creation~\cite{steinhauer2021analogue}.
When compared to ultracold experiments, one of the advantages of thermal vapours is that they are relatively cheap and much simpler to set up. Another advantage comes up when there is the need of obtaining higher densities. Since the susceptibility of a nonlinear medium depends on the density of the medium, the use of a thermal vapour helps to get higher densities.
In ultracold experiments, typical densities vary in the range $10^{11}-10^{12}$ cm$^{-3}$ in a magneto-optical trap (MOT), while in a BEC they range from $10^{13}$ to $10^{15}$ cm$^{-3}$. Meanwhile, hot atomic vapours can have densities orders of magnitude larger than BECs, and are tunable over a much wider range.

The purpose of this work is to characterize self-bound states of light in hot vapours, in analogy to droplet states in binary mixtures of BECs and dipolar systems. 
Quantum droplets consist of small clusters of atoms, self-bounded by the balance of an attractive mean-field energy and repulsive beyond mean-field interactions~\cite{PetrovPRL2015}. Several experiments have successfully observed ultradilute self-bound states in a variety of configurations such as ultracold dipolar systems \cite{schmitt2016self,PhysRevX.6.041039}, Bose-Bose mixture in quasi-$2$D and quasi-$1$D geometries \cite{CabreraScience2018,PhysRevLett.120.135301}, in $3$D geometry \cite{PhysRevLett.120.235301} among others. 

In this study, we consider a nonlinear hot vapour medium modeled as an ensemble of two-level systems. 
Our focus lies on the focusing (attractive) regime. The physical parameter that sets the sign of the interaction is the frequency detuning. 
We investigate three different regimes: a Kerr medium, i.e., the refractive index has a linear dependence with the intensity; the cubic-quintic nonlinearity, that is, up to second order in the intensity and finally, the most general saturating nonlinearity. For each of these situations, we analyse the corresponding energy functional. For this task, we will make use of a variational approximation method and exact numerical simulations ~\cite{anderson_bondeson_lisak_1979,anderson1979variational,Anderson1983,Anderson88,MalomedReview2002}. 

This paper is organized as follows. In section \ref{sec:setting}, we derive the effective NLSE for light propagating through an ensemble of two-level systems with susceptibility $\chi$. In section \ref{sec:energy_functionals}, we compute the energy functionals employing a variational Gaussian ansatz. 
In section \ref{sec:dynamics}, we address the real-time dynamics. We briefly analyse the $\chi^{(3)}$ regime, focusing on the physics of the Townes soliton and commenting on its main features. 
We proceed with the study of collective excitations (the breathing mode), defining the range of parameters for which they can be observed. Also, we establish a connection with the self-evaporation mechanism and its influence on droplet dynamic formation within the $\chi^{(3)}+\chi^{(5)}$ and saturating regimes. Finally, we study the droplet formation for realistic experimental conditions.

\section{\label{sec:setting}Physical system}
We consider an ensemble of two-level systems formed by a ground state $|g \rangle$ and an excited state $|e\rangle$ that can decay into the ground state at a rate $\Gamma$. 
\begin{figure}[ht!]
\centering
\includegraphics[width=0.995\linewidth]{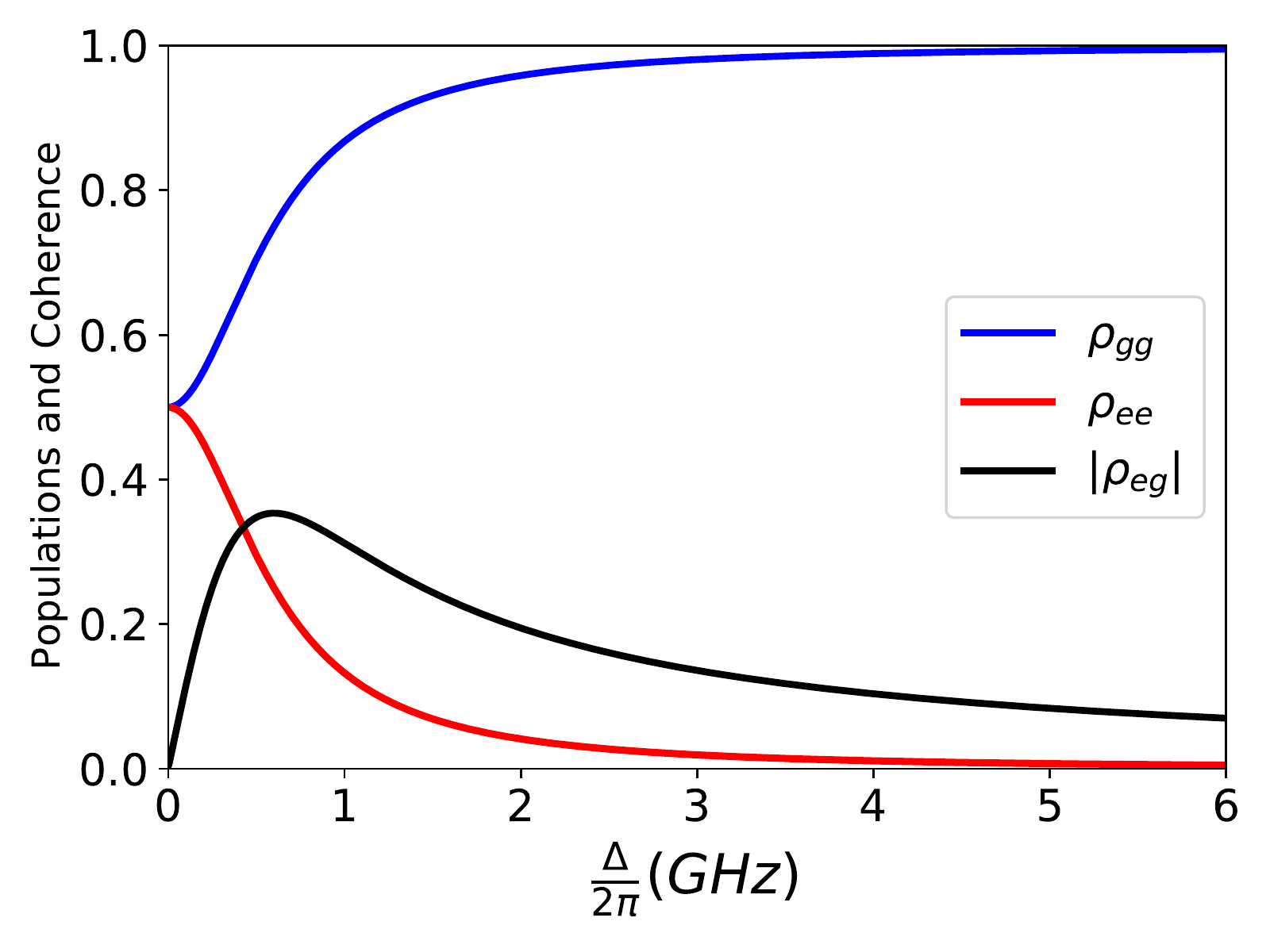}
\caption{\textit{Atomic level description}. Stationary populations and coherence for the $2-$level system as a function of the frequency detuning $\Delta$.}
\label{populations2and4level}
\end{figure}
The optical Bloch equations (OBE) for this configuration are easily obtained, and their steady-state solutions for the populations and the coherence in the steady state are shown in Fig. (\ref{populations2and4level}). 
See Appendix \ref{obe2level} for the full expressions.

The susceptibility is given by:
\begin{equation}
\label{eqs1}
\chi= \frac{\rho_{at}|\mathbf{d}_{eg}|^2}{\hbar \epsilon_{\rm 0}}\frac{-\Delta + i\Gamma /2}{\Delta^{2}+\frac{\Gamma^{2}}{4}+\frac{|\Omega|^2}{2}}
\end{equation}
where $\rho_{at}$ is the atom density and $\hbar\ \Omega=-\mathbf{d}_{eg}\cdot\mathbf{E}$ is the Rabi frequency with $\mathbf{E}$ being the {\it classical} external electric field.

The linear real susceptibility $\chi^{(L)}$ reads
\begin{equation}
\chi^{(L)}=-\frac{\rho_{at}|\mathbf{d}_{eg}|^{2}}{\hbar \epsilon_{0}}\frac{\Delta}{\Delta^{2}+\frac{\Gamma^{2}}{4}}\label{eqs3a}
\end{equation}
while the real nonlinear part of the full susceptibility is obtained after subtracting $\chi^{(L)}$ and taking the real part, yielding
\begin{equation}
\chi^{(NL)}=\frac{\rho_{at}|\mathbf{d}_{eg}|^{2}}{\hbar \epsilon_{0}}\frac{\Delta |\Omega|^{2}/2
}{\left(\Delta^{2}+\frac{\Gamma^{2}}{4}\right)\left(\Delta^{2}+\frac{\Gamma^{2}}{4}+\frac{|\Omega|^2}{2}\right)}\,.
\label{eqs3an}
\end{equation}
In the paraxial approximation, the NLSE for the light field amplitude is given by:
\begin{equation}
 \label{generalNLSE}
 i\frac{\partial \Omega}{\partial z}=-\frac{1}{2k_{0}}\nabla^{2}_{\bot}\Omega-\frac{k_{0}}{2}\chi\Omega\,.
\end{equation}
In this equation, the longitudinal coordinate $z$ plays the role of an effective time while the Laplacian is computed with respect to the transverse coordinates, i.e., $\mathbf{r}\equiv \left(x,y\right)$, and $k_{0}$ is the wave vector. 

Performing the transformation,
\begin{equation}
 \label{eqs6t}
 \bar{\Omega}\left(\mathbf{r},z\right)=\Omega\left(\mathbf{r},z\right)\exp\left(-\rm i\,\frac{k_{0}\chi^{(L)}}{2}z\right),
\end{equation}
we are left with the equation 
\begin{equation}
 \label{nlseSat2}
 i\frac{\partial \,\Omega}{\partial z}=-\frac{1}{2k_{0}}\nabla_{\bot}^{2}\,\Omega-\frac{k_{0}}{2}\chi^{(NL)}\,\Omega.
\end{equation}

This equation can be written in a dimensionless form by performing the scalings
\begin{equation}
 \label{scaling}
 \mathbf{r}^{\prime}=\frac{\mathbf{r}}{r_{0}}\,,\qquad z^{\prime}=\frac{z}{L_{d}}\,,\qquad \psi=\alpha\,\Omega
\end{equation}
where $r_{0}$ is an arbitrary length scale, $L_{d}\left(r_{0}\right)=k_{0}r_{0}^{2}$ the associated diffraction length and $\alpha$ is a parameter depending on the optical parameters whose dimension is inverse of frequency (see below in Eq.(\ref{eqs23a})). Later on, for the numerical simulations, we will consider a Gaussian input beam by setting $r_{0}=w_{0}$, where $w_{0}$ is the {\it initial} beam waist~\cite{FibichGaetaOL2000}, for which the Rayleigh length is defined as $z_{R}=\left(1/2\right)\,L_{d}\left(w_{0}\right)$.  

The dimensionless equation of motion then reads
\begin{eqnarray}
\label{eqs7c}
i\frac{\partial\psi}{\partial z^{\prime}}=-\frac{1}{2}\nabla_{\bot}^{\prime \,2}\psi-\frac{1}{1+2\gamma\left|\psi\right|^2}\left|\psi\right|^{2}\psi\,
\end{eqnarray}
with $\psi$ being the scaled light field amplitude. The Laplacian now has to be computed with respect to the dimensionless transverse coordinates, i.e., $\mathbf{r}^{\prime}\equiv \left(x^{\prime},y^{\prime}\right)$. It is straightforward to check that for $\gamma\left|\psi\right|^{2}\ll 1$, a series expansion of Eq. (\ref{eqs7c}) in $\left|\psi\right|^{2}$ will lead to the cubic NLSE (Kerr medium) in zeroth and to the cubic-quintic NLSE up to first order, respectively. 

The parameter $\alpha$ in Eq. (\ref{scaling}) is related with the strength of the nonlinearity of the system. It is also connected with the dimensionless critical optical power ($\mathcal{P}_{cr}$), a quantity that will be important to classify the regimes for which self-bound state solutions and collapsing behaviour occur. 
For a description beyond the Kerr regime, $\alpha$ will be no longer the only interaction parameter. This role will be shared with the dimensionless parameter $\gamma$, responsible for turning on the terms beyond the cubic NLSE.

The parameters $\alpha$ and $\gamma$ take on the following expressions,
\begin{subequations}
\label{allequations} 
\begin{align}
 \alpha=&\frac{k_0 r_0}{2\Gamma}\sqrt{\eta \frac{\Delta/\Gamma}{\left[\left(\frac{\Delta}{\Gamma}\right)^2 + \frac{1}{4}\right]^2}}, \label{eqs23a}
 \\
 \gamma=&\left[\left(k_0 r_0\right)^2 \eta \frac{\Delta/\Gamma}{\left(\frac{\Delta}{\Gamma}\right)^2+\frac{1}{4}}\right]^{-1}\label{eqs23b}\,
\end{align}
\end{subequations}
where $\Gamma$ is the natural line width and the dimensionless coefficient $\eta$ is related with the density, and consequently with the temperature of the atomic cloud~\cite{steck2001rubidium}. Its expression is given by,
\begin{equation}
 \label{eqs25}
 \eta=\frac{\rho_{at}\left|\mathbf{d}_{eg}\right|^{2}}{\hbar \epsilon_{o}\Gamma}
\end{equation}
where $\mathbf{d}_{eg}$ is the transition dipole moment between the ground state manifold $\rm g$ and the excited state manifold $\rm e$. Alternatively, $\eta$ can be rewritten in terms of the wavelength $\lambda$ employing the spontaneous emission rate $\Gamma$ in vacuum, reading ~\cite{GRIMM200095}
\begin{equation*}
    \Gamma = \frac{\omega_{0}^3}{3\pi\epsilon_{0}\hbar c^3}\left|\mathbf{d}_{eg}\right|^{2}\,,
\end{equation*}
which leads to
\begin{equation}
 \label{eqs25new}
 \eta=\frac{3\,\rho_{at}\,\lambda^{3}}{8\,\pi^{2}}\,.
\end{equation}
The parameter $\alpha$ can be  recast to include negative values of $\Delta$ by taking the absolute value and flipping the global sign of the nonlinear interaction term in Eq. (\ref{eqs7c}). In our numerical analysis, the values for the quantities were chosen to consider the specific case of a thermal vapour of $\,^{85}\rm Rb$ whose natural line width is $\Gamma=2\pi\times 6.06\,$MHz for the $\rm D 2$ line ($5^{2}\,S_{1/2}\,\rightarrow\,5^{2}\,P_{3/2}$ transition). By considering a $\pi-$polarized light, it follows that the value of the effective far-detuned dipole moment is $2.069\times 10^{-29}\,$C$\cdot$m~ while the saturation intensity is equal to $25\,$W$/$m$^2$~\cite{steck2001rubidium}. 

\section{\label{sec:energy_functionals}Energy functionals and the Gaussian variational ansatz}

In this section, we study the stationary properties of the system by means of the analysis of the energy functionals describing stationary configurations. 
From the general saturating nonlinearity, we derive the energy functional for the $\chi^{(3)}$ and $\chi^{(3)}+\chi^{(5)}$ regimes upon Taylor expansion in the parameter $\gamma\left|\psi\right|^{2}$. 
We then evaluate the resulting expressions using a Gaussian ansatz for the dimensionless light field amplitude $\psi$. 
The use of a variational approach allows us to derive analytical results to assess the static and dynamic behaviour of the system close to the stationary configurations, similar to the case of BECs~\cite{pethick_smith_2008,MalomedReview2002}. 

\subsection{Energy functionals}
The energy functional for the saturating regime here denoted by $E^{(sat)}$ can be obtained directly from Eq. (\ref{eqs7c}), which yields
\begin{align}
 \label{EDFgeneral}
 E^{(sat)}=\frac{1}{2}\int \left|\nabla_{\bot} ^{\prime}\psi\left(\mathbf{r}^{\prime}\right)\right|^{2}\,d^{2}\,\mathbf{r}^{\prime}-\frac{1}{2\,\gamma}\int \left|\psi\left(\mathbf{r}^{\prime}\right)\right|^{2}\,d^{2}\,\mathbf{r}^{\prime}\nonumber\\
 +\frac{1}{4\,\gamma^{2}}\int \ln\left[1+2\,\gamma\left|\psi\left(\mathbf{r}^{\prime}\right)\right|^{2}\right]\,d^{2}\,\mathbf{r}^{\prime}\,.\nonumber\\
\end{align}
We start by considering the following Gaussian input profile
~\cite{cappellaro2017equation, MacriCapellaroLucaPRA2018,PhysRevA.102.053303}
\begin{equation}
 \label{eqs9}
 \psi\left(\mathbf{r}^{\prime}\right)=\sqrt{\frac{\mathcal{P}}{\pi\sigma^{\prime 2}}}\exp\left(-\frac{r^{\prime\, 2}}{2\sigma^{\prime 2}}\right)\,
\end{equation}
where the dimensionless width $\sigma^{\prime}=\sigma/r_{\rm 0}$ is the variational parameter, and $\mathcal{P}$ is the dimensionless power which depends on the optical parameters of the system  through $\alpha$ (see Eq. (\ref{scaling})). Although $r_{0}$ is among the parameters contained in the definition of $\alpha$, the values of the physical quantities will be independent of its choice.

We compute $E^{(sat)}$ using the ansatz given in Eq. (\ref{eqs9}), which yields
\begin{eqnarray}
E^{(sat)}=\frac{1}{2\sigma^{\prime 2}}\mathcal{P}-\frac{1}{2\gamma}\mathcal{P}-\frac{\pi}{4\gamma^{2}}\sigma^{\prime 2}\,\rm Li_{2}\left(-\frac{2\gamma}{\pi\sigma^{\prime 2}}\mathcal{P}\right)\nonumber,\\
\label{eqs10c}
\end{eqnarray}
where $\rm Li_{2}\left(x\right)$ is the polylogarithmic function of order $2$. 

For $\gamma\left|\psi\right|^{2}\ll 1$, we can perform a Taylor expansion in the logarithmic term of Eq. (\ref{EDFgeneral}). Truncation to the first order produces the $\chi^{(3)}$ regime, whereas the second order leads to the $\chi^{(3)}+\chi^{(5)}$ regime. 
The expressions for $E^{(3)}$ and $E^{(5)}$ are
\begin{subequations}
\label{allequations}
\begin{align}
E^{(3)}=&\frac{1}{2}\int \left|\nabla_{\bot} ^{\prime}\psi\left(\mathbf{r}^{\prime},z^{\prime}\right)\right|^{2}\,d^{2}\,\mathbf{r}^{\prime}-\frac{1}{2}\int \left|\psi\left(\mathbf{r}^{\prime},z^{\prime}\right)\right|^{4}\,d^{2}\,\mathbf{r}^{\prime}\,,\nonumber
\\
\label{eqs9a}
\\
\nonumber\\
E^{(5)}=&\frac{1}{2}\int \left|\nabla_{\bot} ^{\prime}\psi\left(\mathbf{r}^{\prime},z^{\prime}\right)\right|^{2}\,d^{2}\,\mathbf{r}^{\prime}-\frac{1}{2}\int \left|\psi\left(\mathbf{r}^{\prime},z^{\prime}\right)\right|^{4}\,d^{2}\,\mathbf{r}^{\prime}\nonumber
\\
&+\frac{2\gamma}{3}\,\int \left|\psi\left(\mathbf{r}^{\prime},z^{\prime}\right)\right|^{6}\,d^{2}\,\mathbf{r}^{\prime}.
\label{eqs9b}
\end{align}
\end{subequations}
Employing the Gaussian ansatz from Eq. (\ref{eqs9}), we obtain
\begin{subequations}
\label{allequations}
\begin{align}
E^{(3)}=&\frac{1}{2\sigma^{\prime 2}}\mathcal{P}-\frac{1}{4\pi\sigma^{\prime 2}}\mathcal{P}^{2}\,,\label{eqs10a}
\\
E^{(5)}=&\frac{1}{2\sigma^{\prime 2}}\mathcal{P}-\frac{1}{4\pi\sigma^{\prime 2}}\mathcal{P}^{2}+\frac{2 \gamma}{9\pi^{2}\sigma^{\prime 4}}\mathcal{P}^{3}\label{eqs10b}\,.
\end{align}
\end{subequations}

\subsection{$\chi^{(3)}+\chi^{(5)}$: the cubic-quintic nonlinearity}
We now analyse the cubic-quintic nonlinearity. The equation of motion Eq. (\ref{eqs7c}) up to first order in $\gamma\left|\psi\right|^{2}\ll 1$ leads to
\begin{equation}
    \label{nlseChi3Chi5}
    i\frac{\partial\psi}{\partial z^{\prime}}=-\frac{1}{2}\nabla_{\bot}^{\prime \,2}\psi-\left|\psi\right|^{2}\psi+2\gamma\left|\psi\right|^{4}\psi\,.
\end{equation}
The $\chi^{(3)}+\chi^{(5)}$ regime provides the suitable conditions for creating self-bound states due to the competition between the focusing (attractive) $\chi^{(3)}$ and defocusing (repulsive) $\chi^{(5)}$ nonlinearities \cite{PhysRevE.65.066604,Michinel_PRA2009,FeijooPRL2014}.
The conditions leading to self-bound states of light were pointed out in \cite{PazAlonsoPRL2006}. 
There, it was shown that for a four-level system, an adequate choice of the parameters for an electromagnetic-induced transparency scheme may lead to a giant response for both the coefficients of the cubic and quintic nonlinearities 
(with different signs) stabilizing two-dimensional droplets. 
More recently, it was shown that bound states with finite angular momentum with liquid-like properties can arise when considering a nonlocal photon fluid with a focusing, long-range
nonlinearity generated in the transverse plane
of a laser beam propagating in a thermo-optic medium \cite{westerberg2018self,wilson2018observation}. 

We start with the analysis of the energy functional $E^{(5)}$ given in Eq. (\ref{eqs10b}). 
Setting its derivative with respect to $\sigma^{\prime}$ to zero leads to
\begin{equation}
 \label{sigma3}
 \sigma^{\prime}_{c}=\frac{4\mathcal{P}\sqrt{\gamma}}{3\sqrt{\pi}\sqrt{-2\pi+\mathcal{P}}}\,,
\end{equation}
where $\mathcal{P}>2\pi$ must be satisfied. 
Taking the second derivative  $d^{2}E^{(5)}/d\sigma^{\prime 2}$ at $\sigma^{\prime}=\sigma^{\prime}_{c}$ one can show that Eq.(\ref{sigma3}) is a minimum.
Moreover, this is a global minimum, since the energy is negative at $\sigma^{\prime}=\sigma^{\prime}_{c}$, excluding the presence of metastable minima.
\subsection{{\label{sec:satN}}Saturating nonlinearity}

Stable self-bound states for the saturating regime were investigated, for instance, in \cite{Vakhitov1973,J.JuulRasmussen_1986}. 
For the saturating regime, an analytical expression for the stationary value $\sigma^{\prime}$ is not available. 
For the parameters used in the simulations and in most analysis throughout this work (unless specifically stated otherwise), we set the beam waist $w_{0}$ equal to $7\times 10^{-4}\,$m and the coefficient $\eta$ to unity, leading to an atom density of $\rho_{at} = 8.30\times10^{19}\,m^{-3}$. 

\begin{figure}[ht!]
\centering
\includegraphics[width=1.0\linewidth]{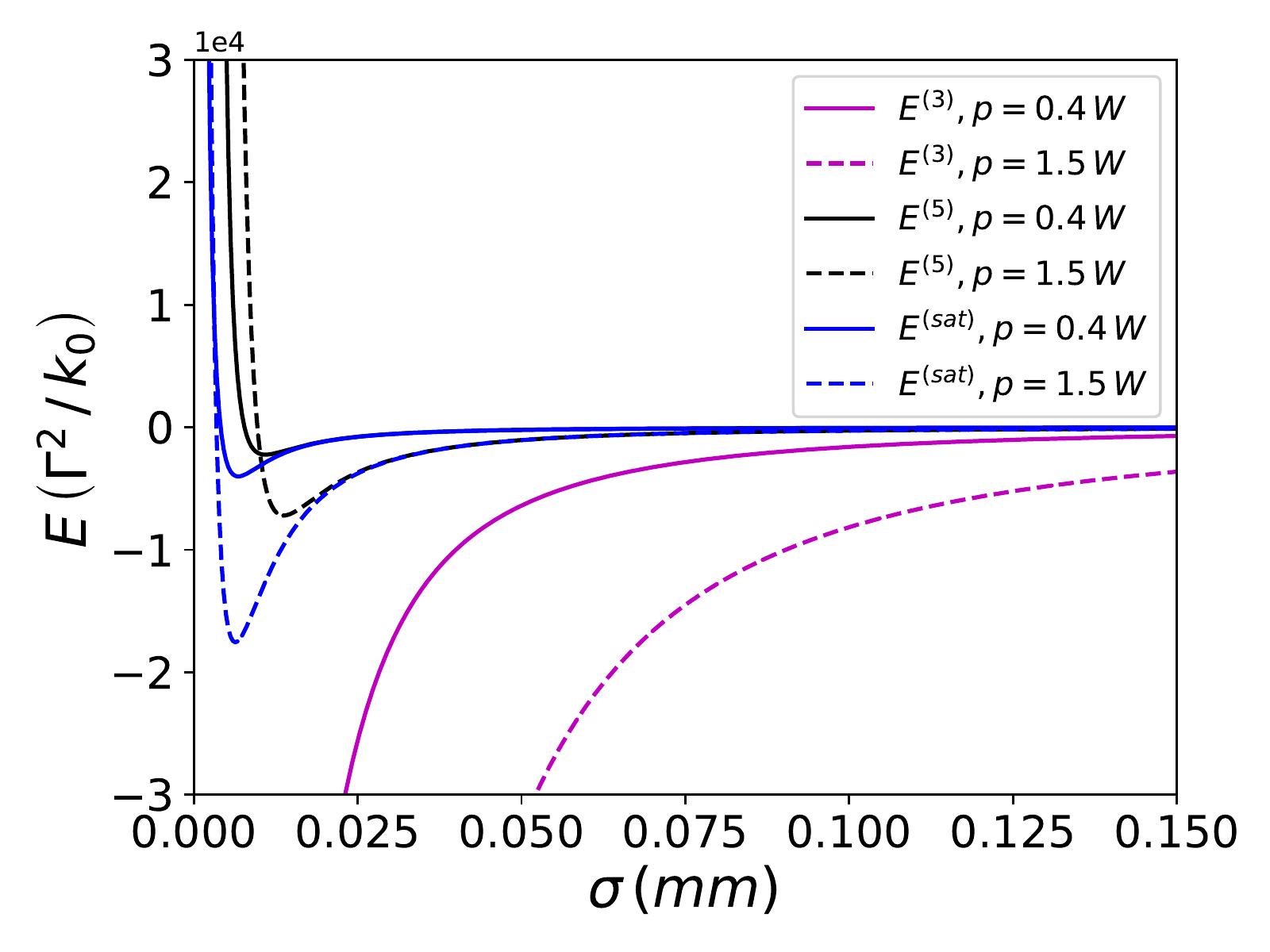}
\caption{\textit{Variational approach to ground state}. Energy computed for the $\chi^{(3)}$ (purple), $\chi^{(3)}+\chi^{(5)}$ (black) and saturating (blue) regimes for $p=0.4$ (solid lines) and $1.5\,W$ (dashed lines), and the frequency detuning equal to $\Delta=2\pi\times 3.0\,$GHz.}
\label{energyCOMP}
\end{figure}

In Fig. (\ref{energyCOMP}), we show the energy as a function of $\sigma$ for the $\chi^{(3)}$, $\chi^{(3)} + \chi^{(5)}$, and saturating regimes for two different values of the incident power $p$ and a detuning $\Delta=2\pi\times 3.0$ GHz. 
We observe that the energy displays a minimum for the $\chi^{(3)} + \chi^{(5)}$ and saturating regimes for a wide range of the powers. 
Similarly, the $\chi^{(3)}$ curve does not hold a minimum: 
The energy either decreases or increases indefinitely, depending on whether the focusing term dominates or the diffraction (kinetic) takes over, respectively. Notwithstanding, for very high values of $\Delta$ the nonlinearity becomes irrelevant when compared to diffraction for all regimes.   

Next, we investigate the values of the width  
$\sigma_{r}\left(z\right)=\sqrt{\left\langle r^{2}\right\rangle -\left\langle r\right\rangle ^{2}}$, a quantity that provides an estimate of the droplet's radius $\sigma$.
We then run numerical simulations of Eqs. (\ref{eqs7c}) and (\ref{nlseChi3Chi5}) using imaginary time evolution to reach the minimum energy state. 
 
\begin{figure}[ht]
\centering
\subfloat{\includegraphics[width=0.9\linewidth]{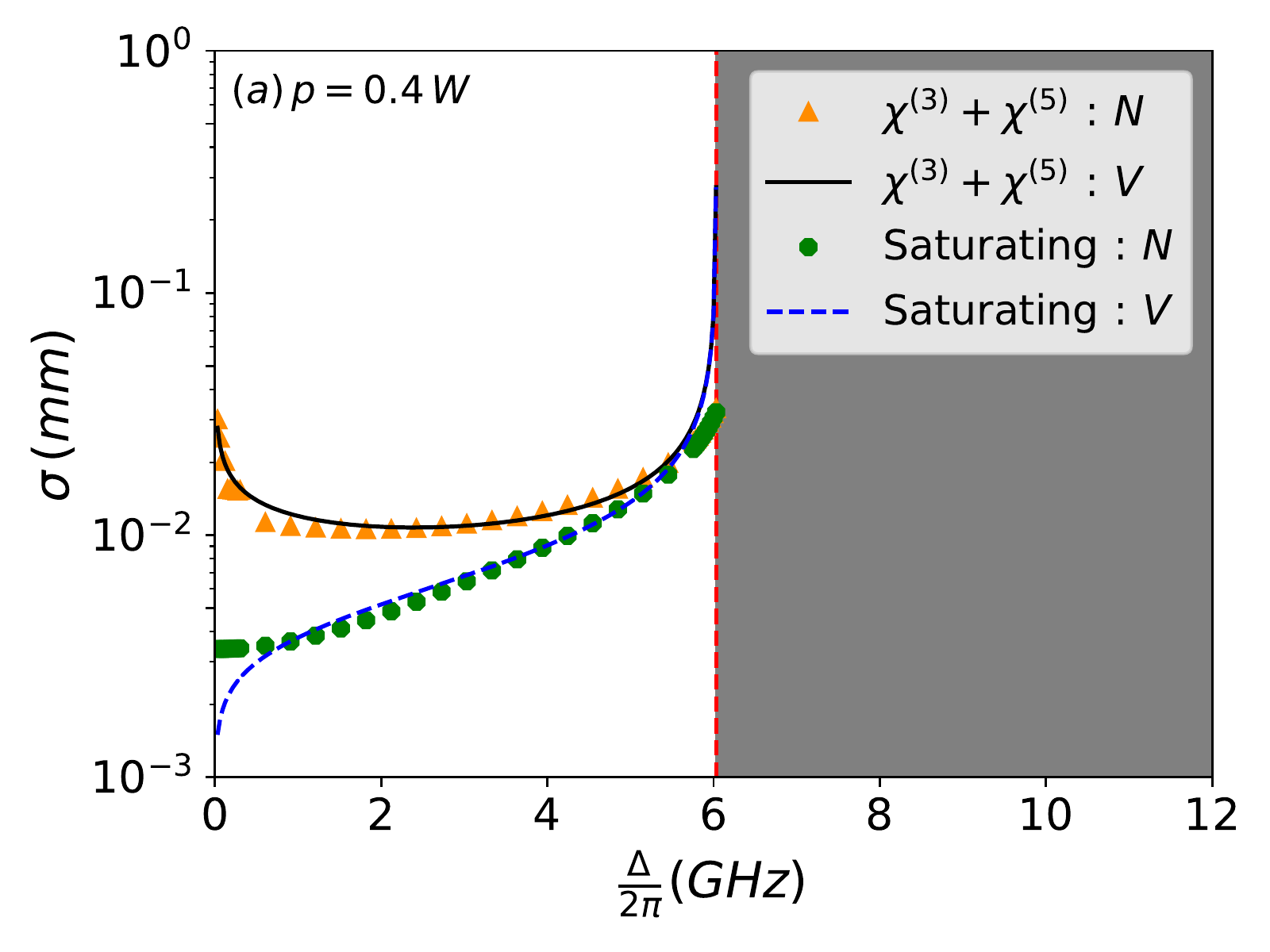}}\\ 
\subfloat{\includegraphics[width=0.9\linewidth]{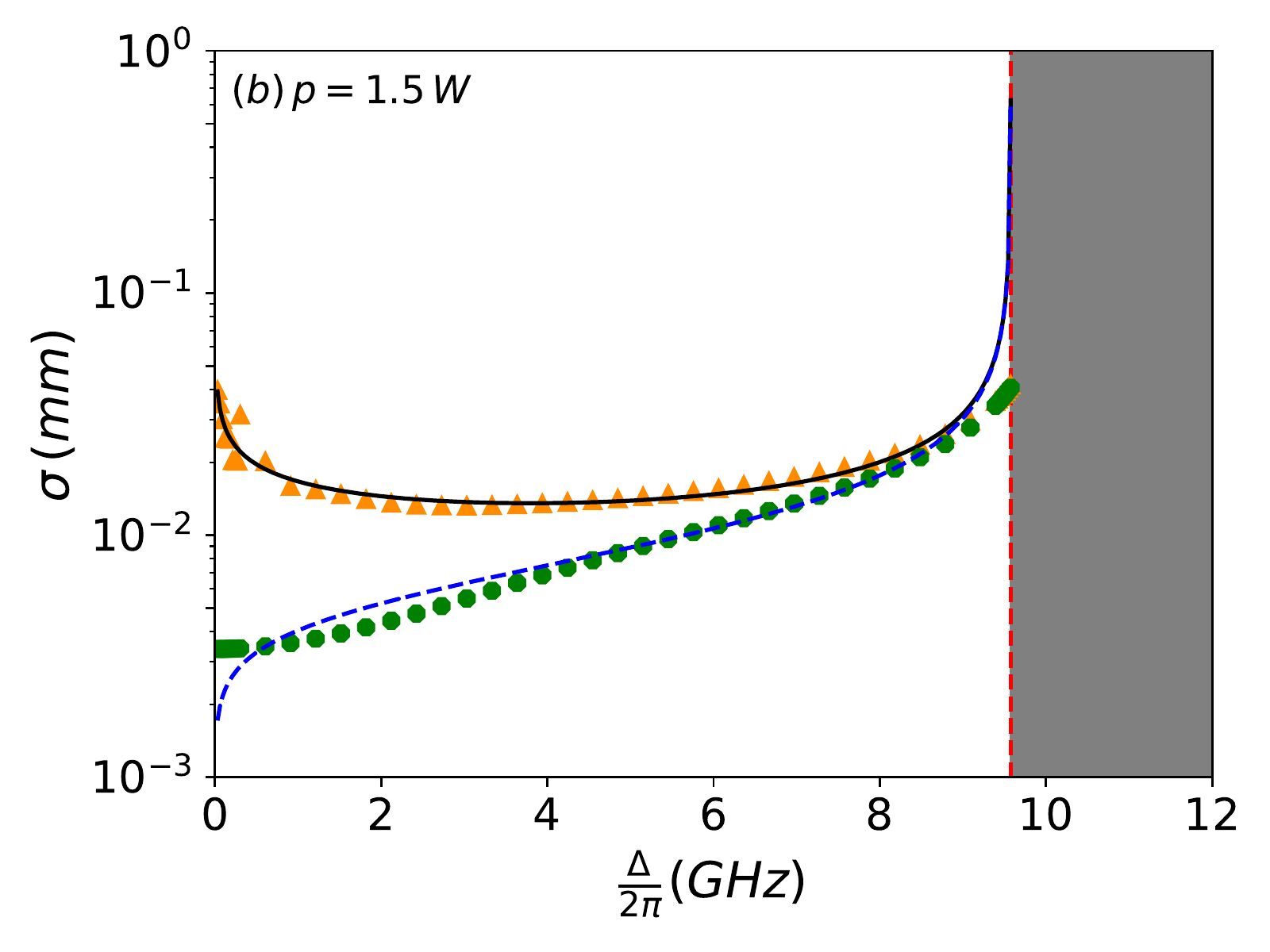}} 
\caption{RMS deviation of the radial position as a function of the frequency detuning $\Delta$. The values of the incident powers are (a) $p=0.4\,$W and (b) $p=1.5\,$W, respectively. The curves with markers refer to the numerical results (N) obtained from the NLSE : orange triangles for the $\chi^{(3)}+\chi^{(5)}$ and green octagons for the saturating regime.  For the variational results (V), the solid black curve represents the $\chi^{(3)}+\chi^{(5)}r$ regime, while the dashed blue line shows the results obtained for the saturating regime.}
\label{sigmachi3chi5Sat70}
\end{figure}

Fig. (\ref{sigmachi3chi5Sat70}) shows the comparison between the variational approach and the numerical results.
There is an upper limit for $\Delta$ which depends on the constraint over the values of $\mathcal{P}$ in Eq. (\ref{sigma3}). This constraint follows from Eq. (\ref{sigma3}) and determines a maximum detuning for a fixed incident power $p$. 
Concurrently, increasing of $\Delta$ makes the nonlinearity weaker, so the droplet states will not sustain. 
The rectangular gray regions in the plots display the forbidden range of values for $\Delta$. For $p=0.4\,$W, the upper limit is for $\Delta \approx 2\pi\times6.0\,$GHz, while for $p=1.5\,$W, the limiting value is $2\pi\times9.6\,$GHz (dashed red lines).

These results reveal a good agreement between the values obtained through the variational Gaussian ansatz and those obtained from the numerical simulations of Eqs. (\ref{eqs7c}) and (\ref{nlseChi3Chi5}). We see that the values for $\sigma$ in the $\chi^{(3)}+\chi^{(5)}$ and saturating regimes will converge for a wide range of increasing values of the frequency detuning, except close to the upper and lower bounds of the frequency detuning. In these two regions, we observe that the beam cannot be approximated by a Gaussian profile, and therefore, we expect a disagreement between the numerical and variational results. 
\subsection{Ground state phase diagram}
\begin{figure*}[t!] 
\includegraphics[width=1.0\linewidth]{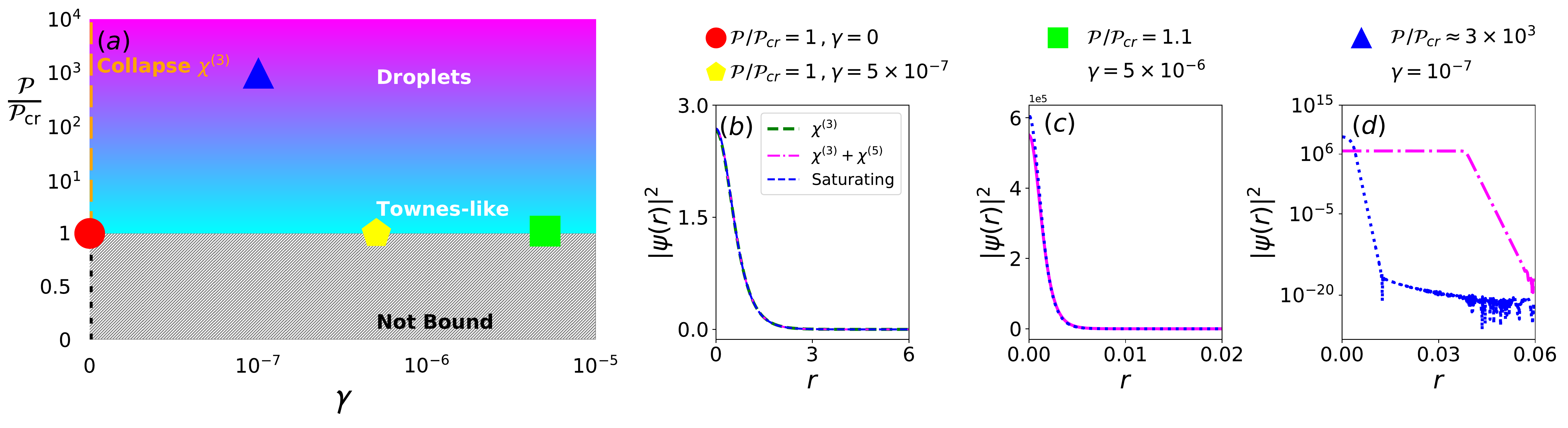}
\caption{\textit{Ground state of the system}. (a) Phase diagram: after an interval in which diffraction dominates over the nonlinear interaction (Not bound region), from cyan to magenta, one transits from the weakly nonlinear regime (Townes soliton) to the strong nonlinear regime (droplets). For the pure $\chi^{(3)}$ regime, if $\mathcal{P}\,/\,\mathcal{P}_{cr}>1$ then one enters the collapsing region as pictured by the orange dotted line for $\gamma=0$. When considering the $\chi^{(3)}+\chi^{(5)}$ and saturating regimes, for any $\gamma > 0$, the collapse is arrested. In the subplots (b-d), we show the intensity profiles $\left|\psi\left(\mathbf{r}\right)\right|^{2}$ for different values of the ratio $\mathcal{P}\,/\,\mathcal{P}_{cr}$ and $\gamma$: (b) $\mathcal{P}\,/\,\mathcal{P}_{cr}=1$, $\gamma=0$ (red circle) and $\gamma=5\times 10^{-7}$ (yellow pentagon); (c) $\mathcal{P}\,/\,\mathcal{P}_{cr}=1.1$, $\gamma=5\times 10^{-6}$ (limegreen square), and (c) $\mathcal{P}\,/\,\mathcal{P}_{cr}\approx 3\times 10^{3}$, $\gamma=10^{-7}$ (blue triangle). }
\label{phase_diagram_GS}
\end{figure*}

We now investigate the phase diagram of the system. 
We run numerical simulations in imaginary time for the different regimes. The results obtained for the intensity profiles (here considering the dimensionless quantities) are shown in the plots of Fig. (\ref{phase_diagram_GS}). 

In two dimensions, it is known that the focusing cubic NLSE admits the Townes solution for a specific value of the dimensionless power that we here denote by $\mathcal{P}_{cr}$, which equals $\mathcal{P}_{cr} = 5.8504$ \cite{FibichGaetaOL2000,fibich2015nonlinear}. The Townes soliton is only one of the stationary solutions that this equation possesses. Higher-order stationary solutions (all of them with $E=0$ alike the Townes soliton) will present nodes in addition to having an associated power greater than $\mathcal{P}_{cr}$~\cite{HAUS,Yankauskas1966}. 
For values of the dimensionless power smaller than $\mathcal{P}_{cr}$, the nonlinear interaction is too weak, and thus the contribution of the transverse Laplacian dominates, leading to a spreading of the intensity profile. 
In the phase diagram, this situation corresponds to the shaded gray region, and it is valid for any value of $\gamma$ (the dotted black line corresponds to $\gamma=0$, that is, the pure Kerr nonlinearity). As $\mathcal{P}$ slightly increases, we eventually reach the critical value for the Townes solution, $\mathcal{P}=\mathcal{P}_{cr}$. This point is represented by the red circle in the phase diagram and by the dashed green line in the subplot (\ref{phase_diagram_GS}b). 
In this point, we have $\mathcal{P}\,/\,\mathcal{P}_{cr}=1$ and $\gamma=0$. Further details on the physics of the Townes solution will be discussed in Section \ref{sec:dynamics}. 
For the yellow pentagon (\ref{phase_diagram_GS}b), 
the higher order nonlinear terms are still irrelevant when compared to the leading order $\chi^{(3)}$ interaction, and the intensity profiles remain the same. However, this picture dramatically changes when the ratio $\mathcal{P}\,/\,\mathcal{P}_{cr}$  increases. 
In this case, the NLSE containing only the Kerr term will lead to collapsing solutions, which is indicated by the vertical dotted orange line along $\gamma=0$. When considering the other regimes, this collapse is arrested and stable configurations can be obtained, as shown in the subplot (\ref{phase_diagram_GS}\,c) for the $\chi^{(3)}+\chi^{(5)}$ and saturating regimes, although the intensity profiles are practically equal for these two situations. For this collapsing region, we did not represent the $\chi^{(3)}$ regime because it would require an extremely fine spatial grid. 
Here, the system suffers a very strong focusing effect, so the peak intensity takes on very high values, which characterizes the collapsing behaviour. 
By considering even higher values of the dimensionless power, we start seeing some differences between the $\chi^{(3)}+\chi^{(5)}$ and saturating regimes. The nonlinearity gets stronger closer to resonance, and concomitantly, we eventually see the formation of flat-top profiles for the $\chi^{3}+\chi^{5}$ regime while its saturating counterpart displays a Gaussian-like shape as shown in subplot (\ref{phase_diagram_GS}\,d).

\section{\label{sec:dynamics}real-time dynamics: general aspects, breathing mode and realistic experimental conditions}
In this section, we investigate the real-time dynamics of a Gaussian beam for the nonlinearities presented above.
In subsection \ref{sec:chi3}, we review the real-time dynamics for the $\chi^{(3)}$ regime and comment on the physics of the Townes soliton. 
We proceed in subsections \ref{sec:breathingmode} and  \ref{sec:realistic} with the investigation of the breathing mode and the real-time dynamics for a Gaussian beam under realistic experimental conditions for both $\chi^{(3)}+ \chi^{(5)}$ and saturating regimes, respectively. 

\subsection{\label{sec:chi3}Review of the Kerr nonlinearity and the dynamics of the Townes soliton ($\chi^{(3)}$ regime)}

We start with the analysis of the cubic NLSE which can be obtained from Eq. (\ref{eqs7c}) for $\gamma\left|\psi\right|^{2}\ll 1$ in zeroth order, yielding
\begin{equation}
    \label{nlseChi3}
    i\frac{\partial\psi}{\partial z^{\prime}}=-\frac{1}{2}\nabla_{\bot}^{\prime \,2}\psi-\left|\psi\right|^{2}\psi\,
\end{equation}
If we assume waveguide solutions of the form $\psi=R(r^{\prime})\exp\left(i\,z^{\prime}\right)$, the stationary equation of motion reduces to 
\begin{equation}
 \label{townes1}
 \frac{1}{2}\nabla^{\prime \,2}R-R+R^{3}=0,\quad R^{\,\prime}\left(0\right)=0,\quad R\left(\infty\right)=0.
\end{equation}
The solution $R\left(r^{\,\prime}\right)$ above a critical power is a monotonically decreasing function, the Townes soliton \cite{ChiaoGarmireTownes64,FibichGaetaOL2000}. 

\begin{figure}[ht]
\centering
\includegraphics[width=0.995\linewidth]{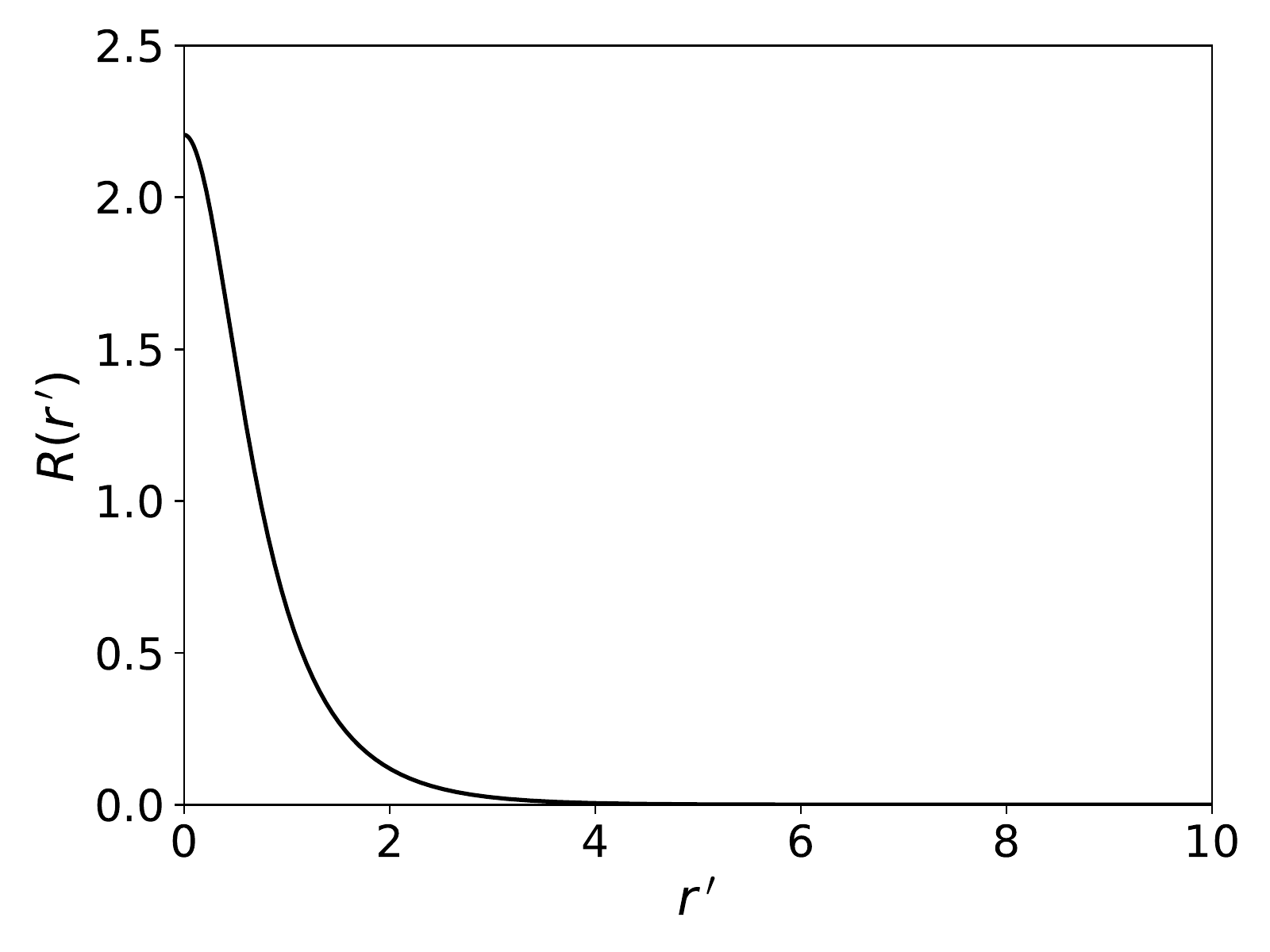}
\caption{\textit{Townes profile}. Numerical solution for the Townes soliton profile, $R(r^{\prime})$, using the shooting method.}
\label{fig:variancechi3}
\end{figure}

Before delving into the physics of the Townes soliton, we review some important aspects of the cubic NLSE.
In the context of cold atoms, solitons were investigated in several experiments~\cite{Donley2001,Salomon,Strecker2002}.
We note that the dimensionality of the system plays a crucial role. To illustrate this, let us consider a wave-packet of size $l$ whose energy functional is given by Eq. (\ref{eqs9a}). In $D$ spatial dimensions, the cubic NLSE leads to a kinetic term that scales as $E_{kin}\propto l^{-2}$ while the interaction term as $E_{int}\propto l^{-D}$. 
For $D=1$ we know that the energy displays a stable minimum and then bright solitons exist for any interaction strength and atom numbers~\cite{Salomon,Strecker2002}. However, for $D=3$, the system is dynamic unstable, and no solitons can be conceived for this specific case of a Kerr nonlinearity~\cite{Donley2001,Eigen}.
Lastly, for $D=2$, the system does not possess a characteristic length scale. A stationary solution is available only for a discrete value of the interaction strength that makes $E_{kin}$ and $E_{int}$ to perfectly balance each other. In our optical system, this value is converted into the critical value of the dimensionless power. This stationary solution is exactly the one obtained by Townes, 
whose energy is zero and its chemical potential is negative.
Solving Eq. (\ref{townes1}) numerically for an input Gaussian beam using the shooting method, we obtain the Townes soliton shown in Fig. (\ref{fig:variancechi3}).

From the numerical solution of Eq. (\ref{townes1}), we computed the dimensionless critical power, $\mathcal{P}_{cr}$, and obtained the value:
\begin{equation}
 \mathcal{P}_{cr}=\int \left|R\right|^{2}d^2\,\mathbf{r^{\,\prime}} = 5.8504.
 \label{pc}
\end{equation}
The solutions of Eq. (\ref{nlseChi3}) do not blow up provided the incident power 
(i.e., $\int \left|\psi\left(r^{\,\prime},0\right)\right|^{2}d^2\,\mathbf{r^{\,\prime}}$) is strictly below $\mathcal{P}_{cr}$ \cite{weinstein1982nonlinear}.

\begin{figure}[ht!]
\centering
\includegraphics[width=0.995\linewidth]{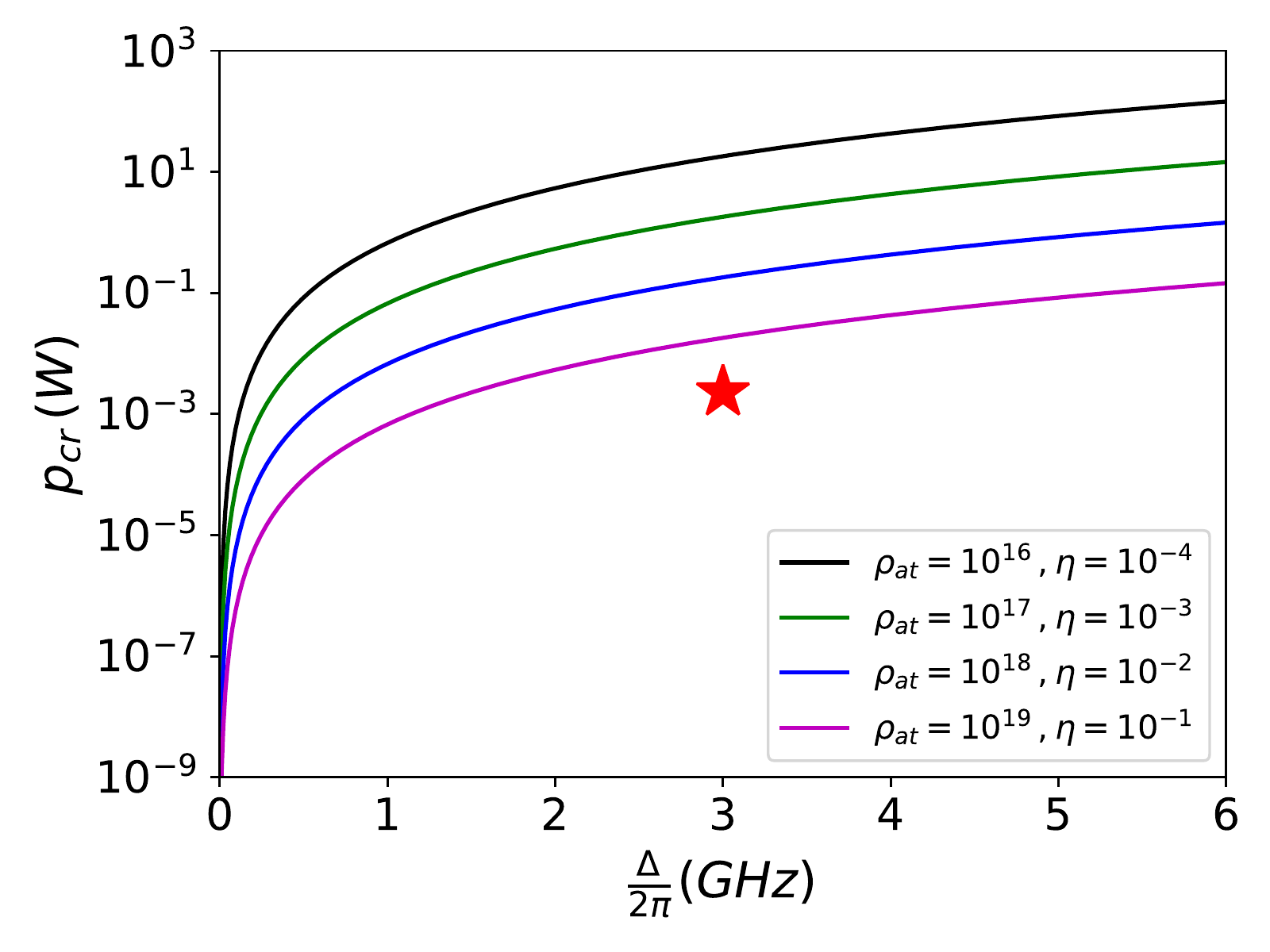}
\caption{\textit{Matching the Townes soliton condition}. Critical power as a function of the frequency detuning $\Delta$. Each curve is obtained for different values of the dimensionless parameter $\eta$, and consequently of the atom density, as identified in the labels.
The star point is computed at $\Delta = 2\pi\times 3.0\,$GHz and $\rho_{at}= 8.3\times10^{19}\,m^{-3}$ (see main text).} 
\label{townespower}
\end{figure}

With our simplified two-level description, we can estimate experimental accessible parameters for realizing a Townes soliton in a hot vapour setup. Restoring the units, and using the relation between the Rabi frequency and the intensity, we derive an expression for the dimensional critical power which matches the condition given in Eq. (\ref{pc}).  In doing so, we obtain
\begin{equation}
     p_{cr}=8\,I_{sat}\,\left(\frac{5.8504}{k_{0}^{2}\,\eta}\right)\left\{\frac{\Delta/\Gamma}{\left[\left(\frac{\Delta}{\Gamma}\right)^{2}+\frac{1}{4}\right]^{2}}\right\}^{-1}\,,
 \label{pc8}
 \end{equation}
where $I_{sat}$ is the saturation intensity. 
In Fig. \ref{townespower}, we show a plot of $p_{cr}$ as a function of $\Delta$ for different values of the prefactor $\eta$ (see Eq. (\ref{eqs25})) which, in turn, depends on the vapour density.
As an example, let us consider $\Delta = 2\pi\times 3.0\,$GHz and $\rho_{at}= 8.3\times10^{19}\,m^{-3}$. For this case, a power close to $2.3\,$mW would be needed to match the Townes condition given in Eq. (\ref{pc}). This configuration is represented by the red star in the plot. We computed the value of the nonlinear refractive index $n_2$ for the same frequency detuning, saturation intensity and atom density, and obtained $n_2 = 1.2\times 10^{-10}$m$^{2}\,/\,$W, in good agreement with the experimental result obtained in ~\cite{pierreThesis}.
In general, we observe that $p_{cr}$ grows as the value of the prefactor $\eta$ decreases. In other words, the critical experimental power has higher values as the atom density of the system decreases.

\begin{figure}[ht!]
\centering
\subfloat{\includegraphics[width=0.85\linewidth]{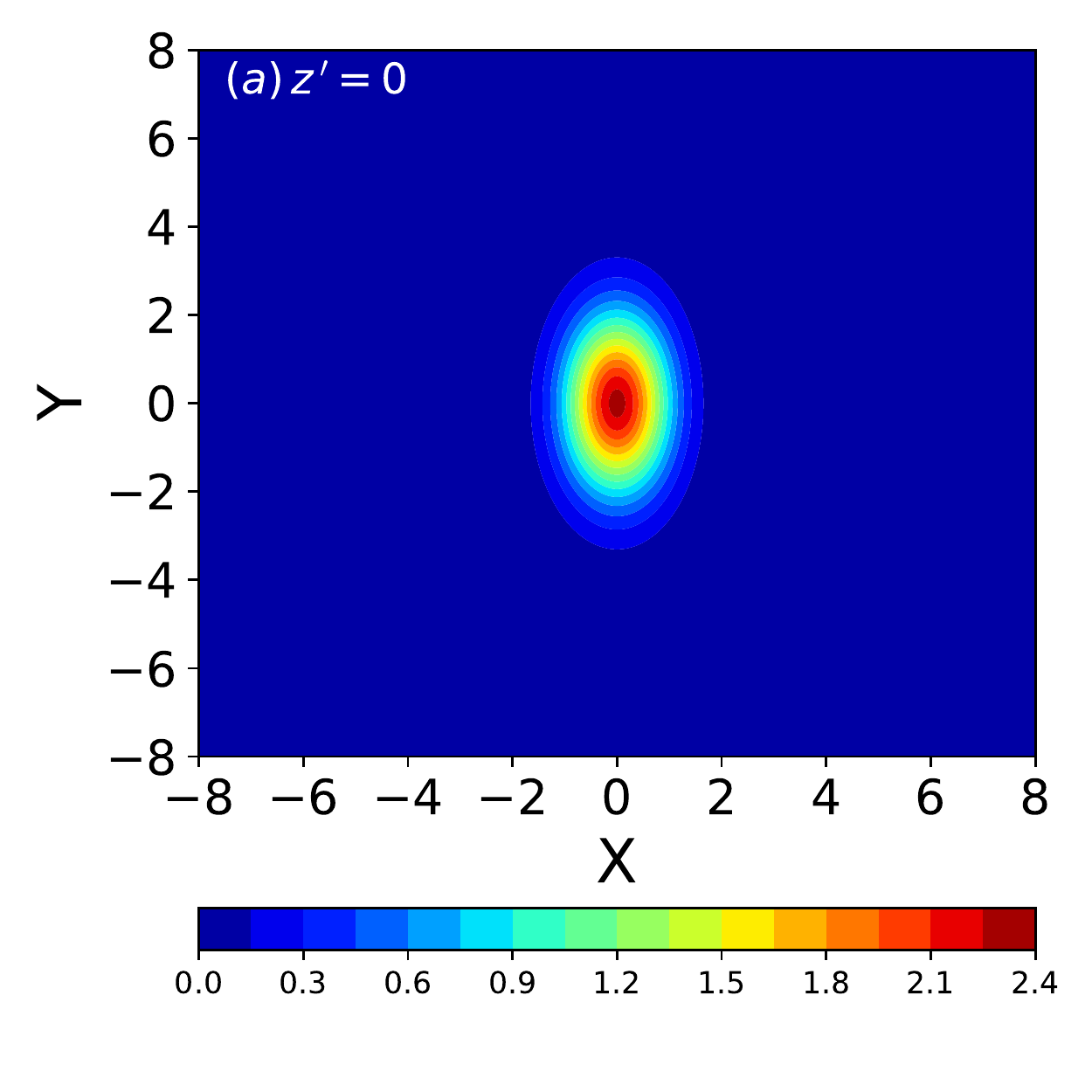}}\\
\subfloat{\includegraphics[width=0.85\linewidth]{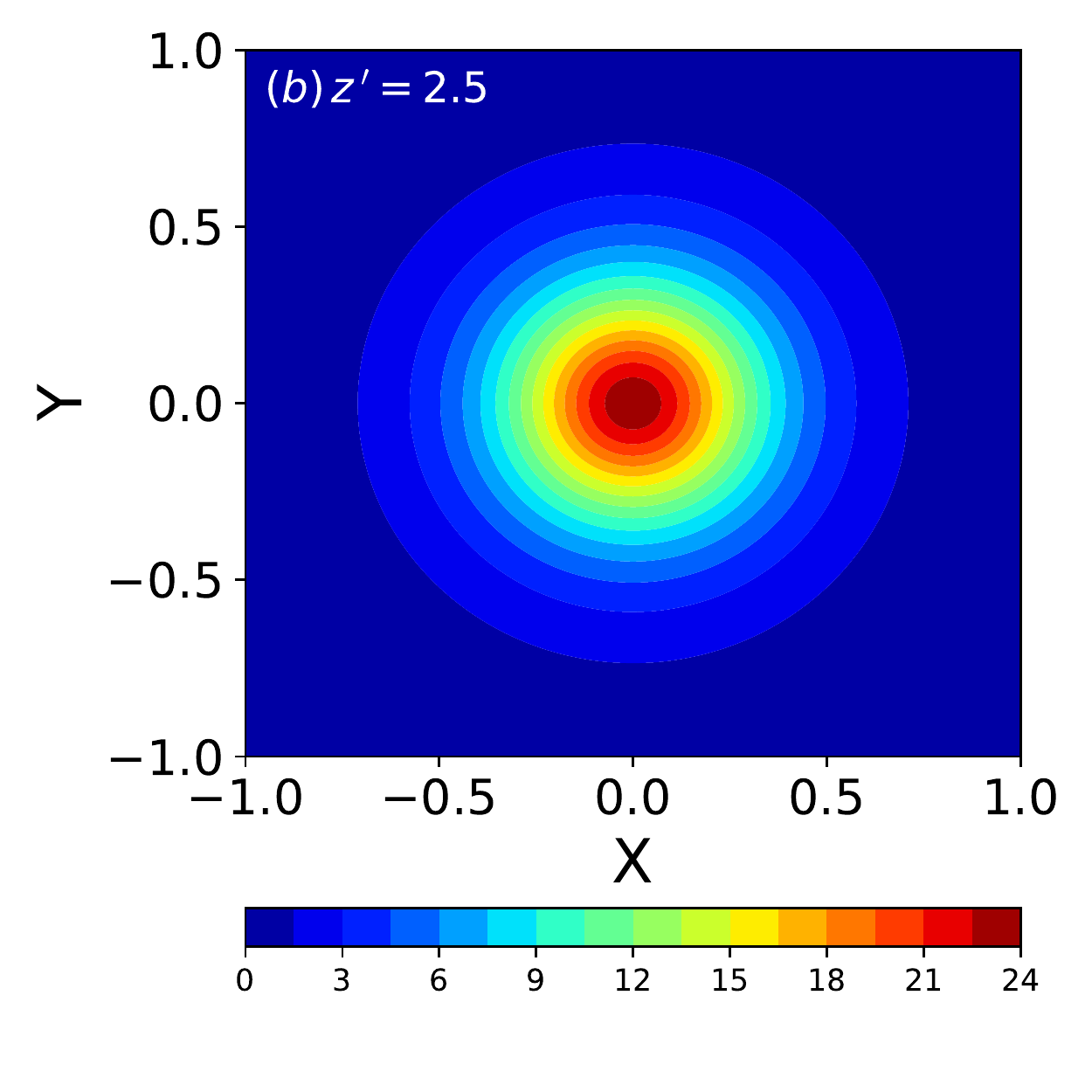}}
\caption{\textit{$\chi^{(3)}$ regime}. Propagation of an elliptically
shaped input beam ($z^{\,\prime}=0$) (a) as it evolves with the effective interaction time. As the beam propagates, it self-focuses, and a circularly symmetric Townes profile is formed (b) at an intermediate state ($z^{\,\prime}=2.5$) as the beam collapses. To guarantee a better visibility of the Townes profile, the part (b) has been
zoomed.}
\label{TownesFibichGaeta}
\end{figure}

An additional interesting feature of the cubic NLSE is that for self-focusing beams, the dynamic evolution naturally makes the initial configuration to evolve towards the Townes profile regardless of the initial shape of the beam~\cite{ChiaoGarmireTownes64}. A neat example is the case of an elliptically shaped input beam, verified experimentally in \cite{MollGaetaFibich}. We consider the following elliptical input beam:
\begin{equation}
 \label{ellipticalTownes}
 \psi \left(x\,, y\right) = \sqrt{\frac{\mathcal{P}}{2\pi}}\exp\left(-\frac{ x ^{2}}{8}-\frac{ y ^{2}}{2}\right),
\end{equation}
with scaled power $\mathcal{P}=14.5$, well above the critical value for beams with this shape. The plots of the intensity profile (I) in Fig. (\ref{TownesFibichGaeta}) display the initial and the intermediate state at $z^{\prime}=2.5$  of the real-time evolution, respectively. In this case, the spatial profile of the collapsing elliptical input beam evolves to the circularly symmetric shape profile which characterizes the Townes soliton. Recently, the self-similar evolution related with the Townes soliton physics has been investigated in BECs with two components~\cite{bakkali2021realization} and through the use of a Feshbach resonance~\cite{chen2021observation}. Another remarkable property of the Townes soliton is the scale-invariance, recently verified in a $2$D Bose gas~\cite{Cheng2021}. 
From a given stationary solution $R\left(r^{\prime}\right)$, we can build a family of solutions with the same critical power through a dilation operation~\cite{fibich2015nonlinear}. Higher nonlinearities explicitly break scale-invariance. Nevertheless, in the limit where the stationary solution is characterized by a power slightly above the critical value, the higher-order nonlinearities can be neglected, and the solutions resemble the Townes soliton profile for a long propagation.

Finally, we investigate some aspects related to the collapsing behaviour of this solution: an exact Townes shape remains stable in time evolution, but perturbations in shape will lead to unstable solutions. Following the approach in \cite{BERGE1998259}, we derive an expression for the critical distance for the collapsing of the wave packet. 

Consider the following ansatz for the wave function $\psi\left(\mathbf{r},z\right)$,
\begin{equation}
    \label{wc2mt}
    \psi\left(\mathbf{r},z\right)=A\left(z\right)L\left(\frac{r}{a\left(z\right)}\right)\exp\left[i\theta\left(r,z\right)\right]\,,
\end{equation}
where $A\left(z\right)$ is a complex-valued amplitude, $a\left(z\right)$ is the wave radius and $L\left(r/a\left(z\right)\right)$ is a real function describing the profile. 
The general expression for $a\left(z\right)$ can be found by following a variational procedure which involves obtaining the Lagrangian and later the equations of motion of the system, and solving them for the variable $a\left(z\right)$. Following this procedure, we obtained the following expression:
\begin{equation}
    \label{wc8mt}
    a\left(z\right)=a_{0}\left[\left(z\sqrt{2\mu}\right)^2\left(1-\nu/\mu\right)+1\right]^{1/2}
\end{equation}
where $a_{0}\equiv a\left(0\right)\neq 0$ is the initial wave radius  and $\dot{a}\left(0\right)=0$. The parameters $\mu$ and $\nu$ are given in terms of the initial conditions and integrals of the profile $L\left(r/a\left(z\right)\right)$. 
See appendix \ref{wave_colapse}  for a detailed derivation.

From Eq. (\ref{wc8mt}) we determine the collapse distance
\begin{equation}
    \label{wc9mt}
    z_{cr}=\frac{1}{\sqrt{2\left(\nu-\mu\right)}}\,.
\end{equation}
We provide explicit results for 
two different test functions: a Gaussian form (GS) and the hyperbolic secant (HS) 
\begin{equation}
    L\left(\frac{r}{a\left(z\right)}\right)=
    \begin{cases}
        \exp\left(-\frac{r^{2}}{2a^{2}}\right)\,, & \text{GS}\\
        \sqrt{2}\sech\left(\frac{r}{a}\right)\,. & \text{HS}
    \end{cases}
\end{equation}

\begin{figure}[ht!]
\centering
\centering
\includegraphics[width=0.995\linewidth]{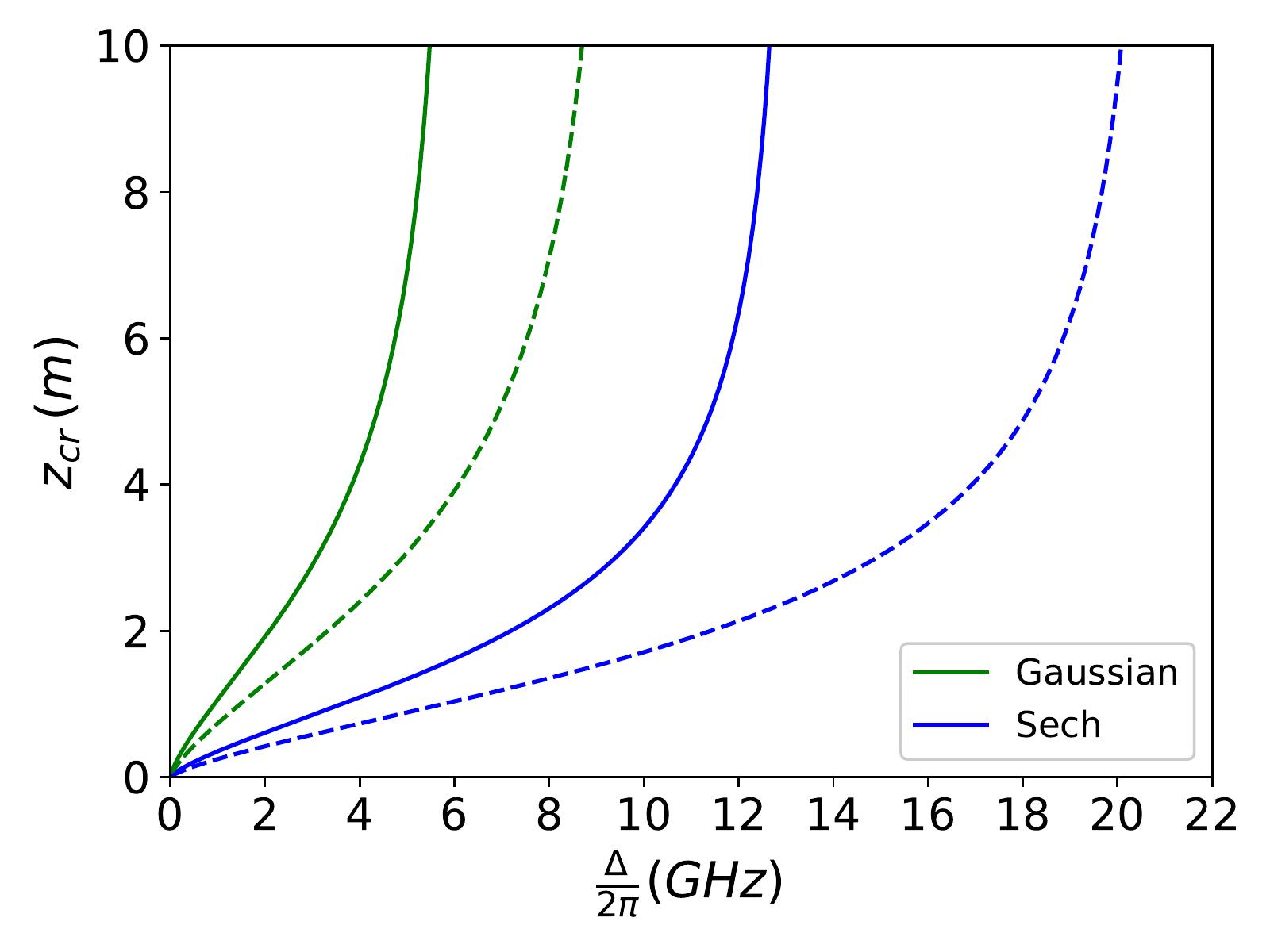}
\caption{\textit{Collapse analysis}. Critical distance $z_{cr}$ as a function of the frequency detuning $\Delta$ considering different profiles for the test function $L\left(r/a\left(z\right)\right)$. The solid curves correspond to the incident power $p=0.4\,$W while the dashed ones to $p=1.5\,$W.}
\label{criticaldistancechi3}
\end{figure}

For each of these test functions, we computed the integrals for the coefficients $\mu$ and $\nu$, leading to the following expressions for $a\left(z\right)$ and $z_{cr}$:

\begin{eqnarray}
    a\left(z\right)=
    \begin{cases}
        a_{0}\sqrt{\frac{z^2}{a_{0}^8}\left(1-\frac{\mathcal{P}a_{0}^{2}}{2\pi}\right)+1}\,, & \text{GS}\nonumber\\
        \nonumber\\
        \frac{a_{0}}{3}\sqrt{\frac{2z^2}{\zeta\left(3\right)a_{0}^8}\left[\frac{2\left(1+\ln 4\right)}{3}-\frac{\mathcal{P}\left(1+\ln 16\right)a_{0}^{2}}{\pi\ln 2}\right]+1}\,, & \text{HS}\nonumber\\
    \end{cases}\\
\end{eqnarray}
In Fig. (\ref{criticaldistancechi3}) we show the critical distance $z_{cr}$ as a function of the frequency detuning. The values considered for the incident power are equal to $p=0.4\,$ and $1.5\,$W while the initial wave radius $a_{0}$ was chosen to be equal the beam waist, $w_{0}=7\times 10^{-4}\,$m. 
The results show that the collapse distance $z_\text{cr}$ decreases upon increasing the incident power. Also, we observe that the detuning range is consistent with the allowed ranges of Fig. (\ref{sigmachi3chi5Sat70}). 

\subsection{\label{sec:breathingmode}
The dynamics of the breathing mode}

In this section, we analyse the breathing mode in the droplet dynamics. 
In the context of ultracold Bose-Bose mixtures, there have been recent studies aiming at understanding the relevant processes involved in the formation of metastable droplets from out-of-equilibrium mixtures~\cite{SemeghiniTerradas,PhysRevE.102.062217,Sachdeva,FortModugno}. 
In certain regimes, droplets cannot sustain any discrete excitation since all the excited energy states are higher in energy than the particle emission threshold, hence the name self-evaporation process. Because of that, the droplet is able to dissipate any excess of energy by releasing atoms or breaking up into smaller pieces. 

For the optical system considered in this work, we derive the analytical expressions for the chemical potential for both the $\chi^{3}+\chi^{5}$ and saturating regimes.
Considering the Gaussian variational ansatz introduced in Section \ref{sec:energy_functionals}, the breathing frequency $\omega$ is obtained by expanding the energy in the vicinity of $\sigma^{\prime}=\sigma^{\prime}_{c}$. 

For the $\chi^{3}+\chi^{5}$ regime, one can show that
\begin{equation}
\label{breathing3}
    \omega=\sqrt{-\frac{81\pi\left(2\pi-\mathcal{P}\right)^{3}}{256\mathcal{P}^{3}\gamma^{2}}}
\end{equation}
with $\mathcal{P}>2\pi$. 
The chemical potential is obtained from the time-dependent NLSE given in Eq. (\ref{nlseChi3Chi5}) which results in
\begin{widetext}
\begin{equation*}
\mu^{(5)} = \left[\frac{1}{2}\int \left|\nabla\psi\left(\mathbf{r}^{\prime},z^{\prime}\right)\right|^{2}d^{2}\mathbf{r}^{\prime}-\int \left|\psi\left(\mathbf{r}^{\prime},z^{\prime}\right)\right|^{4}d^{2}\mathbf{r}^{\prime}+2\gamma\int \left|\psi\left(\mathbf{r}^{\prime},z^{\prime}\right)\right|^{6}d^{2}\mathbf{r}^{\prime}\right]\Bigg/\int \left|\psi\left(\mathbf{r}^{\prime},z^{\prime}\right)\right|^{2}d^{2}\mathbf{r}^{\prime}
\end{equation*}
\end{widetext}
In the Gaussian approximation of Eq. (\ref{eqs9}), we obtain
\begin{equation}
 \label{breathing4}
\mu^{(5)} = \frac{1}{2\sigma^{\prime\,2}}-\frac{\mathcal{P}}{2\pi\sigma^{\prime\,2}}+\frac{2\gamma\mathcal{P}^{2}}{3\pi^{2}\sigma^{\prime\, 4}}\,. 
\end{equation}
For the saturating regime, an analytical expression for $\omega$ is not available. The chemical potential is computed from Eq. (\ref{eqs7c})
\begin{widetext}
\begin{equation*}
\mu^{(sat)} = \left[\frac{1}{2}\int \left|\nabla\psi\left(\mathbf{r}^{\prime}\right)\right|^{2}d^{2}\mathbf{r}^{\prime}-\int \frac{\left|\psi\left(\mathbf{r}^{\prime}\right)\right|^{4}}{1+2\,\gamma\left|\psi\left(\mathbf{r}^{\prime}\right)\right|^{2}}d^{2}\mathbf{r}^{\prime}\right]\Bigg/\int \left|\psi\left(\mathbf{r}^{\prime}\right)\right|^{2}d^{2}\mathbf{r}^{\prime}.
\end{equation*}
\end{widetext}
Within the Gaussian approximation the expression above reads
\begin{equation}
 \label{breathing5}
\mu^{(sat)} = \frac{1}{2\sigma^{\prime\,2}}-\frac{1}{2\,\gamma}+\frac{\pi\sigma^{\prime\,2}}{4\,\gamma^2\,\mathcal{P}}\ln\left(1+\frac{2\,\gamma}{\pi\sigma^{\prime\,2}}\mathcal{P}\right)\,. 
\end{equation}
With these expressions, we define the range of parameters to observe the breathing mode or the self-evaporation.

To identify the behaviours of interest, we look at the cases in which $\omega/\left|\mu\right|\geq 1$ and $\omega/\left|\mu\right|< 1$. 
The former indicates the region where no monopole excitation can be observed, while the latter is the case where monopole excitations are present. 
In the numerical simulations, the breathing mode is excited by changing the intensity, i.e., $\left|\psi\right|^{2}$ of the ground state by a factor of $1.05$. 
In the context of a BEC, this would be equivalent to a slight increase of the particle number.
Subsequently, the frequency $\omega$ was obtained numerically through the Fourier analysis of the droplet width $\sigma_{r^{\prime}}(z)$. In order to avoid spurious reflections of the wave function that may take place at the boundary of the computational domain, we used absorbing boundary conditions.

\begin{figure}[ht!]
\centering
\includegraphics[width=0.995\linewidth]{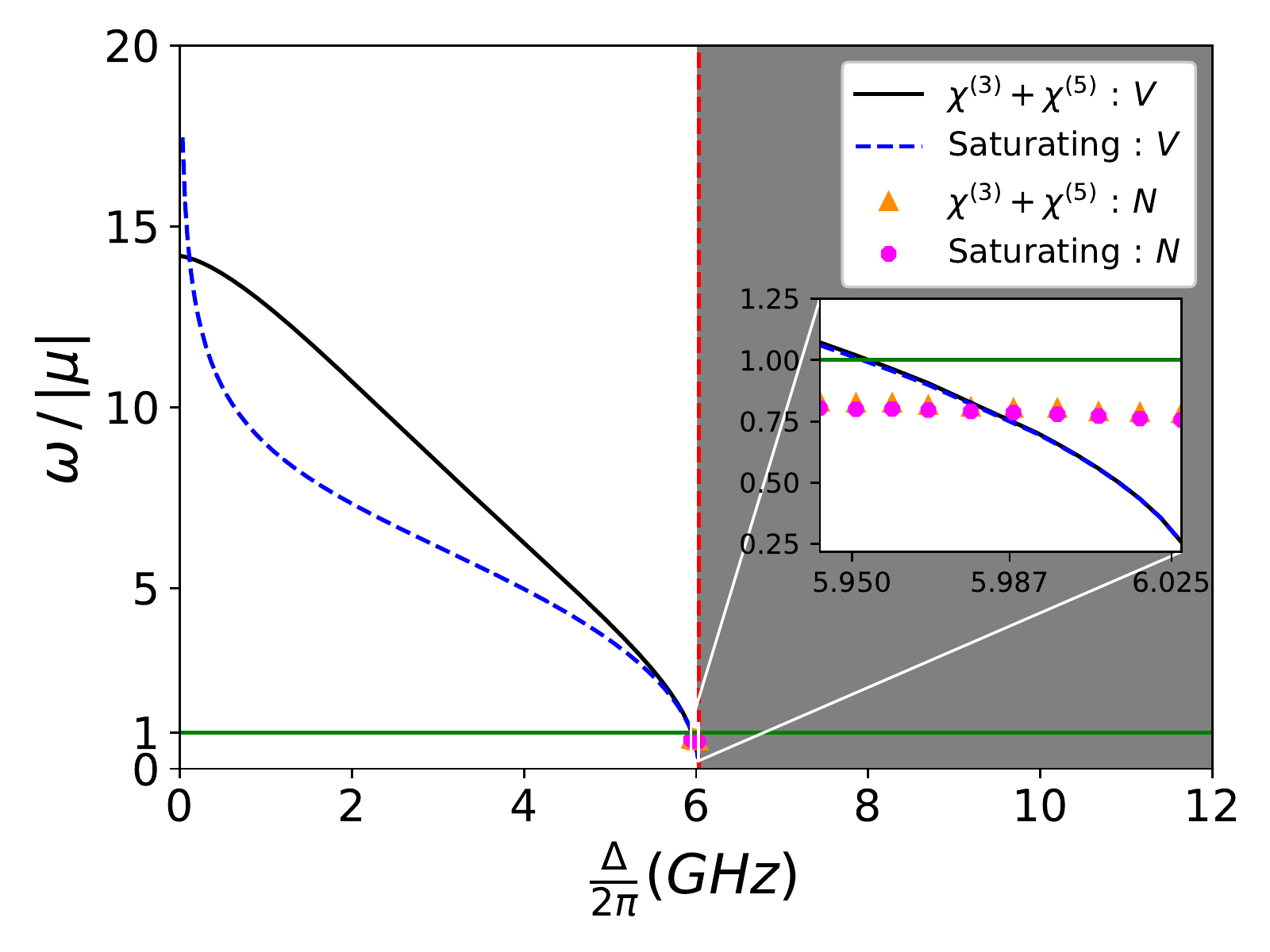}
\caption{Ratio between the excitation energy $\omega$ and the absolute value of the chemical potential $\left|\mu\right|$ for the $2-$ (a) and $4-$level (b) systems. The value of the incident power is $p=0.4\,W$. The curves with markers refer to the numerical results (N) while the variational (V) ones are those given by the solid black $\left(\chi^{(3)} + \chi^{(5)}\right)$ and dashed blue (saturating) curves. The inset shows the narrow region where $\omega/\left|\mu\right|< 1$, which represents the sector where the breathing mode can be found.}
\label{ratio_omegachempot}
\end{figure}

In the plot of Fig. (\ref{ratio_omegachempot}), we display the ratio of the breathing frequency and the chemical potential for different regimes, following the predictions from the Gaussian ansatz and the results obtained from numerical simulations at an incident power $p=0.4\,$W. 
\begin{figure}[ht!]
\centering
\subfloat{\includegraphics[width=1.0\linewidth]{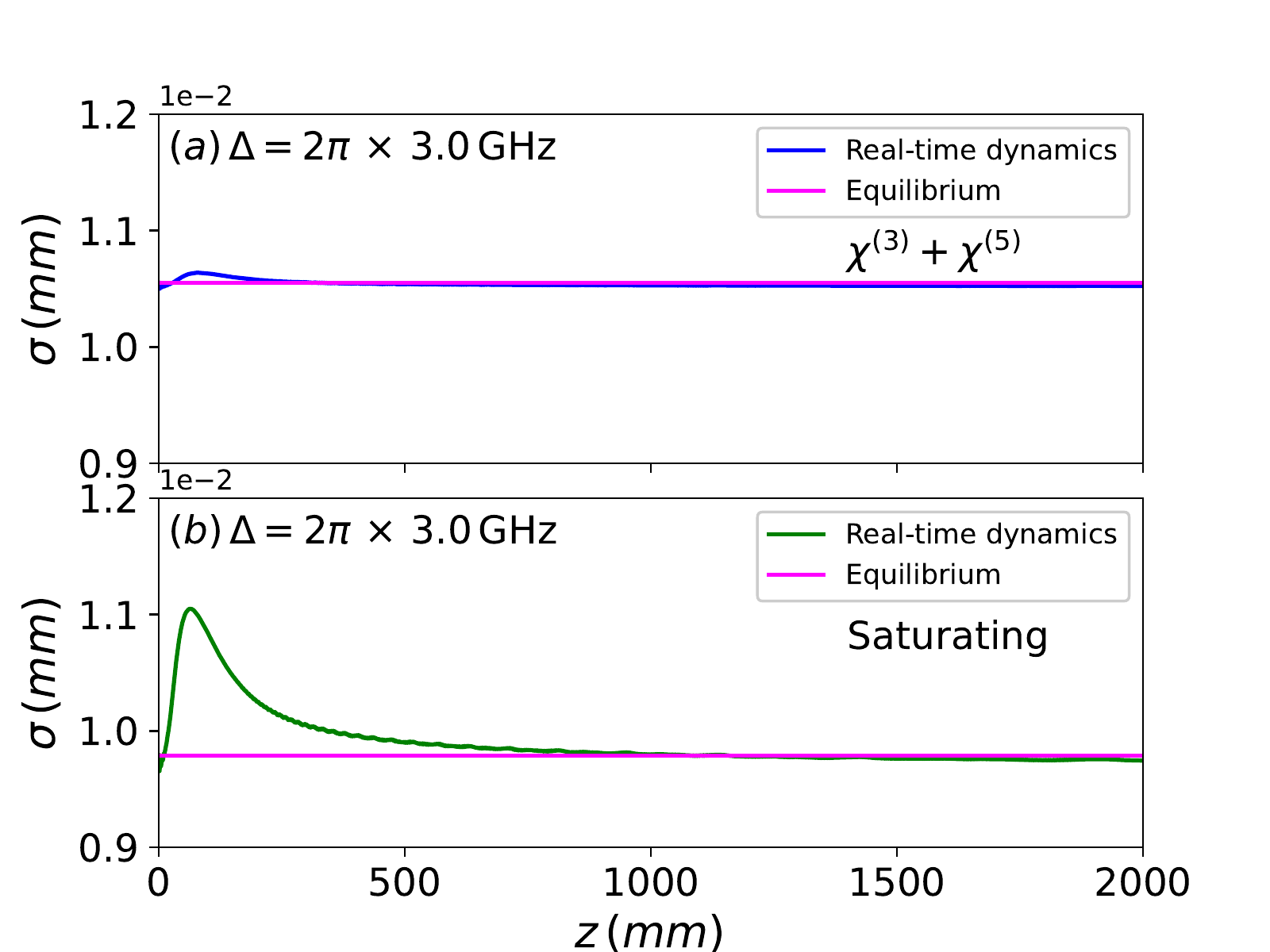}}
\\
\subfloat{\includegraphics[width=1.0\linewidth]{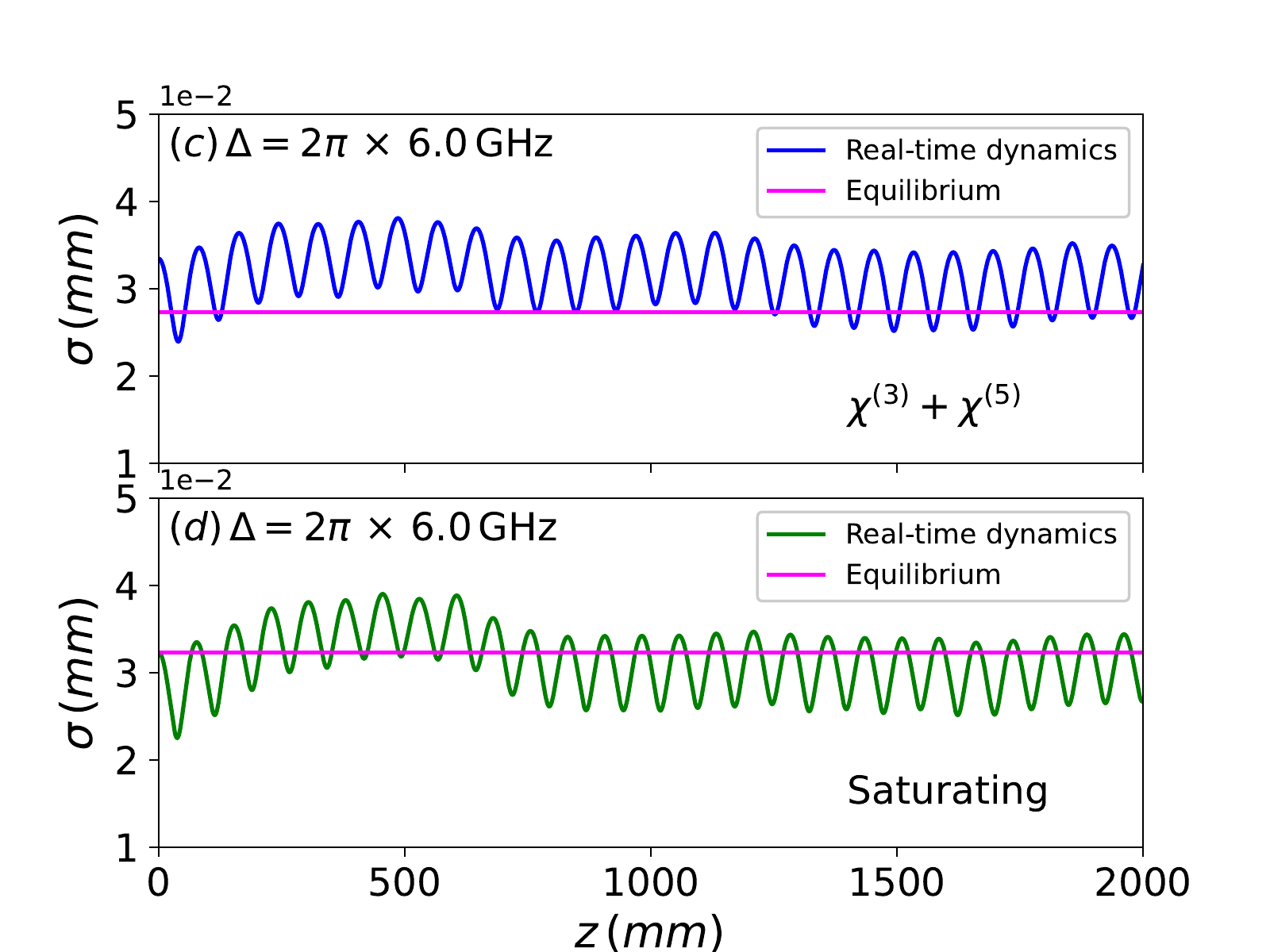}}
\caption{Evolution of the droplet width, $\sigma_{r}(z)$, for the $2-$level system for both the $\chi^{(3)} + \chi^{(5)}$ (solid blue curves, subplots (a) and (c)) and saturating (solid green curves, subplots (b) and (d)) regimes. The horizontal lines in magenta show the stationary equilibrium values. The power is chosen to be $p=0.4\,W$.}
\label{droplet_width_NOR}
\end{figure}
The system does not support a breathing mode, as the excitation energy is greater than the chemical potential for a wide range of allowed values of $\Delta$. 
There, the droplet width does not undergo sinusoidal oscillations, see Figs. (\ref{droplet_width_NOR} a) and (\ref{droplet_width_NOR} b) for the $\chi^{(3)}+\chi^{(5)}$ and saturating regimes, respectively: $\sigma_{r}(z)$ has an initial increase and then decays approaching the value of equilibrium, that is, the droplet width for the ground state.
This is the behaviour for the entire window of $\Delta$ in which $\omega/\left|\mu\right|\geq 1$.

However, there will still be a very limited region which is characterized by the values of $\Delta$ whose curves are below the green horizontal line which represents the case that $\omega=\left|\mu\right|$ and this can be seen with more details in the inset of the plot in Fig. (\ref{ratio_omegachempot}). In this very narrow region, $\sigma_{r}(z)$ shows sinusoidal oscillations, as can be seen in Figs. (\ref{droplet_width_NOR}c) and (\ref{droplet_width_NOR}d). 

\begin{figure}[ht!]
\centering
\includegraphics[width=0.995\linewidth]{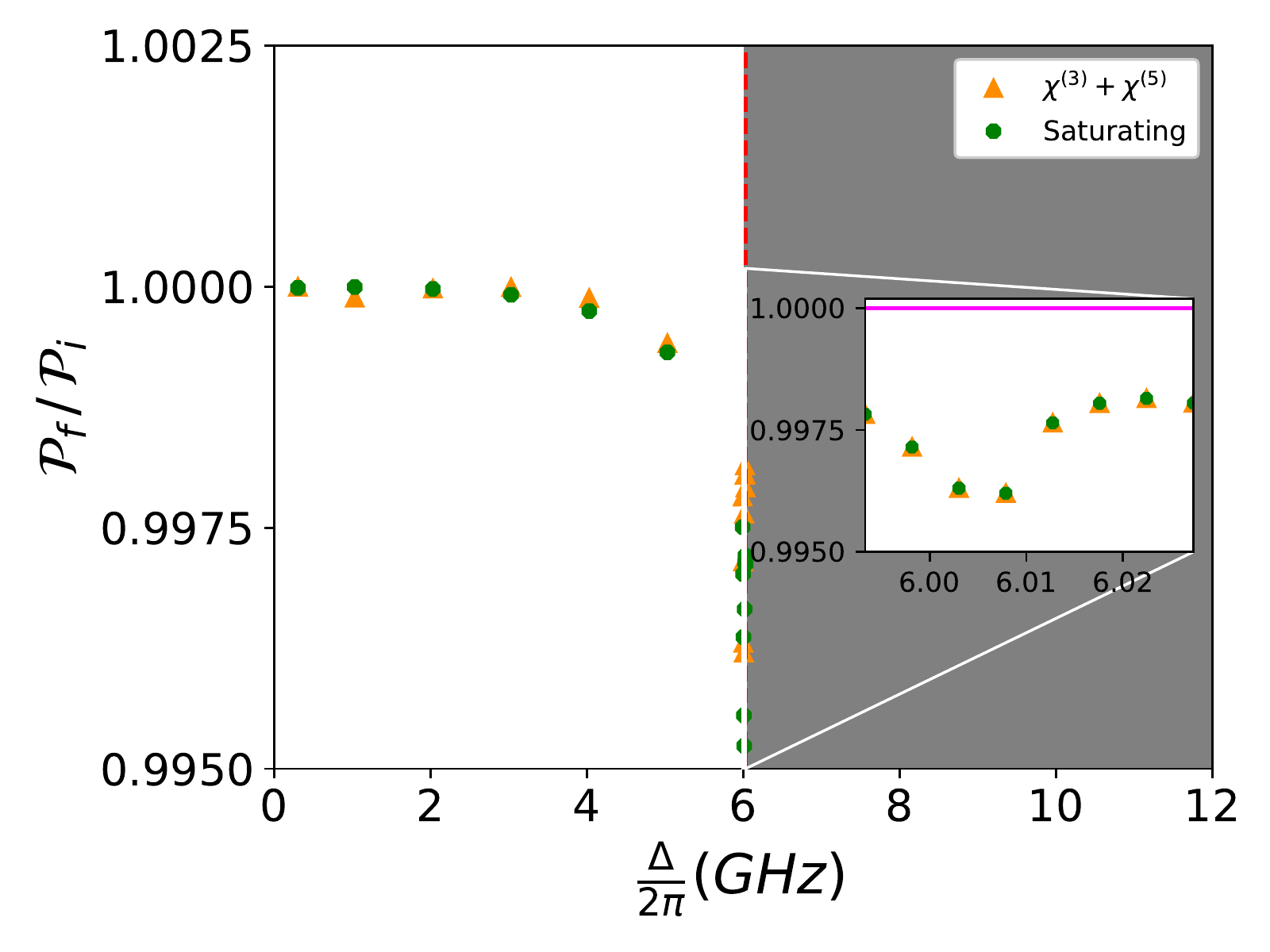}
\caption{Fraction of power lost due to the self-evaporation mechanism for the $2-$ (a) and $4-$level (b) systems obtained from the numerical simulations of the NLSE. The value of the incident power is $p=0.4\,W$. The inset shows the zoomed region where more points were analysed, and for which the Gaussian variational approach predicts to be within the region in which the breathing mode can exist.}
\label{fraction_selfevap}
\end{figure}

Finally, we turn our attention to the self-evaporation mechanism by looking at the fraction of power that is lost, so the system can then relax to its equilibrium state. The plot in Fig. (\ref{fraction_selfevap}) show the numerical results for the ratio between the final $\left(\mathcal{P}_{f}\right)$ and the initial power $\left(\mathcal{P}_{i}\right)$, revealing that a tiny fraction of power is lost through self evaporation in this regime.   

\subsection{\label{sec:realistic}Dynamics for a Gaussian input beam under realistic experimental conditions}

We now analyze the dynamics for the realistic experimental case for a Gaussian input beam. 

We show the results obtained from $2$D numerical simulations of the Eqs. (\ref{eqs7c}) and (\ref{nlseChi3Chi5}) for the intensity profiles of the system setting $p=0.4\,$W,  frequency detunings $2\pi\times 3.0\,$GHz and $2\pi\times 6.0\,$GHz, and initial beam waist $0.1\,$mm. We let the system evolve for a distance $z_\text{max}\approx 2.0\,$m.

\begin{figure}[ht!]
\centering
\subfloat{\includegraphics[width=0.47\linewidth]{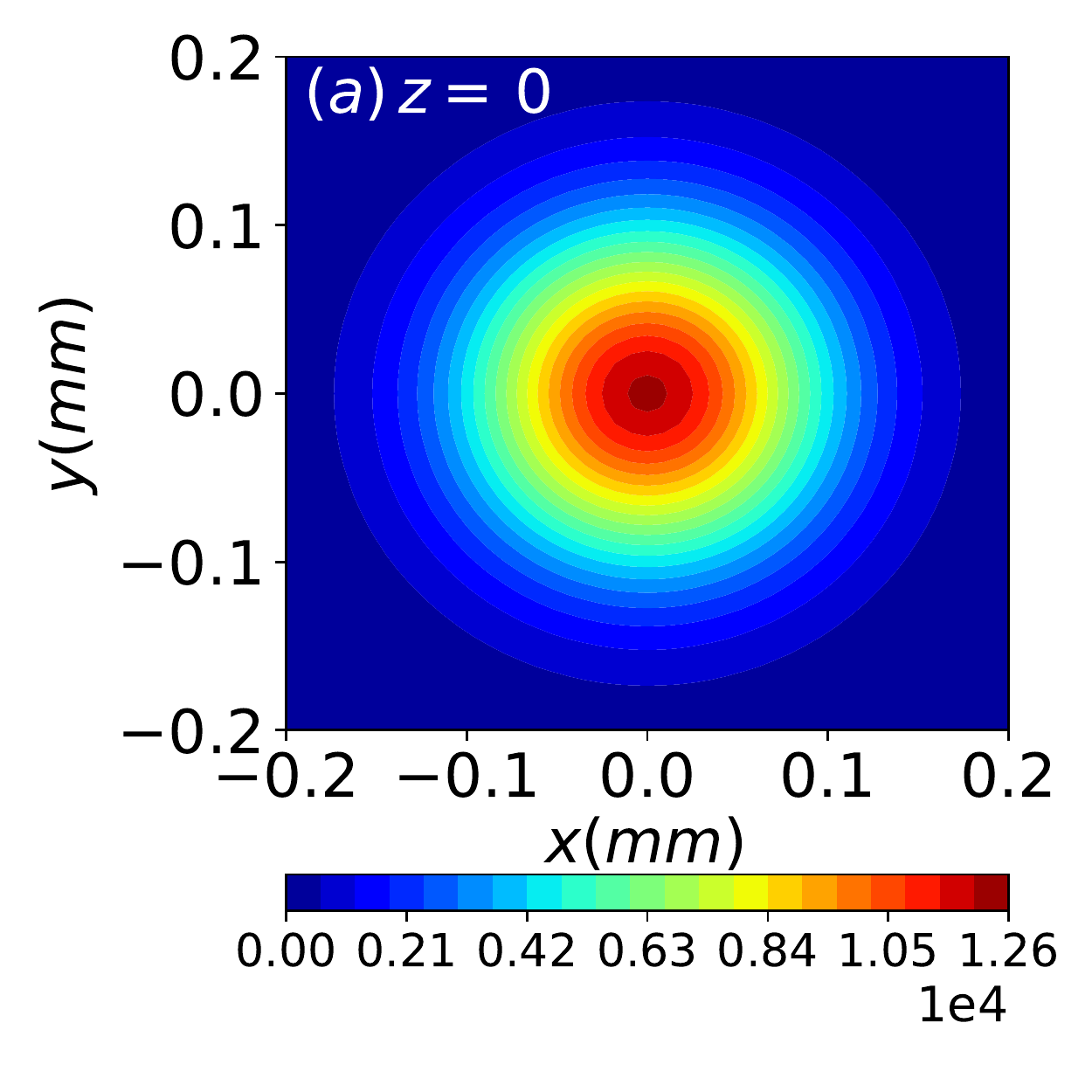}}
\subfloat{\includegraphics[width=0.47\linewidth]{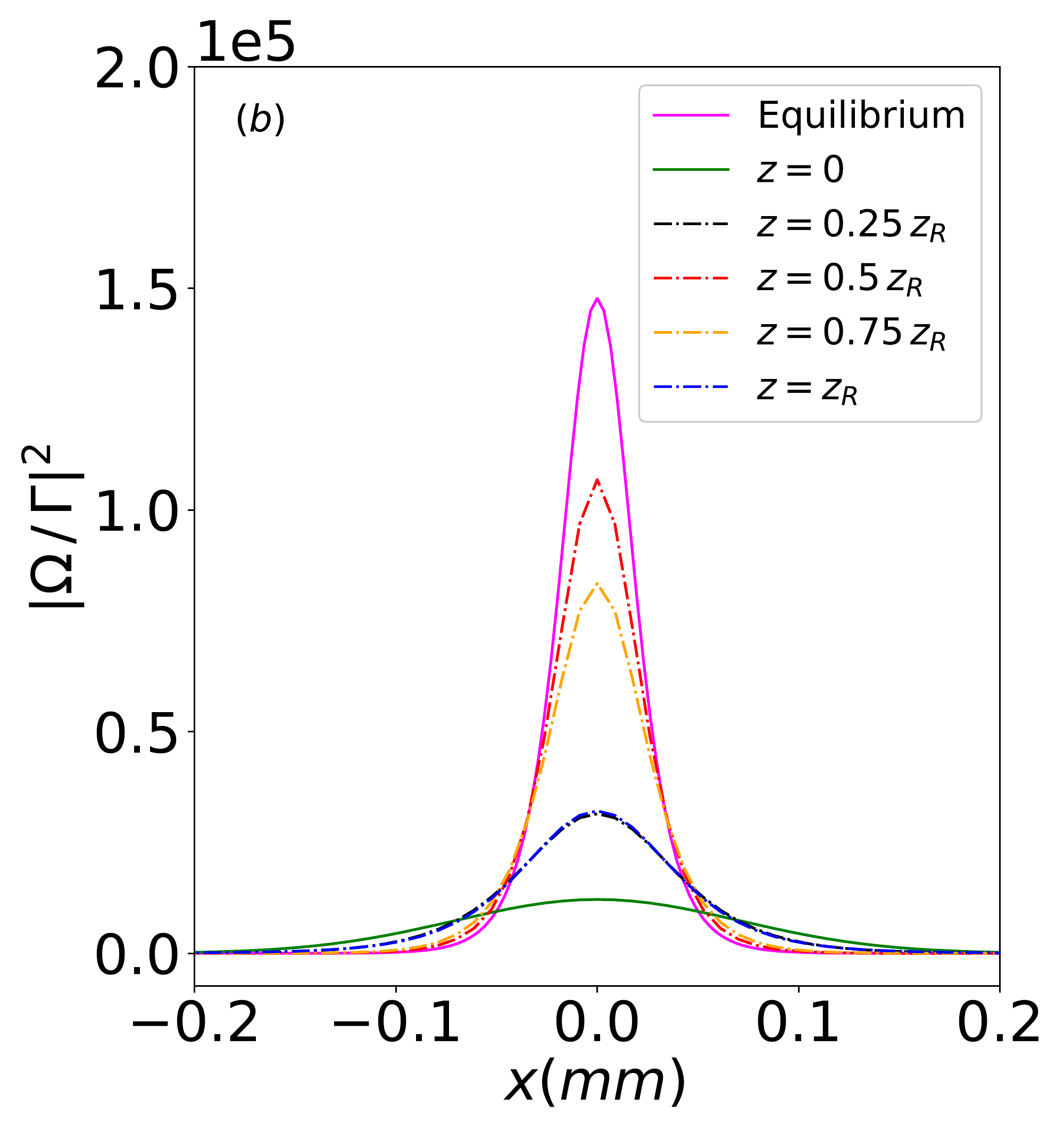}}
\caption{(a) Density plot showing the input beam for the $\chi^{(3)}+\chi^{(5)}$ regime when $\Delta=2\pi\times 6.0$ GHz. In (b), it is shown the cut along the $y-$axis, $\left|\Omega\left(z,x,0\right)/\,\Gamma\right|^{2}$, displaying the evolution of input beam at different effective lengths.}
\label{2DNS_chi3chi5_995}
\end{figure}

\begin{figure}[ht!]
\centering
\subfloat{\includegraphics[width=0.47\linewidth]{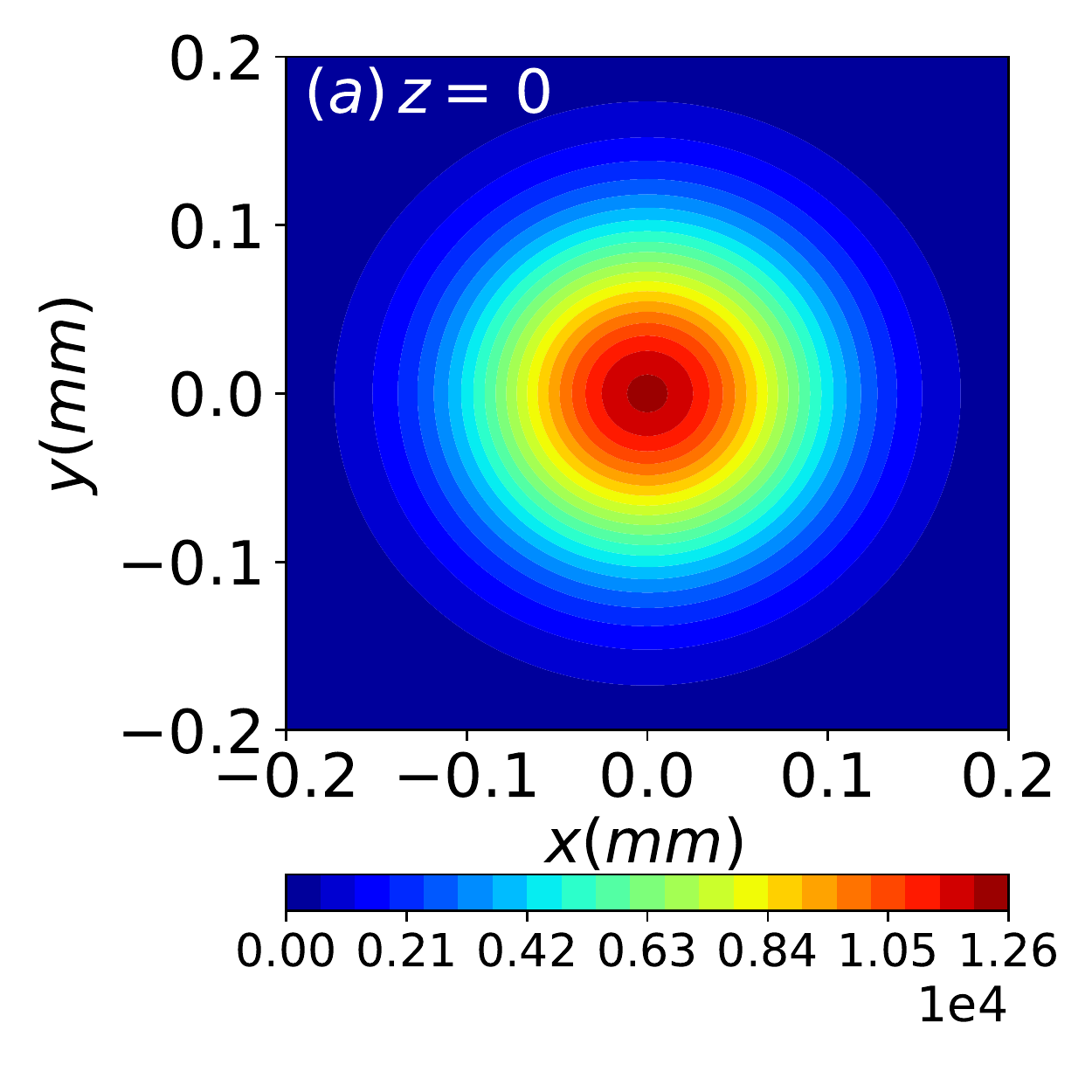}}
\subfloat{\includegraphics[width=0.47\linewidth]{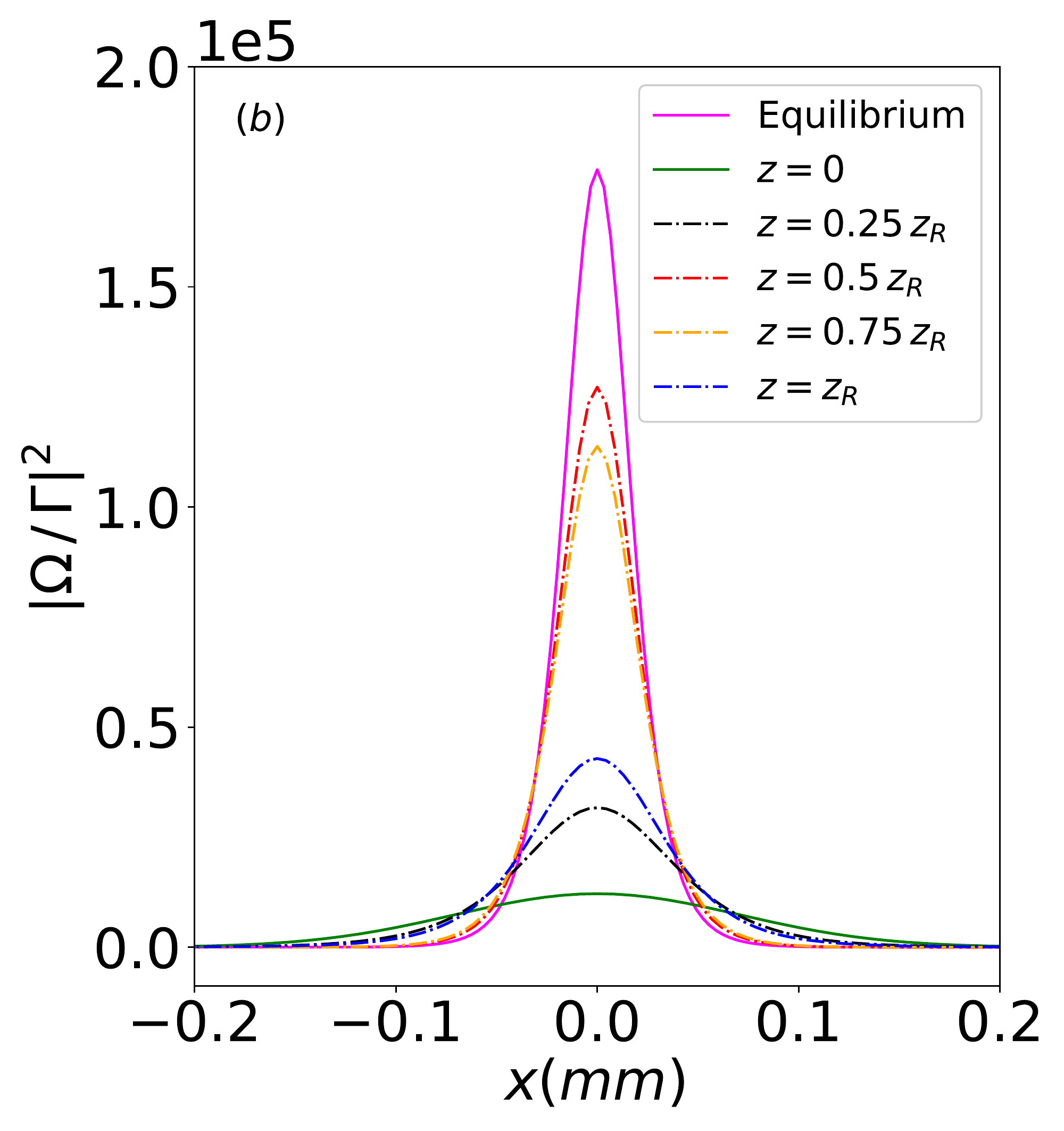}}
\caption{(a) Density plot showing the input beam for the saturating regime when $\Delta=2\pi\times 6.0$ GHz. In (b), it is shown the cut along the $y-$axis, $\left|\Omega\left(z,x,0\right)/\,\Gamma\right|^{2}$, displaying the evolution of input beam at different effective lengths.}
\label{2DNS_sat_995}
\end{figure}

In the density plots of Figs. (\ref{2DNS_chi3chi5_995}a) and (\ref{2DNS_sat_995}a), we show the initial state of the system, with $\Delta=2\pi\times\,6.0\,$GHz for the $\chi^{(3)}+\chi^{(5)}$ and saturating regimes, respectively. 
As the beam travels through the nonlinear medium, we observe that the system oscillates with decreasing amplitude as the time evolves. 
This oscillatory behaviour around the equilibrium configuration of the system is depicted in the plots of Figs. (\ref{2DNS_chi3chi5_995}b) and (\ref{2DNS_sat_995}b), which show a cut along $y=0$. 
The equilibrium states were obtained through 
imaginary-time simulations of Eqs. (\ref{eqs7c}) and (\ref{nlseChi3Chi5}). Due to absorbing boundary conditions, the curves obtained through propagation in real-time shown in these plots do not overlap perfectly the ones of the equilibrium states. 
Before moving on, we briefly comment two examples of previous experiments with cold atoms~\cite{Labeyrie:11} and hot vapours~\cite{PhysRevA.104.013515}. We verify whether the range of experimental parameters used in them were optimal for observing droplets of light. For the former, it was shown that, for a near-resonant propagating beam, a large cloud of cold $^{87}$Rb atoms acts as a saturable Kerr medium and produces self-trapping of light, that is, the waist remains stationary for an appropriate choice of parameters. For that experiment, the value of the ratio $\mathcal{P}\,/\,\mathcal{P}_{cr}$ is approximately $1.9$ and $\gamma=0.20$, that is, a highly saturated regime. Much of the cloud is contained in the droplet region based on our phase diagram in Fig. (\ref{phase_diagram_GS}). 
Nevertheless, if we compute the ratio $\omega/\left|\mu\right|$, it is greater than unity, so we will not be able to see the manifestation of the breathing mode. For the latter, it was analysed the out-of-equilibrium dynamics of a two-dimensional paraxial fluid of light using a near-resonant laser propagating through a hot atomic vapour and the formation of shock waves. For this experiment, $\mathcal{P}\,/\,\mathcal{P}_{cr}>1$ for a broad range of the frequency detuning, and we can eventually reach conditions by increasing $\Delta$ for which the observation of the breathing mode is possible, although, at this point, it is uncertain whether this would occur for realistic propagation distances or not, and consequently a more careful analysis supported by numerical simulations would be required.

\begin{figure}[ht!]
\centering
\subfloat{\includegraphics[width=0.47\linewidth]{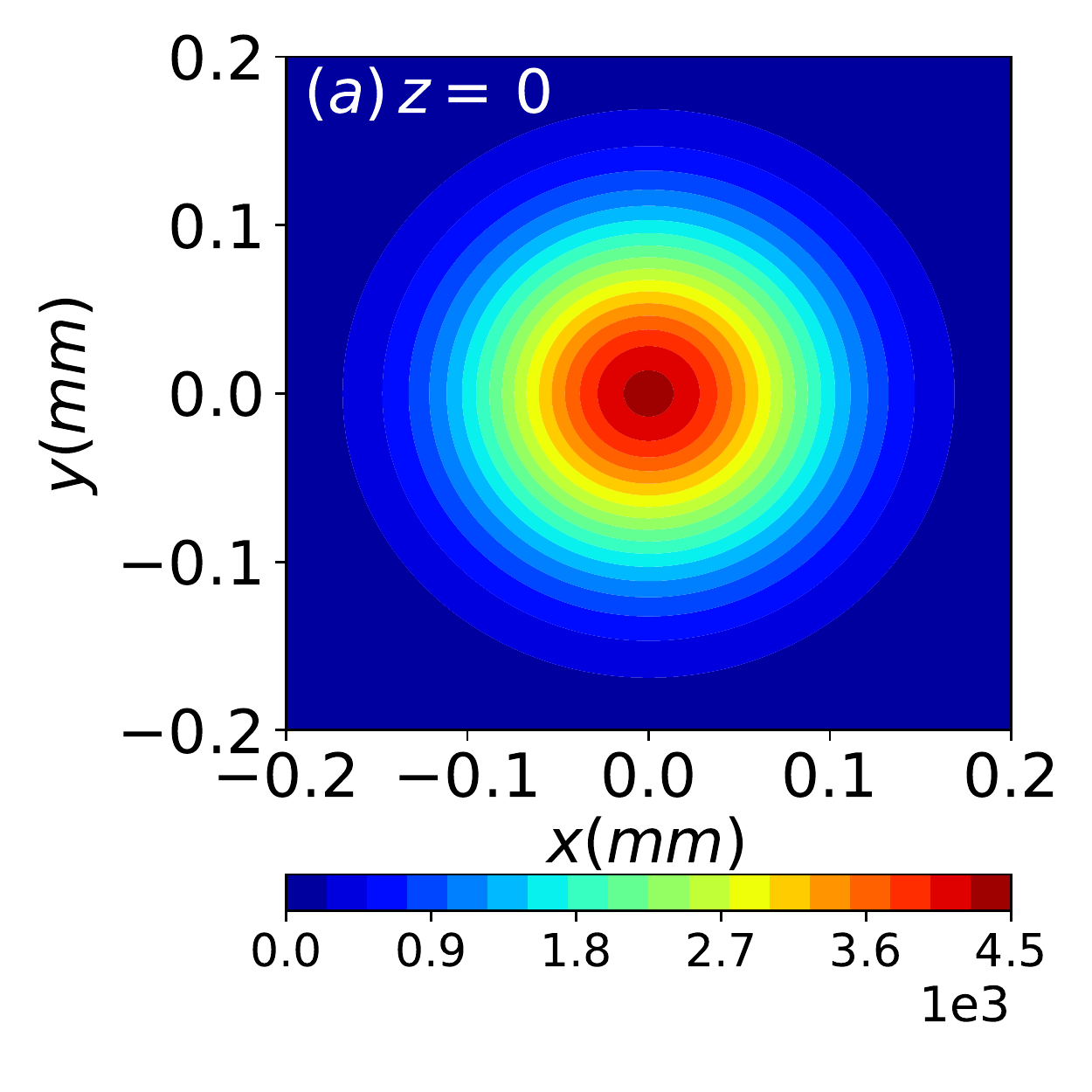}}
\subfloat{\includegraphics[width=0.47\linewidth]{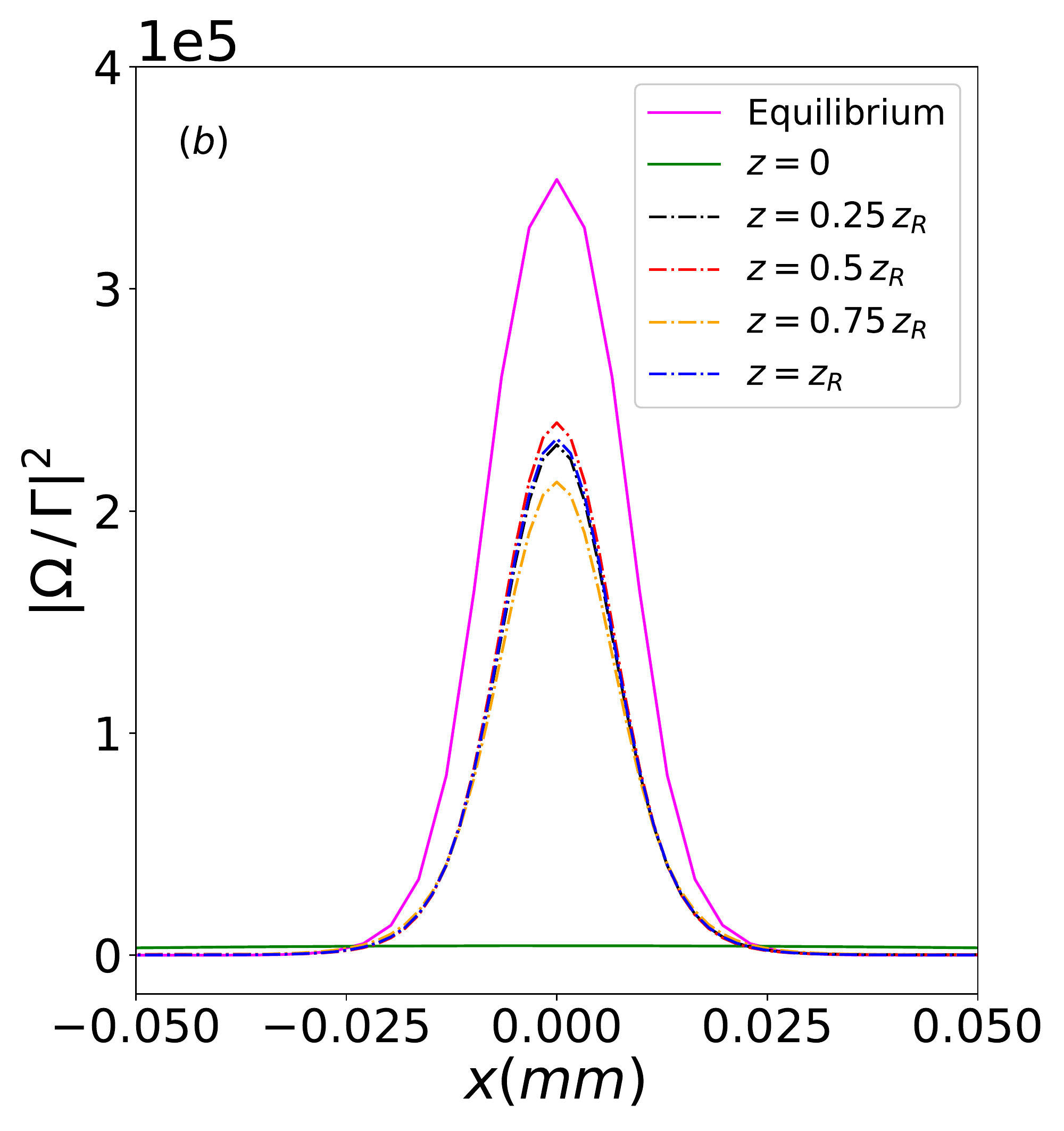}}
\caption{\textit{$2$D Numerical simulations:} (a) Density plot showing the input beam for the $\chi^{(3)}+\chi^{(5)}$ regime when $\Delta=2\pi\times 3.0$ GHz. In (b), it is shown the cut along the $y-$axis, $\left|\Omega\left(z,x,0\right)/\,\Gamma\right|^{2}$, displaying the evolution of input beam at different effective lengths.}
\label{2DNS_chi3chi5_500}
\end{figure}

\begin{figure}[ht!]
\centering
\subfloat{\includegraphics[width=0.47\linewidth]{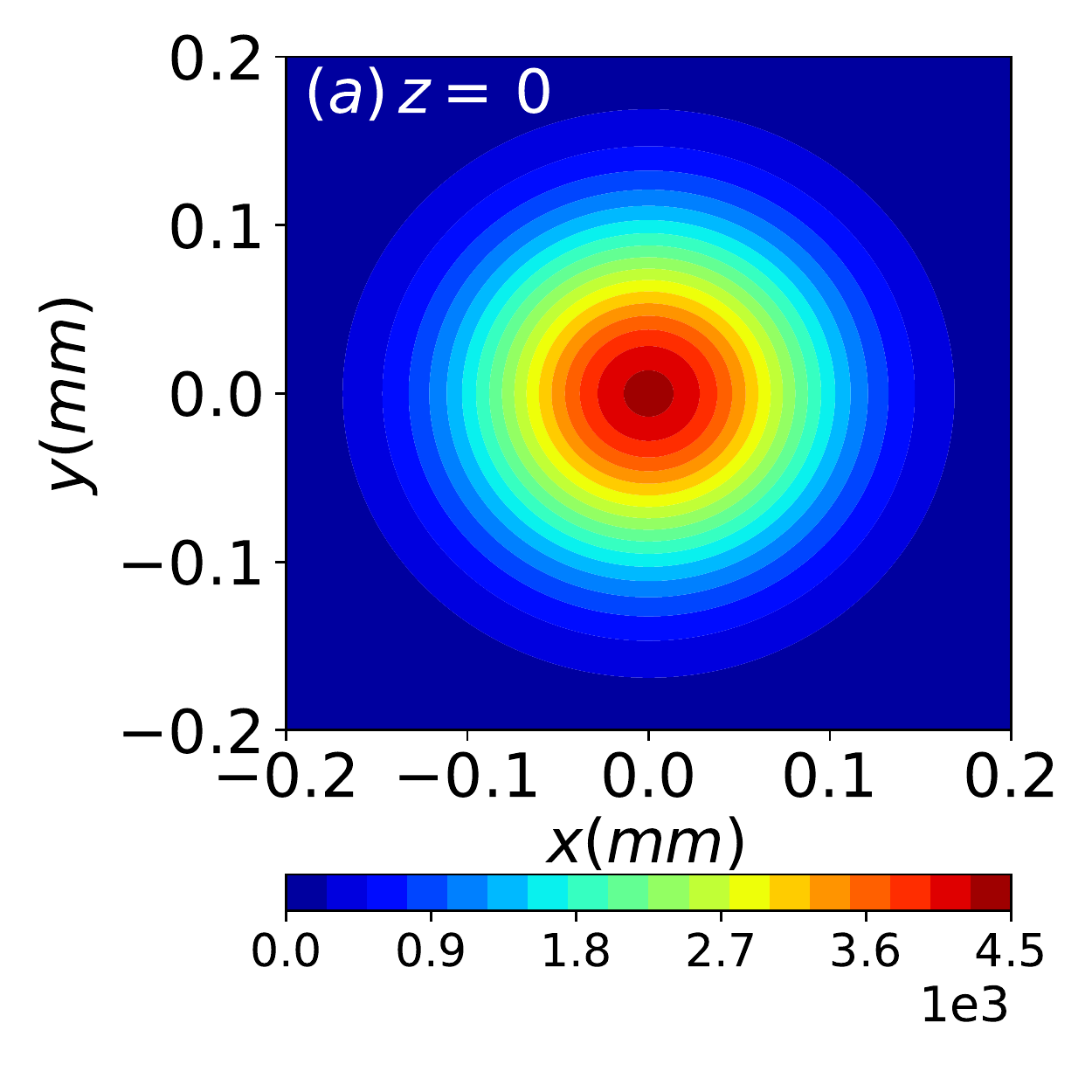}}
\subfloat{\includegraphics[width=0.47\linewidth]{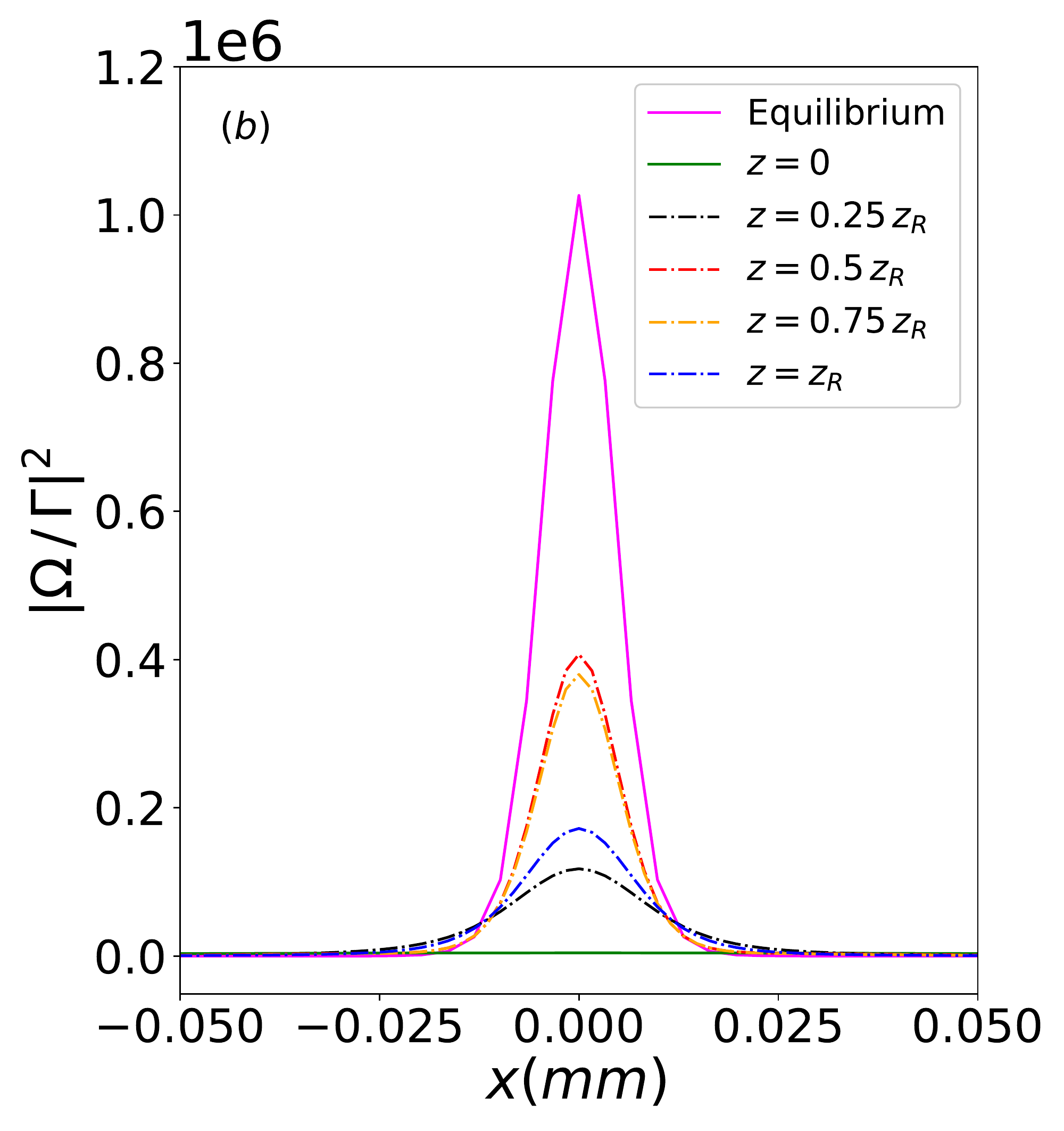}}
\caption{\textit{$2$D Numerical simulations:} (a) Density plot showing the input beam for the saturating regime when $\Delta=2\pi\times 3.0$ GHz. In (b), it is shown the cut along the $y-$axis, $\left|\Omega\left(z,x,0\right)/\,\Gamma\right|^{2}$, displaying the evolution of input beam at different effective lengths.}
\label{2DNS_sat_500}
\end{figure}

In Figs. (\ref{2DNS_chi3chi5_500}) and (\ref{2DNS_sat_500}) are displayed the results for $\Delta=2\pi\times\,3.0\,$GHz for the $\chi^{(3)}+\chi^{(5)}$ and saturating regimes, respectively. Again, the oscillatory behavior around the equilibrium is present. Nevertheless, it is clear that the shrinkage of the input beam is much more pronounced than that observed for $\Delta=2\pi\times\,6.0\,$GHz, and this only gets more accentuated as the frequency detuning is decreased (strength of the nonlinearity increases). 

Actually, this strong effect of the nonlinearity that makes the input beam to shrink to a point leads to difficulties in the numerical simulations, as problems with spatial resolution start to appear. Furthermore, as we approach the resonance, the radial symmetry starts to break down and because of that, $1$D and $2$D numerical simulations display very opposite behaviour for this system in this region. In fact, $1$D simulations are not adequate to represent the dynamics of the system close to resonance. A faithful representation of the dynamics of the system in this region is only possible and reliable through full $2$D numerical simulations. To give an estimate of at what point the $1$D and $2$D numerical simulations stop agreeing, we compared the curves obtained for the intensity and found that there is a quite fair overlap until $\Delta=2\pi\times 3.0\,$GHz. 

This inconsistency can be made more clear if we check the outcomes of numerical simulations for lower values of $\Delta$. For the full $2$D case, different dynamical behaviours can emerge. One of them is that the input beam, due to the strong nonlinearity, can eventually shrink dramatically such that it breaks completely into small fragments, even for short propagation effective lengths. This fragmentation process in the optical system considered here could be linked to the results of the physics investigated in ~\cite{chen2021observation}, in which the universal nonequilibrium dynamics in degenerate $2
$D Bose gases was investigated.
The authors considered an initially large sample, and then perform a quench from repulsive (defocusing) to attractive (focusing) interactions through the use of a Feshbach resonance. 
If the appropriate regime is reached, which means the value of the product between particle number and interacting strength, $N\,\left|g\right|$  close to the Townes threshold, then this quench procedure makes possible the observation of the dynamic formation of Townes solitons from modulational instability (MI). 
The MI breaks up the initial sample into fragments, universally around the Townes threshold. 
We leave the detailed investigation of fragmentation within our model for a future work.

\begin{figure}[ht]
\centering
\includegraphics[width=0.995\linewidth]{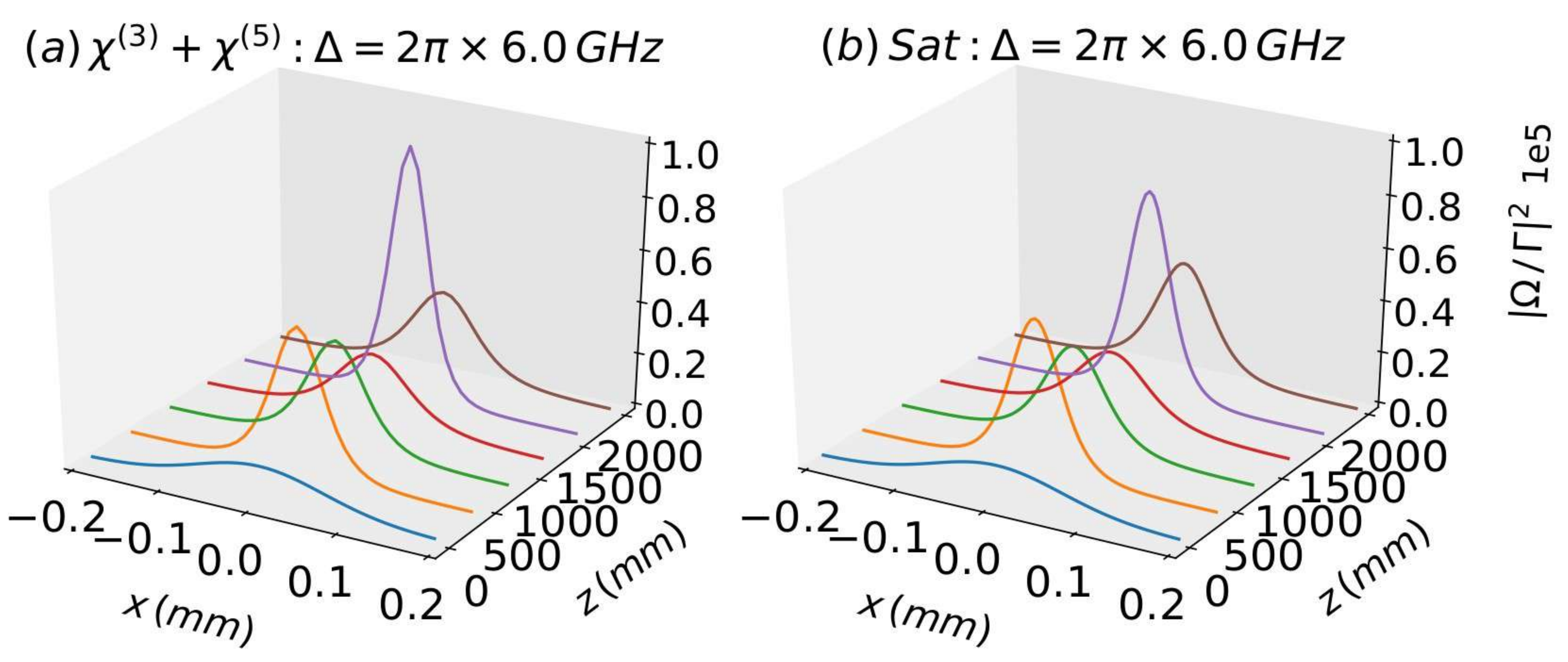}
\caption{\textit{$2$D Numerical simulations:} Cut along the transversal $y$ direction of the intensity profile for the $\chi^{(3)}+\chi^{(5)}$ (a) and the saturating regime (b) for $\Delta=2\pi\times\,6.0\,$GHz.}
\label{density_y_cut_995}
\end{figure}

\begin{figure}[ht]
\centering
\includegraphics[width=0.995\linewidth]{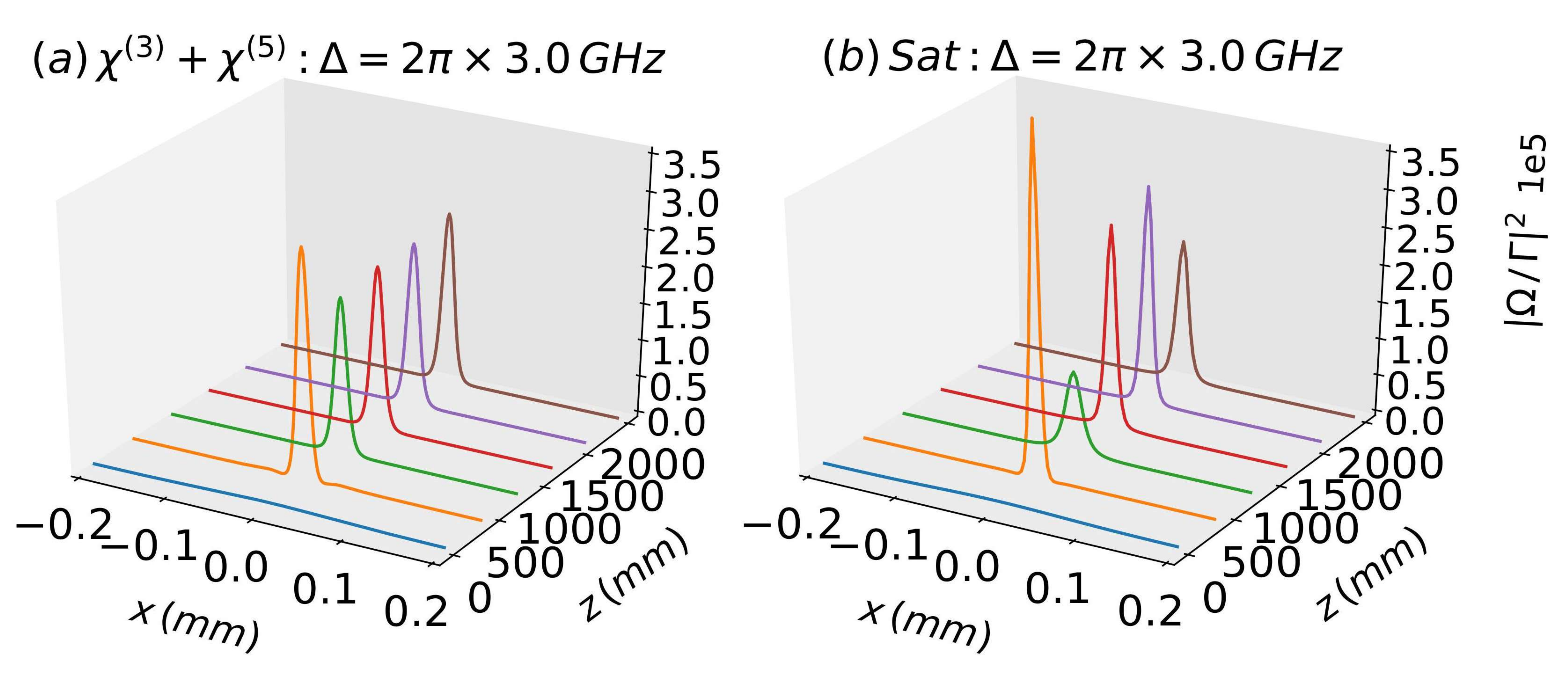}
\caption{\textit{$2$D Numerical simulations:} Cut along the transversal $y$ direction of the intensity profile for the $\chi^{(3)}+\chi^{(5)}$ (a) and the saturating regime (b) for $\Delta=2\pi\times\,3.0\,$GHz.}
\label{density_y_cut_500}
\end{figure}

\begin{figure}[ht]
\centering
\includegraphics[width=0.995\linewidth]{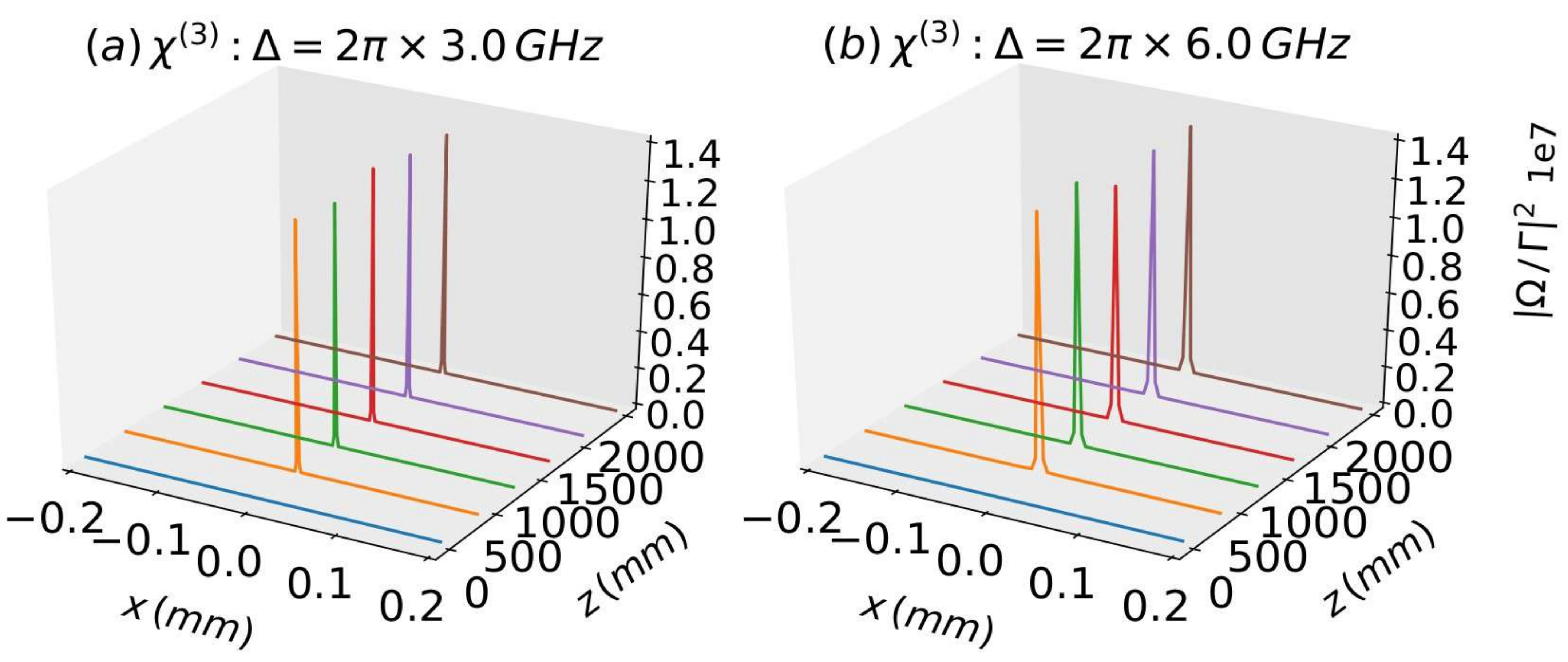}
\caption{\textit{$2$D Numerical simulations:} Cut along the transversal $y$ direction of the intensity profile for the $\chi^{(3)}$ regime considering $\Delta=2\pi\times 3.0\,$GHz (a) and $\Delta=2\pi\times\,6.0\,$GHz (b).}
\label{density_y_cut_chi3}
\end{figure}

Finally, we compare the dynamics for the different regimes in this realistic scenario. In Figs. (\ref{density_y_cut_995}) and (\ref{density_y_cut_500}) we show the integrated beam profile for the $\chi^{(3)}+\chi^{(5)}$ and saturating regimes for $\Delta=2\pi\times 3.0\,$ and $2\pi\times 6.0\,$GHz, while in Fig. (\ref{density_y_cut_chi3}) we display the results for the $\chi^{(3)}$ regime. 
It is clear how the effect of self-focusing makes the width of the intensity profile in the $\chi^{(3)}$ regime shrink dramatically when compared to the other two regimes, in which the defocusing effect of $\chi^{(5)}$ and the saturation hinder this focusing behaviour.

\section{Conclusions}
Motivated by recent experiments on quantum fluids of light well described within the paraxial approximation and their analogy with Bose-Einstein condensates (BECs) described by the Gross-Pitaesvskii equation (GPE), 
we investigated the conditions to observe droplets of light in nonlinear optical media. 
We reviewed the cubic focusing NLSE and the physics of the Townes soliton. 
For this regime, it is known that self-bound states cannot be formed as there is no mechanism to compensate the combined repulsive effects due to diffraction and the focusing (attractive) Kerr nonlinearity. 
When anomalous second-order dispersion is taken into account, the system becomes $3+1$ dimensional with the proper time playing the role of a third spatial coordinate. In this scenario, the existence of optical bullets was foreseen~\cite{Silberberg:90}, though these structures would be unstable. We then concentrated our efforts on the $\chi^{(3)}+\chi^{(5)}$ and saturating nonlinearities. 
For the cubic-quintic NLSE, several works had already pointed out the possibility of having self-bound states~\cite{PhysRevE.65.066604} while for the saturating regime only the optical bullets were predicted in the presence of anomalous dispersion \cite{akhmediev1992modulation}. 
By using the variational Gaussian ansatz approach, we obtained an analytical expression for the radial width of the droplet of light for the $\chi^{(3)}+\chi^{(5)}$ regime.
The variational results were then compared with numerical simulations. We found a good agreement, certifying the reliability of the variational method. 
We also investigated the dynamics of the breathing mode and how the self-evaporation mechanism can influence the dynamical process of formation of droplets of light for both the $\chi^{(3)}+\chi^{(5)}$ and saturating regimes. 
Our study revealed that the system may sustain collective excitations only for a very limited region of the frequency detuning, while overdamped oscillations are present in a wide range of $\Delta$.
For realistic experimental conditions, we employed $1$D and $2$D numerical simulations. The former predicted that close to resonance a flat-top profile is formed for the $\chi^{(3)}+\chi^{(5)}$ and an oscillatory behaviour in the saturating case. However, refined $2$D numerical simulations revealed that close to resonance, the radial symmetry no longer holds, with the input beam becoming susceptible to fragmentation for the regimes of interest. 
On the other hand, if we consider increasing frequency detuning, the behaviour shown by the system becomes very similar for both regimes with identical $1$D or $2$D simulation results.
In conclusion, this study enabled us to establish the values of experimental parameters for which the droplet states can be observed in platforms such as hot atomic vapours and to predict some important aspects of formation of such self-bound light states. 
Extensions of this work may include a detailed study of the stability of these droplets, 
the investigation of finite-temperature effects that can become relevant for the region closer to resonance in which Doppler effect becomes significant, as well as advance in the understanding of the self-evaporation mechanism, and the  effects of effective nonlocal nonlinearities \cite{https://doi.org/10.48550/arxiv.2109.01063}. 
The presence of nonlocality might determine the formation of nontrivial patterns similarly to the case of ultracold Bose gases with magnetic \cite{https://doi.org/10.48550/arxiv.2201.02672,PhysRevLett.119.215302} or soft-core interactions \cite{macri2014ground,Cinti2014,PhysRevA.89.011402,PhysRevA.96.043605}.

\textit{Note added.} We thank J. Dalibard for pointing out a related work in the context of $2$D Bose mixtures ~\cite{jdalibard}.

\begin{acknowledgments}
We gratefully acknowledge insightul discussions with P. Azam, A. Marini, F. Maucher, T. Pohl, H. F. Silva,  V. Zampronio, and B. Bakkali-Hassani and J. Beugnon for critical reading of the manuscript.  This study was financed in part by the Coordenação de Aperfeiçoamento de Pessoal de Nível Superior –
Brasil (CAPES) – Finance Code 001.
T.M. acknowledges the hospitality of the Institut de Physique de Nice, Valbonne, 
where this work was initiated.
This work was supported by the Serrapilheira Institute 
(grant number Serra-1812-27802),
CAPES-NUFFIC project number 88887.156521/2017-00.
This research was developed with the help of
XMDS2 software package \cite{XMDS2013}. We thank the High Performance Computing Center (NPAD) at UFRN for providing computational resources.
\end{acknowledgments}

\appendix
\section{The Optical Bloch equations for the two-level system.}\label{obe2level}
This appendix is devoted to the presentation of the Optical Bloch equations for the two-level system considered in this work and their corresponding solutions for the coherences and populations in the steady state.

The characterization of this system was given in Section \ref{sec:setting} in the main text. The OBE for the two-level configuration after having applied the rotating-wave approximation (RWA) are given by:
\begin{subequations}
\label{allequations} 
\begin{eqnarray}
\dot{\rho}_{gg}=i\frac{\Omega}{2}\rho_{ge}-i\frac{\Omega^{*}}{2}\rho_{eg}+\Gamma\rho_{ee}\,,\\
\dot{\rho}_{ee}=-i\frac{\Omega}{2}\rho_{ge}+i\frac{\Omega^{*}}{2}\rho_{eg}-\Gamma\rho_{ee}\,,\\
\dot{\rho}_{ge}=-i\Delta\rho_{ge}-i\frac{\Omega^{*}}{2}\left(\rho_{ee}-\rho_{gg}\right)-\frac{\Gamma}{2}\rho_{ge}\,,\\
\dot{\rho}_{eg}=i\Delta\rho_{eg}+i\frac{\Omega}{2}\left(\rho_{ee}-\rho_{gg}\right)-\frac{\Gamma}{2}\rho_{eg}
\label{eq:twolevel1}
\end{eqnarray}
\end{subequations}
where $\Delta\equiv\omega-\left(\omega_{2}-\omega_{1}\right)$ is the frequency detuning. If we solve for the steady-state, then the expressions for the populations and coherences are simply:
\begin{subequations}
\label{allequations}
\begin{eqnarray}
 \label{pop2leveleqsAPP}
 \rho_{gg}=\frac{\Gamma^{2}+4\Delta^{2}+\left|\Omega\right|^{2}}{\Gamma^{2}+4\Delta^{2}+2\left|\Omega\right|^{2}}\,,\\
 \rho_{ee}=\frac{\left|\Omega\right|^{2}}{\Gamma^{2}+4\Delta^{2}+2\left|\Omega\right|^{2}}\,,\\
 \rho_{ge}=\frac{\left(\rm i \Gamma+2\Delta\right)\Omega^{*}}{\Gamma^{2}+4\Delta^{2}+2\left|\Omega\right|^{2}}\,,\\
 \rho_{eg}=\frac{\left(-\rm i \Gamma+2\Delta\right)\Omega}{\Gamma^{2}+4\Delta^{2}+2\left|\Omega\right|^{2}}\,.
\end{eqnarray}
\end{subequations}

\section{Wave collapse in the $\chi^{(3)}$ regime: a variational approach}\label{wave_colapse}
In this appendix, we derive the expression of the collapsing distance, $z_{cr}$, for the cubic NLSE. This derivation follows a similar analysis in~\cite{BERGE1998259}. 

The Lagrangian density for the cubic NLSE given in Eq. (\ref{nlseChi3}) is
\begin{equation}
    \label{wc1}
    \mathcal{L}=\frac{i}{2}\left(\psi^{*}\partial_{z}\psi-\psi\partial_{z}\psi^{*}\right)-\mathcal{H},\qquad\,\mathcal{H}=\left|\frac{\partial \psi}{\partial r}\right|^{2}-\frac{\left|\psi\right|^{4}}{2}\,.
\end{equation}
Next, we build an ansatz for the wave function $\psi(r,z)$. 
We assume a  self-similar trial function
\begin{equation}
    \label{wc2}
    \psi\left(\mathbf{r},z\right)=A\left(z\right)R\left(\frac{r}{a\left(z\right)}\right)\exp\left[i\theta\left(r,z\right)\right]\,,
\end{equation}
where $R\left(r/a\left(z\right)\right)$ is an arbitrary real profile function only depending on the spatial coordinate $\xi=r/a\left(z\right)$ rescaled with respect to the wave radius $a\left(z\right)$, and $A(z)$ is a normalization factor at distance $z$. Plugging this ansatz into Eq. (\ref{wc1}) and integrating over the radial coordinates, it yields
\begin{equation}
    \label{wc3}
    L=-\left|A\right|^{2}a^{2}\left\{\alpha_{0}\dot{\theta}_{0}+\alpha_{2}a^{2}\left[\dot{\theta}_{2}+2\theta_{2}^{2}\right]+\frac{\lambda}{2a^{2}}-\frac{\left|A\right|^{2}\beta}{2}\right\}\,.
\end{equation}

The coefficients
\begin{equation}
    \label{wc4}
    \alpha_{m}\equiv \left|\left| r^{m/2}R\right|\right|_{2}\,\,,\,\,\beta\equiv \left|\left| R\right|\right|_{4}\,\,,\,\,\lambda\equiv \left|\left| \partial_{r}R\right|\right|_{2}
\end{equation}
where we define 
$\parallel f \parallel _{p} = 2\pi\int_{0}^{+\infty}
f^p ( r ) r \, dr$. 
Let us now consider $A\left(z\right)$, $\theta_{0}\left(z\right)$, $\theta_{2}\left(z\right)$ and $a\left(z\right)$ as canonical variables. 
We are then led to the following dynamical equations
\begin{eqnarray}
    \begin{cases}
			\left|A\right|^{2}a^{2}=\text{P}=\text{const.}\\
            \theta_{2}\left(z\right)=\frac{\dot{a}}{2a}\\
            \frac{d^2 a}{dz^{2}}=\frac{\lambda}{\alpha_{2}a^{3}}-\frac{P\beta}{\alpha_{2}a^{3}}
\end{cases},
\label{wc5}
\end{eqnarray}
and by multiplying both sides of the equation for $a\left(z\right)$ by $\dot{a}(z)$, the remaining expression can be easily integrated, and it results into
\begin{equation}
    \label{wc6}
    \frac{1}{2}\left[\frac{d}{dz}\left(\frac{a}{a_{0}}\right)\right]^{2}+\Pi\left(\frac{a}{a_{0}}\right)=0
\end{equation}
with 
\begin{equation}
\label{wc7}
    \Pi\left(x\right)\equiv \left(\mu-\nu\right)\left(\frac{1}{x^2}-1\right)\,\,,\,\,\mu\equiv \frac{2\lambda}{\alpha_{2}a_{0}^{4}}\,\,,\,\,\nu\equiv \frac{\left|A_{0}\right|^{2}\beta}{\alpha_{2}a_{0}^{2}}
\end{equation}
which applies to the case in which $a_{0}\equiv a\left(z=0\right)\neq 0$  and $\dot{a}\left(0\right)=\dot{a}\left(z=0\right)=0$. 
It follows that the solution for Eq. (\ref{wc6}) becomes,
\begin{equation}
    \label{wc8}
    a\left(z\right)=a_{0}\left[\left(z\sqrt{2\mu}\right)^2\left(1-\nu/\mu\right)+1\right]^{1/2}
\end{equation}
which predicts that the wave collapses with $a\left(z\right)\rightarrow 0$ at the finite distance
\begin{equation}
    \label{wc9}
    z_{cr}=\frac{1}{\sqrt{2\left(\nu-\mu\right)}},
\end{equation}
under the constraint that $\nu/\mu>1$. 
In the main text, this expression was applied, considering different forms for the function $R\left(r/a\left(z\right)\right)$ and with $a_{0}$ being the beam waist. 

\bibliographystyle{apsrev4-2}
\bibliography{refermod}

\begin{thebibliography}{74}%
\makeatletter
\providecommand \@ifxundefined [1]{%
 \@ifx{#1\undefined}
}%
\providecommand \@ifnum [1]{%
 \ifnum #1\expandafter \@firstoftwo
 \else \expandafter \@secondoftwo
 \fi
}%
\providecommand \@ifx [1]{%
 \ifx #1\expandafter \@firstoftwo
 \else \expandafter \@secondoftwo
 \fi
}%
\providecommand \natexlab [1]{#1}%
\providecommand \enquote  [1]{``#1''}%
\providecommand \bibnamefont  [1]{#1}%
\providecommand \bibfnamefont [1]{#1}%
\providecommand \citenamefont [1]{#1}%
\providecommand \href@noop [0]{\@secondoftwo}%
\providecommand \href [0]{\begingroup \@sanitize@url \@href}%
\providecommand \@href[1]{\@@startlink{#1}\@@href}%
\providecommand \@@href[1]{\endgroup#1\@@endlink}%
\providecommand \@sanitize@url [0]{\catcode `\\12\catcode `\$12\catcode
  `\&12\catcode `\#12\catcode `\^12\catcode `\_12\catcode `\%12\relax}%
\providecommand \@@startlink[1]{}%
\providecommand \@@endlink[0]{}%
\providecommand \url  [0]{\begingroup\@sanitize@url \@url }%
\providecommand \@url [1]{\endgroup\@href {#1}{\urlprefix }}%
\providecommand \urlprefix  [0]{URL }%
\providecommand \Eprint [0]{\href }%
\providecommand \doibase [0]{https://doi.org/}%
\providecommand \selectlanguage [0]{\@gobble}%
\providecommand \bibinfo  [0]{\@secondoftwo}%
\providecommand \bibfield  [0]{\@secondoftwo}%
\providecommand \translation [1]{[#1]}%
\providecommand \BibitemOpen [0]{}%
\providecommand \bibitemStop [0]{}%
\providecommand \bibitemNoStop [0]{.\EOS\space}%
\providecommand \EOS [0]{\spacefactor3000\relax}%
\providecommand \BibitemShut  [1]{\csname bibitem#1\endcsname}%
\let\auto@bib@innerbib\@empty
\bibitem [{\citenamefont {Carusotto}\ and\ \citenamefont
  {Ciuti}(2013)}]{CarusottoRevModPhys.85.299}%
  \BibitemOpen
  \bibfield  {author} {\bibinfo {author} {\bibfnamefont {I.}~\bibnamefont
  {Carusotto}}\ and\ \bibinfo {author} {\bibfnamefont {C.}~\bibnamefont
  {Ciuti}},\ }\href {https://doi.org/10.1103/RevModPhys.85.299} {\bibfield
  {journal} {\bibinfo  {journal} {Rev. Mod. Phys.}\ }\textbf {\bibinfo {volume}
  {85}},\ \bibinfo {pages} {299} (\bibinfo {year} {2013})}\BibitemShut
  {NoStop}%
\bibitem [{\citenamefont {Glorieux}\ \emph {et~al.}(2022)\citenamefont
  {Glorieux}, \citenamefont {Aladjidi}, \citenamefont {Lett},\ and\
  \citenamefont {Kaiser}}]{ReviewQG}%
  \BibitemOpen
  \bibfield  {author} {\bibinfo {author} {\bibfnamefont {Q.}~\bibnamefont
  {Glorieux}}, \bibinfo {author} {\bibfnamefont {T.}~\bibnamefont {Aladjidi}},
  \bibinfo {author} {\bibfnamefont {P.~D.}\ \bibnamefont {Lett}},\ and\
  \bibinfo {author} {\bibfnamefont {R.}~\bibnamefont {Kaiser}},\ }\href@noop {}
  {\bibinfo {title} {Hot atomic vapors for nonlinear and quantum optics}}
  (\bibinfo {year} {2022}),\ \Eprint {https://arxiv.org/abs/2209.04622}
  {arXiv:2209.04622 [quant-ph]} \BibitemShut {NoStop}%
\bibitem [{\citenamefont {Connaughton}\ \emph {et~al.}(2005)\citenamefont
  {Connaughton}, \citenamefont {Josserand}, \citenamefont {Picozzi},
  \citenamefont {Pomeau},\ and\ \citenamefont
  {Rica}}]{PicoziConaughtonPRL2005}%
  \BibitemOpen
  \bibfield  {author} {\bibinfo {author} {\bibfnamefont {C.}~\bibnamefont
  {Connaughton}}, \bibinfo {author} {\bibfnamefont {C.}~\bibnamefont
  {Josserand}}, \bibinfo {author} {\bibfnamefont {A.}~\bibnamefont {Picozzi}},
  \bibinfo {author} {\bibfnamefont {Y.}~\bibnamefont {Pomeau}},\ and\ \bibinfo
  {author} {\bibfnamefont {S.}~\bibnamefont {Rica}},\ }\href
  {https://doi.org/10.1103/PhysRevLett.95.263901} {\bibfield  {journal}
  {\bibinfo  {journal} {Phys. Rev. Lett.}\ }\textbf {\bibinfo {volume} {95}},\
  \bibinfo {pages} {263901} (\bibinfo {year} {2005})}\BibitemShut {NoStop}%
\bibitem [{\citenamefont {Santic}\ \emph {et~al.}(2018)\citenamefont {Santic},
  \citenamefont {Fusaro}, \citenamefont {Salem}, \citenamefont {Garnier},
  \citenamefont {Picozzi},\ and\ \citenamefont {Kaiser}}]{RobinPRL2018}%
  \BibitemOpen
  \bibfield  {author} {\bibinfo {author} {\bibfnamefont {N.}~\bibnamefont
  {Santic}}, \bibinfo {author} {\bibfnamefont {A.}~\bibnamefont {Fusaro}},
  \bibinfo {author} {\bibfnamefont {S.}~\bibnamefont {Salem}}, \bibinfo
  {author} {\bibfnamefont {J.}~\bibnamefont {Garnier}}, \bibinfo {author}
  {\bibfnamefont {A.}~\bibnamefont {Picozzi}},\ and\ \bibinfo {author}
  {\bibfnamefont {R.}~\bibnamefont {Kaiser}},\ }\href
  {https://doi.org/10.1103/PhysRevLett.120.055301} {\bibfield  {journal}
  {\bibinfo  {journal} {Phys. Rev. Lett.}\ }\textbf {\bibinfo {volume} {120}},\
  \bibinfo {pages} {055301} (\bibinfo {year} {2018})}\BibitemShut {NoStop}%
\bibitem [{\citenamefont {Baudin}\ \emph {et~al.}(2020)\citenamefont {Baudin},
  \citenamefont {Fusaro}, \citenamefont {Krupa}, \citenamefont {Garnier},
  \citenamefont {Rica}, \citenamefont {Millot},\ and\ \citenamefont
  {Picozzi}}]{BaudinPRL2020}%
  \BibitemOpen
  \bibfield  {author} {\bibinfo {author} {\bibfnamefont {K.}~\bibnamefont
  {Baudin}}, \bibinfo {author} {\bibfnamefont {A.}~\bibnamefont {Fusaro}},
  \bibinfo {author} {\bibfnamefont {K.}~\bibnamefont {Krupa}}, \bibinfo
  {author} {\bibfnamefont {J.}~\bibnamefont {Garnier}}, \bibinfo {author}
  {\bibfnamefont {S.}~\bibnamefont {Rica}}, \bibinfo {author} {\bibfnamefont
  {G.}~\bibnamefont {Millot}},\ and\ \bibinfo {author} {\bibfnamefont
  {A.}~\bibnamefont {Picozzi}},\ }\href
  {https://doi.org/10.1103/PhysRevLett.125.244101} {\bibfield  {journal}
  {\bibinfo  {journal} {Phys. Rev. Lett.}\ }\textbf {\bibinfo {volume} {125}},\
  \bibinfo {pages} {244101} (\bibinfo {year} {2020})}\BibitemShut {NoStop}%
\bibitem [{\citenamefont {Carusotto}(2014)}]{carusotto2014superfluid}%
  \BibitemOpen
  \bibfield  {author} {\bibinfo {author} {\bibfnamefont {I.}~\bibnamefont
  {Carusotto}},\ }\href
  {https://doi.org/https://doi.org/10.1098/rspa.2014.0320} {\bibfield
  {journal} {\bibinfo  {journal} {Proceedings of the Royal Society A:
  Mathematical, Physical and Engineering Sciences}\ }\textbf {\bibinfo {volume}
  {470}},\ \bibinfo {pages} {20140320} (\bibinfo {year} {2014})}\BibitemShut
  {NoStop}%
\bibitem [{\citenamefont {Fontaine}\ \emph {et~al.}(2018)\citenamefont
  {Fontaine}, \citenamefont {Bienaim\'e}, \citenamefont {Pigeon}, \citenamefont
  {Giacobino}, \citenamefont {Bramati},\ and\ \citenamefont
  {Glorieux}}]{QuentinFontainePRL2018}%
  \BibitemOpen
  \bibfield  {author} {\bibinfo {author} {\bibfnamefont {Q.}~\bibnamefont
  {Fontaine}}, \bibinfo {author} {\bibfnamefont {T.}~\bibnamefont
  {Bienaim\'e}}, \bibinfo {author} {\bibfnamefont {S.}~\bibnamefont {Pigeon}},
  \bibinfo {author} {\bibfnamefont {E.}~\bibnamefont {Giacobino}}, \bibinfo
  {author} {\bibfnamefont {A.}~\bibnamefont {Bramati}},\ and\ \bibinfo {author}
  {\bibfnamefont {Q.}~\bibnamefont {Glorieux}},\ }\href
  {https://doi.org/10.1103/PhysRevLett.121.183604} {\bibfield  {journal}
  {\bibinfo  {journal} {Phys. Rev. Lett.}\ }\textbf {\bibinfo {volume} {121}},\
  \bibinfo {pages} {183604} (\bibinfo {year} {2018})}\BibitemShut {NoStop}%
\bibitem [{\citenamefont {Fontaine}\ \emph {et~al.}(2020)\citenamefont
  {Fontaine}, \citenamefont {Larr\'e}, \citenamefont {Lerario}, \citenamefont
  {Bienaim\'e}, \citenamefont {Pigeon}, \citenamefont {Faccio}, \citenamefont
  {Carusotto}, \citenamefont {Giacobino}, \citenamefont {Bramati},\ and\
  \citenamefont {Glorieux}}]{QuentinFontainePRR2020}%
  \BibitemOpen
  \bibfield  {author} {\bibinfo {author} {\bibfnamefont {Q.}~\bibnamefont
  {Fontaine}}, \bibinfo {author} {\bibfnamefont {P.-E.}\ \bibnamefont
  {Larr\'e}}, \bibinfo {author} {\bibfnamefont {G.}~\bibnamefont {Lerario}},
  \bibinfo {author} {\bibfnamefont {T.}~\bibnamefont {Bienaim\'e}}, \bibinfo
  {author} {\bibfnamefont {S.}~\bibnamefont {Pigeon}}, \bibinfo {author}
  {\bibfnamefont {D.}~\bibnamefont {Faccio}}, \bibinfo {author} {\bibfnamefont
  {I.}~\bibnamefont {Carusotto}}, \bibinfo {author} {\bibfnamefont
  {E.}~\bibnamefont {Giacobino}}, \bibinfo {author} {\bibfnamefont
  {A.}~\bibnamefont {Bramati}},\ and\ \bibinfo {author} {\bibfnamefont
  {Q.}~\bibnamefont {Glorieux}},\ }\href
  {https://doi.org/10.1103/PhysRevResearch.2.043297} {\bibfield  {journal}
  {\bibinfo  {journal} {Phys. Rev. Research}\ }\textbf {\bibinfo {volume}
  {2}},\ \bibinfo {pages} {043297} (\bibinfo {year} {2020})}\BibitemShut
  {NoStop}%
\bibitem [{\citenamefont {Azam}\ \emph {et~al.}(2022)\citenamefont {Azam},
  \citenamefont {Griffin}, \citenamefont {Nazarenko},\ and\ \citenamefont
  {Kaiser}}]{azam2022vortex}%
  \BibitemOpen
  \bibfield  {author} {\bibinfo {author} {\bibfnamefont {P.}~\bibnamefont
  {Azam}}, \bibinfo {author} {\bibfnamefont {A.}~\bibnamefont {Griffin}},
  \bibinfo {author} {\bibfnamefont {S.}~\bibnamefont {Nazarenko}},\ and\
  \bibinfo {author} {\bibfnamefont {R.}~\bibnamefont {Kaiser}},\ }\href
  {https://doi.org/10.1103/PhysRevA.105.043510} {\bibfield  {journal} {\bibinfo
   {journal} {Phys. Rev. A}\ }\textbf {\bibinfo {volume} {105}},\ \bibinfo
  {pages} {043510} (\bibinfo {year} {2022})}\BibitemShut {NoStop}%
\bibitem [{\citenamefont {Isoard}\ \emph {et~al.}(2019)\citenamefont {Isoard},
  \citenamefont {Kamchatnov},\ and\ \citenamefont
  {Pavloff}}]{KamchatnovPRA2019}%
  \BibitemOpen
  \bibfield  {author} {\bibinfo {author} {\bibfnamefont {M.}~\bibnamefont
  {Isoard}}, \bibinfo {author} {\bibfnamefont {A.~M.}\ \bibnamefont
  {Kamchatnov}},\ and\ \bibinfo {author} {\bibfnamefont {N.}~\bibnamefont
  {Pavloff}},\ }\href {https://doi.org/10.1103/PhysRevA.99.053819} {\bibfield
  {journal} {\bibinfo  {journal} {Phys. Rev. A}\ }\textbf {\bibinfo {volume}
  {99}},\ \bibinfo {pages} {053819} (\bibinfo {year} {2019})}\BibitemShut
  {NoStop}%
\bibitem [{\citenamefont {Ivanov}\ \emph {et~al.}(2020)\citenamefont {Ivanov},
  \citenamefont {Suchorski}, \citenamefont {Kamchatnov}, \citenamefont
  {Isoard},\ and\ \citenamefont {Pavloff}}]{KamchatnovPRE2020}%
  \BibitemOpen
  \bibfield  {author} {\bibinfo {author} {\bibfnamefont {S.~K.}\ \bibnamefont
  {Ivanov}}, \bibinfo {author} {\bibfnamefont {J.-E.}\ \bibnamefont
  {Suchorski}}, \bibinfo {author} {\bibfnamefont {A.~M.}\ \bibnamefont
  {Kamchatnov}}, \bibinfo {author} {\bibfnamefont {M.}~\bibnamefont {Isoard}},\
  and\ \bibinfo {author} {\bibfnamefont {N.}~\bibnamefont {Pavloff}},\ }\href
  {https://doi.org/10.1103/PhysRevE.102.032215} {\bibfield  {journal} {\bibinfo
   {journal} {Phys. Rev. E}\ }\textbf {\bibinfo {volume} {102}},\ \bibinfo
  {pages} {032215} (\bibinfo {year} {2020})}\BibitemShut {NoStop}%
\bibitem [{\citenamefont {Simmons}\ \emph {et~al.}(2020)\citenamefont
  {Simmons}, \citenamefont {Bayocboc}, \citenamefont {Pillay}, \citenamefont
  {Colas}, \citenamefont {McCulloch},\ and\ \citenamefont
  {Kheruntsyan}}]{Queensland2PRL020}%
  \BibitemOpen
  \bibfield  {author} {\bibinfo {author} {\bibfnamefont {S.~A.}\ \bibnamefont
  {Simmons}}, \bibinfo {author} {\bibfnamefont {F.~A.}\ \bibnamefont
  {Bayocboc}}, \bibinfo {author} {\bibfnamefont {J.~C.}\ \bibnamefont
  {Pillay}}, \bibinfo {author} {\bibfnamefont {D.}~\bibnamefont {Colas}},
  \bibinfo {author} {\bibfnamefont {I.~P.}\ \bibnamefont {McCulloch}},\ and\
  \bibinfo {author} {\bibfnamefont {K.~V.}\ \bibnamefont {Kheruntsyan}},\
  }\href {https://doi.org/10.1103/PhysRevLett.125.180401} {\bibfield  {journal}
  {\bibinfo  {journal} {Phys. Rev. Lett.}\ }\textbf {\bibinfo {volume} {125}},\
  \bibinfo {pages} {180401} (\bibinfo {year} {2020})}\BibitemShut {NoStop}%
\bibitem [{\citenamefont {Azam}\ \emph {et~al.}(2021)\citenamefont {Azam},
  \citenamefont {Fusaro}, \citenamefont {Fontaine}, \citenamefont {Garnier},
  \citenamefont {Bramati}, \citenamefont {Picozzi}, \citenamefont {Kaiser},
  \citenamefont {Glorieux},\ and\ \citenamefont
  {Bienaim\'e}}]{PhysRevA.104.013515}%
  \BibitemOpen
  \bibfield  {author} {\bibinfo {author} {\bibfnamefont {P.}~\bibnamefont
  {Azam}}, \bibinfo {author} {\bibfnamefont {A.}~\bibnamefont {Fusaro}},
  \bibinfo {author} {\bibfnamefont {Q.}~\bibnamefont {Fontaine}}, \bibinfo
  {author} {\bibfnamefont {J.}~\bibnamefont {Garnier}}, \bibinfo {author}
  {\bibfnamefont {A.}~\bibnamefont {Bramati}}, \bibinfo {author} {\bibfnamefont
  {A.}~\bibnamefont {Picozzi}}, \bibinfo {author} {\bibfnamefont
  {R.}~\bibnamefont {Kaiser}}, \bibinfo {author} {\bibfnamefont
  {Q.}~\bibnamefont {Glorieux}},\ and\ \bibinfo {author} {\bibfnamefont
  {T.}~\bibnamefont {Bienaim\'e}},\ }\href
  {https://doi.org/10.1103/PhysRevA.104.013515} {\bibfield  {journal} {\bibinfo
   {journal} {Phys. Rev. A}\ }\textbf {\bibinfo {volume} {104}},\ \bibinfo
  {pages} {013515} (\bibinfo {year} {2021})}\BibitemShut {NoStop}%
\bibitem [{\citenamefont {Bienaim\'e}\ \emph {et~al.}(2021)\citenamefont
  {Bienaim\'e}, \citenamefont {Isoard}, \citenamefont {Fontaine}, \citenamefont
  {Bramati}, \citenamefont {Kamchatnov}, \citenamefont {Glorieux},\ and\
  \citenamefont {Pavloff}}]{TomBienaimePRL2021}%
  \BibitemOpen
  \bibfield  {author} {\bibinfo {author} {\bibfnamefont {T.}~\bibnamefont
  {Bienaim\'e}}, \bibinfo {author} {\bibfnamefont {M.}~\bibnamefont {Isoard}},
  \bibinfo {author} {\bibfnamefont {Q.}~\bibnamefont {Fontaine}}, \bibinfo
  {author} {\bibfnamefont {A.}~\bibnamefont {Bramati}}, \bibinfo {author}
  {\bibfnamefont {A.~M.}\ \bibnamefont {Kamchatnov}}, \bibinfo {author}
  {\bibfnamefont {Q.}~\bibnamefont {Glorieux}},\ and\ \bibinfo {author}
  {\bibfnamefont {N.}~\bibnamefont {Pavloff}},\ }\href
  {https://doi.org/10.1103/PhysRevLett.126.183901} {\bibfield  {journal}
  {\bibinfo  {journal} {Phys. Rev. Lett.}\ }\textbf {\bibinfo {volume} {126}},\
  \bibinfo {pages} {183901} (\bibinfo {year} {2021})}\BibitemShut {NoStop}%
\bibitem [{\citenamefont {Abuzarli}\ \emph {et~al.}(2021)\citenamefont
  {Abuzarli}, \citenamefont {Bienaim{\'{e}}}, \citenamefont {Giacobino},
  \citenamefont {Bramati},\ and\ \citenamefont {Glorieux}}]{Abuzarli_2021}%
  \BibitemOpen
  \bibfield  {author} {\bibinfo {author} {\bibfnamefont {M.}~\bibnamefont
  {Abuzarli}}, \bibinfo {author} {\bibfnamefont {T.}~\bibnamefont
  {Bienaim{\'{e}}}}, \bibinfo {author} {\bibfnamefont {E.}~\bibnamefont
  {Giacobino}}, \bibinfo {author} {\bibfnamefont {A.}~\bibnamefont {Bramati}},\
  and\ \bibinfo {author} {\bibfnamefont {Q.}~\bibnamefont {Glorieux}},\ }\href
  {https://doi.org/10.1209/0295-5075/134/24001} {\bibfield  {journal} {\bibinfo
   {journal} {{EPL} (Europhysics Letters)}\ }\textbf {\bibinfo {volume}
  {134}},\ \bibinfo {pages} {24001} (\bibinfo {year} {2021})}\BibitemShut
  {NoStop}%
\bibitem [{\citenamefont {Martone}\ \emph {et~al.}(2021)\citenamefont
  {Martone}, \citenamefont {Bienaim\'e},\ and\ \citenamefont
  {Cherroret}}]{MartonePRA2021}%
  \BibitemOpen
  \bibfield  {author} {\bibinfo {author} {\bibfnamefont {G.~I.}\ \bibnamefont
  {Martone}}, \bibinfo {author} {\bibfnamefont {T.}~\bibnamefont
  {Bienaim\'e}},\ and\ \bibinfo {author} {\bibfnamefont {N.}~\bibnamefont
  {Cherroret}},\ }\href {https://doi.org/10.1103/PhysRevA.104.013510}
  {\bibfield  {journal} {\bibinfo  {journal} {Phys. Rev. A}\ }\textbf {\bibinfo
  {volume} {104}},\ \bibinfo {pages} {013510} (\bibinfo {year}
  {2021})}\BibitemShut {NoStop}%
\bibitem [{\citenamefont {Steinhauer}\ \emph {et~al.}(2022)\citenamefont
  {Steinhauer}, \citenamefont {Abuzarli}, \citenamefont {Aladjidi},
  \citenamefont {Bienaim{\'e}}, \citenamefont {Piekarski}, \citenamefont {Liu},
  \citenamefont {Giacobino}, \citenamefont {Bramati},\ and\ \citenamefont
  {Glorieux}}]{steinhauer2021analogue}%
  \BibitemOpen
  \bibfield  {author} {\bibinfo {author} {\bibfnamefont {J.}~\bibnamefont
  {Steinhauer}}, \bibinfo {author} {\bibfnamefont {M.}~\bibnamefont
  {Abuzarli}}, \bibinfo {author} {\bibfnamefont {T.}~\bibnamefont {Aladjidi}},
  \bibinfo {author} {\bibfnamefont {T.}~\bibnamefont {Bienaim{\'e}}}, \bibinfo
  {author} {\bibfnamefont {C.}~\bibnamefont {Piekarski}}, \bibinfo {author}
  {\bibfnamefont {W.}~\bibnamefont {Liu}}, \bibinfo {author} {\bibfnamefont
  {E.}~\bibnamefont {Giacobino}}, \bibinfo {author} {\bibfnamefont
  {A.}~\bibnamefont {Bramati}},\ and\ \bibinfo {author} {\bibfnamefont
  {Q.}~\bibnamefont {Glorieux}},\ }\href
  {https://doi.org/10.1038/s41467-022-30603-1} {\bibfield  {journal} {\bibinfo
  {journal} {Nature Communications}\ }\textbf {\bibinfo {volume} {13}},\
  \bibinfo {pages} {2890} (\bibinfo {year} {2022})}\BibitemShut {NoStop}%
\bibitem [{\citenamefont {Petrov}(2015)}]{PetrovPRL2015}%
  \BibitemOpen
  \bibfield  {author} {\bibinfo {author} {\bibfnamefont {D.~S.}\ \bibnamefont
  {Petrov}},\ }\href {https://doi.org/10.1103/PhysRevLett.115.155302}
  {\bibfield  {journal} {\bibinfo  {journal} {Phys. Rev. Lett.}\ }\textbf
  {\bibinfo {volume} {115}},\ \bibinfo {pages} {155302} (\bibinfo {year}
  {2015})}\BibitemShut {NoStop}%
\bibitem [{\citenamefont {Schmitt}\ \emph {et~al.}(2016)\citenamefont
  {Schmitt}, \citenamefont {Wenzel}, \citenamefont {B{\"o}ttcher},
  \citenamefont {Ferrier-Barbut},\ and\ \citenamefont
  {Pfau}}]{schmitt2016self}%
  \BibitemOpen
  \bibfield  {author} {\bibinfo {author} {\bibfnamefont {M.}~\bibnamefont
  {Schmitt}}, \bibinfo {author} {\bibfnamefont {M.}~\bibnamefont {Wenzel}},
  \bibinfo {author} {\bibfnamefont {F.}~\bibnamefont {B{\"o}ttcher}}, \bibinfo
  {author} {\bibfnamefont {I.}~\bibnamefont {Ferrier-Barbut}},\ and\ \bibinfo
  {author} {\bibfnamefont {T.}~\bibnamefont {Pfau}},\ }\href
  {https://doi.org/10.1038/nature20126} {\bibfield  {journal} {\bibinfo
  {journal} {Nature}\ }\textbf {\bibinfo {volume} {539}},\ \bibinfo {pages}
  {259} (\bibinfo {year} {2016})}\BibitemShut {NoStop}%
\bibitem [{\citenamefont {Chomaz}\ \emph {et~al.}(2016)\citenamefont {Chomaz},
  \citenamefont {Baier}, \citenamefont {Petter}, \citenamefont {Mark},
  \citenamefont {W\"achtler}, \citenamefont {Santos},\ and\ \citenamefont
  {Ferlaino}}]{PhysRevX.6.041039}%
  \BibitemOpen
  \bibfield  {author} {\bibinfo {author} {\bibfnamefont {L.}~\bibnamefont
  {Chomaz}}, \bibinfo {author} {\bibfnamefont {S.}~\bibnamefont {Baier}},
  \bibinfo {author} {\bibfnamefont {D.}~\bibnamefont {Petter}}, \bibinfo
  {author} {\bibfnamefont {M.~J.}\ \bibnamefont {Mark}}, \bibinfo {author}
  {\bibfnamefont {F.}~\bibnamefont {W\"achtler}}, \bibinfo {author}
  {\bibfnamefont {L.}~\bibnamefont {Santos}},\ and\ \bibinfo {author}
  {\bibfnamefont {F.}~\bibnamefont {Ferlaino}},\ }\href
  {https://doi.org/10.1103/PhysRevX.6.041039} {\bibfield  {journal} {\bibinfo
  {journal} {Phys. Rev. X}\ }\textbf {\bibinfo {volume} {6}},\ \bibinfo {pages}
  {041039} (\bibinfo {year} {2016})}\BibitemShut {NoStop}%
\bibitem [{\citenamefont {Cabrera}\ \emph {et~al.}(2018)\citenamefont
  {Cabrera}, \citenamefont {Tanzi}, \citenamefont {Sanz}, \citenamefont
  {Naylor}, \citenamefont {Thomas}, \citenamefont {Cheiney},\ and\
  \citenamefont {Tarruell}}]{CabreraScience2018}%
  \BibitemOpen
  \bibfield  {author} {\bibinfo {author} {\bibfnamefont {C.~R.}\ \bibnamefont
  {Cabrera}}, \bibinfo {author} {\bibfnamefont {L.}~\bibnamefont {Tanzi}},
  \bibinfo {author} {\bibfnamefont {J.}~\bibnamefont {Sanz}}, \bibinfo {author}
  {\bibfnamefont {B.}~\bibnamefont {Naylor}}, \bibinfo {author} {\bibfnamefont
  {P.}~\bibnamefont {Thomas}}, \bibinfo {author} {\bibfnamefont
  {P.}~\bibnamefont {Cheiney}},\ and\ \bibinfo {author} {\bibfnamefont
  {L.}~\bibnamefont {Tarruell}},\ }\href
  {https://doi.org/10.1126/science.aao5686} {\bibfield  {journal} {\bibinfo
  {journal} {Science}\ }\textbf {\bibinfo {volume} {359}},\ \bibinfo {pages}
  {301} (\bibinfo {year} {2018})}\BibitemShut {NoStop}%
\bibitem [{\citenamefont {Cheiney}\ \emph {et~al.}(2018)\citenamefont
  {Cheiney}, \citenamefont {Cabrera}, \citenamefont {Sanz}, \citenamefont
  {Naylor}, \citenamefont {Tanzi},\ and\ \citenamefont
  {Tarruell}}]{PhysRevLett.120.135301}%
  \BibitemOpen
  \bibfield  {author} {\bibinfo {author} {\bibfnamefont {P.}~\bibnamefont
  {Cheiney}}, \bibinfo {author} {\bibfnamefont {C.~R.}\ \bibnamefont
  {Cabrera}}, \bibinfo {author} {\bibfnamefont {J.}~\bibnamefont {Sanz}},
  \bibinfo {author} {\bibfnamefont {B.}~\bibnamefont {Naylor}}, \bibinfo
  {author} {\bibfnamefont {L.}~\bibnamefont {Tanzi}},\ and\ \bibinfo {author}
  {\bibfnamefont {L.}~\bibnamefont {Tarruell}},\ }\href
  {https://doi.org/10.1103/PhysRevLett.120.135301} {\bibfield  {journal}
  {\bibinfo  {journal} {Phys. Rev. Lett.}\ }\textbf {\bibinfo {volume} {120}},\
  \bibinfo {pages} {135301} (\bibinfo {year} {2018})}\BibitemShut {NoStop}%
\bibitem [{\citenamefont {Semeghini}\ \emph {et~al.}(2018)\citenamefont
  {Semeghini}, \citenamefont {Ferioli}, \citenamefont {Masi}, \citenamefont
  {Mazzinghi}, \citenamefont {Wolswijk}, \citenamefont {Minardi}, \citenamefont
  {Modugno}, \citenamefont {Modugno}, \citenamefont {Inguscio},\ and\
  \citenamefont {Fattori}}]{PhysRevLett.120.235301}%
  \BibitemOpen
  \bibfield  {author} {\bibinfo {author} {\bibfnamefont {G.}~\bibnamefont
  {Semeghini}}, \bibinfo {author} {\bibfnamefont {G.}~\bibnamefont {Ferioli}},
  \bibinfo {author} {\bibfnamefont {L.}~\bibnamefont {Masi}}, \bibinfo {author}
  {\bibfnamefont {C.}~\bibnamefont {Mazzinghi}}, \bibinfo {author}
  {\bibfnamefont {L.}~\bibnamefont {Wolswijk}}, \bibinfo {author}
  {\bibfnamefont {F.}~\bibnamefont {Minardi}}, \bibinfo {author} {\bibfnamefont
  {M.}~\bibnamefont {Modugno}}, \bibinfo {author} {\bibfnamefont
  {G.}~\bibnamefont {Modugno}}, \bibinfo {author} {\bibfnamefont
  {M.}~\bibnamefont {Inguscio}},\ and\ \bibinfo {author} {\bibfnamefont
  {M.}~\bibnamefont {Fattori}},\ }\href
  {https://doi.org/10.1103/PhysRevLett.120.235301} {\bibfield  {journal}
  {\bibinfo  {journal} {Phys. Rev. Lett.}\ }\textbf {\bibinfo {volume} {120}},\
  \bibinfo {pages} {235301} (\bibinfo {year} {2018})}\BibitemShut {NoStop}%
\bibitem [{\citenamefont {Anderson}\ \emph {et~al.}(1979)\citenamefont
  {Anderson}, \citenamefont {Bondeson},\ and\ \citenamefont
  {Lisak}}]{anderson_bondeson_lisak_1979}%
  \BibitemOpen
  \bibfield  {author} {\bibinfo {author} {\bibfnamefont {D.}~\bibnamefont
  {Anderson}}, \bibinfo {author} {\bibfnamefont {A.}~\bibnamefont {Bondeson}},\
  and\ \bibinfo {author} {\bibfnamefont {M.}~\bibnamefont {Lisak}},\ }\href
  {https://doi.org/10.1017/S0022377800021826} {\bibfield  {journal} {\bibinfo
  {journal} {Journal of Plasma Physics}\ }\textbf {\bibinfo {volume} {21}},\
  \bibinfo {pages} {259–266} (\bibinfo {year} {1979})}\BibitemShut {NoStop}%
\bibitem [{\citenamefont {Anderson}\ and\ \citenamefont
  {Bonnedal}(1979)}]{anderson1979variational}%
  \BibitemOpen
  \bibfield  {author} {\bibinfo {author} {\bibfnamefont {D.}~\bibnamefont
  {Anderson}}\ and\ \bibinfo {author} {\bibfnamefont {M.}~\bibnamefont
  {Bonnedal}},\ }\href {https://doi.org/https://doi.org/10.1063/1.862445}
  {\bibfield  {journal} {\bibinfo  {journal} {The Physics of Fluids}\ }\textbf
  {\bibinfo {volume} {22}},\ \bibinfo {pages} {105} (\bibinfo {year}
  {1979})}\BibitemShut {NoStop}%
\bibitem [{\citenamefont {Anderson}(1983)}]{Anderson1983}%
  \BibitemOpen
  \bibfield  {author} {\bibinfo {author} {\bibfnamefont {D.}~\bibnamefont
  {Anderson}},\ }\href {https://doi.org/10.1103/PhysRevA.27.3135} {\bibfield
  {journal} {\bibinfo  {journal} {Phys. Rev. A}\ }\textbf {\bibinfo {volume}
  {27}},\ \bibinfo {pages} {3135} (\bibinfo {year} {1983})}\BibitemShut
  {NoStop}%
\bibitem [{\citenamefont {Anderson}\ \emph {et~al.}(1988)\citenamefont
  {Anderson}, \citenamefont {Lisak},\ and\ \citenamefont
  {Reichel}}]{Anderson88}%
  \BibitemOpen
  \bibfield  {author} {\bibinfo {author} {\bibfnamefont {D.}~\bibnamefont
  {Anderson}}, \bibinfo {author} {\bibfnamefont {M.}~\bibnamefont {Lisak}},\
  and\ \bibinfo {author} {\bibfnamefont {T.}~\bibnamefont {Reichel}},\ }\href
  {https://doi.org/10.1364/JOSAB.5.000207} {\bibfield  {journal} {\bibinfo
  {journal} {J. Opt. Soc. Am. B}\ }\textbf {\bibinfo {volume} {5}},\ \bibinfo
  {pages} {207} (\bibinfo {year} {1988})}\BibitemShut {NoStop}%
\bibitem [{\citenamefont {Malomed}(2002)}]{MalomedReview2002}%
  \BibitemOpen
  \bibfield  {author} {\bibinfo {author} {\bibfnamefont {B.~A.}\ \bibnamefont
  {Malomed}},\ }\href@noop {} {\emph {\bibinfo {title} {Variational methods in
  nonlinear fiber optics and related fields}}},\ Progress in Optics\ (\bibinfo
  {publisher} {Elsevier},\ \bibinfo {year} {2002})\BibitemShut {NoStop}%
\bibitem [{\citenamefont {Fibich}\ and\ \citenamefont
  {Gaeta}(2000)}]{FibichGaetaOL2000}%
  \BibitemOpen
  \bibfield  {author} {\bibinfo {author} {\bibfnamefont {G.}~\bibnamefont
  {Fibich}}\ and\ \bibinfo {author} {\bibfnamefont {A.~L.}\ \bibnamefont
  {Gaeta}},\ }\href {https://doi.org/10.1364/OL.25.000335} {\bibfield
  {journal} {\bibinfo  {journal} {Opt. Lett.}\ }\textbf {\bibinfo {volume}
  {25}},\ \bibinfo {pages} {335} (\bibinfo {year} {2000})}\BibitemShut
  {NoStop}%
\bibitem [{\citenamefont {Steck}(2008)}]{steck2001rubidium}%
  \BibitemOpen
  \bibfield  {author} {\bibinfo {author} {\bibfnamefont {D.~A.}\ \bibnamefont
  {Steck}},\ }\href {https://steck.us/alkalidata/rubidium85numbers.pdf}
  {\bibinfo {title} {Rubidium 85 {D} line data}} (\bibinfo {year}
  {2008})\BibitemShut {NoStop}%
\bibitem [{\citenamefont {Grimm}\ \emph {et~al.}(2000)\citenamefont {Grimm},
  \citenamefont {Weidemüller},\ and\ \citenamefont
  {Ovchinnikov}}]{GRIMM200095}%
  \BibitemOpen
  \bibfield  {author} {\bibinfo {author} {\bibfnamefont {R.}~\bibnamefont
  {Grimm}}, \bibinfo {author} {\bibfnamefont {M.}~\bibnamefont
  {Weidemüller}},\ and\ \bibinfo {author} {\bibfnamefont {Y.~B.}\ \bibnamefont
  {Ovchinnikov}},\ }\href@noop {} {\emph {\bibinfo {title} {Optical Dipole
  Traps for Neutral Atoms}}},\ Advances In Atomic, Molecular, and Optical
  Physics\ (\bibinfo  {publisher} {Academic Press},\ \bibinfo {year}
  {2000})\BibitemShut {NoStop}%
\bibitem [{\citenamefont {Pethick}\ and\ \citenamefont
  {Smith}(2008)}]{pethick_smith_2008}%
  \BibitemOpen
  \bibfield  {author} {\bibinfo {author} {\bibfnamefont {C.~J.}\ \bibnamefont
  {Pethick}}\ and\ \bibinfo {author} {\bibfnamefont {H.}~\bibnamefont
  {Smith}},\ }\href@noop {} {\emph {\bibinfo {title} {Bose–Einstein
  Condensation in Dilute Gases}}},\ \bibinfo {edition} {2nd}\ ed.\ (\bibinfo
  {publisher} {Cambridge University Press},\ \bibinfo {year}
  {2008})\BibitemShut {NoStop}%
\bibitem [{\citenamefont {Cappellaro}\ \emph {et~al.}(2017)\citenamefont
  {Cappellaro}, \citenamefont {Macr{\`i}}, \citenamefont {Bertacco},\ and\
  \citenamefont {Salasnich}}]{cappellaro2017equation}%
  \BibitemOpen
  \bibfield  {author} {\bibinfo {author} {\bibfnamefont {A.}~\bibnamefont
  {Cappellaro}}, \bibinfo {author} {\bibfnamefont {T.}~\bibnamefont
  {Macr{\`i}}}, \bibinfo {author} {\bibfnamefont {G.~F.}\ \bibnamefont
  {Bertacco}},\ and\ \bibinfo {author} {\bibfnamefont {L.}~\bibnamefont
  {Salasnich}},\ }\href {https://doi.org/10.1038/s41598-017-13647-y} {\bibfield
   {journal} {\bibinfo  {journal} {Scientific Reports}\ }\textbf {\bibinfo
  {volume} {7}},\ \bibinfo {pages} {13358} (\bibinfo {year}
  {2017})}\BibitemShut {NoStop}%
\bibitem [{\citenamefont {Cappellaro}\ \emph {et~al.}(2018)\citenamefont
  {Cappellaro}, \citenamefont {Macr\`{\i}},\ and\ \citenamefont
  {Salasnich}}]{MacriCapellaroLucaPRA2018}%
  \BibitemOpen
  \bibfield  {author} {\bibinfo {author} {\bibfnamefont {A.}~\bibnamefont
  {Cappellaro}}, \bibinfo {author} {\bibfnamefont {T.}~\bibnamefont
  {Macr\`{\i}}},\ and\ \bibinfo {author} {\bibfnamefont {L.}~\bibnamefont
  {Salasnich}},\ }\href {https://doi.org/10.1103/PhysRevA.97.053623} {\bibfield
   {journal} {\bibinfo  {journal} {Phys. Rev. A}\ }\textbf {\bibinfo {volume}
  {97}},\ \bibinfo {pages} {053623} (\bibinfo {year} {2018})}\BibitemShut
  {NoStop}%
\bibitem [{\citenamefont {Hu}\ and\ \citenamefont
  {Liu}(2020)}]{PhysRevA.102.053303}%
  \BibitemOpen
  \bibfield  {author} {\bibinfo {author} {\bibfnamefont {H.}~\bibnamefont
  {Hu}}\ and\ \bibinfo {author} {\bibfnamefont {X.-J.}\ \bibnamefont {Liu}},\
  }\href {https://doi.org/10.1103/PhysRevA.102.053303} {\bibfield  {journal}
  {\bibinfo  {journal} {Phys. Rev. A}\ }\textbf {\bibinfo {volume} {102}},\
  \bibinfo {pages} {053303} (\bibinfo {year} {2020})}\BibitemShut {NoStop}%
\bibitem [{\citenamefont {Michinel}\ \emph {et~al.}(2002)\citenamefont
  {Michinel}, \citenamefont {Campo-T\'aboas}, \citenamefont
  {Garc\'{\i}a-Fern\'andez}, \citenamefont {Salgueiro},\ and\ \citenamefont
  {Quiroga-Teixeiro}}]{PhysRevE.65.066604}%
  \BibitemOpen
  \bibfield  {author} {\bibinfo {author} {\bibfnamefont {H.}~\bibnamefont
  {Michinel}}, \bibinfo {author} {\bibfnamefont {J.}~\bibnamefont
  {Campo-T\'aboas}}, \bibinfo {author} {\bibfnamefont {R.}~\bibnamefont
  {Garc\'{\i}a-Fern\'andez}}, \bibinfo {author} {\bibfnamefont {J.~R.}\
  \bibnamefont {Salgueiro}},\ and\ \bibinfo {author} {\bibfnamefont {M.~L.}\
  \bibnamefont {Quiroga-Teixeiro}},\ }\href
  {https://doi.org/10.1103/PhysRevE.65.066604} {\bibfield  {journal} {\bibinfo
  {journal} {Phys. Rev. E}\ }\textbf {\bibinfo {volume} {65}},\ \bibinfo
  {pages} {066604} (\bibinfo {year} {2002})}\BibitemShut {NoStop}%
\bibitem [{\citenamefont {Alexandrescu}\ \emph {et~al.}(2009)\citenamefont
  {Alexandrescu}, \citenamefont {Michinel},\ and\ \citenamefont
  {P\'erez-Garc\'{\i}a}}]{Michinel_PRA2009}%
  \BibitemOpen
  \bibfield  {author} {\bibinfo {author} {\bibfnamefont {A.}~\bibnamefont
  {Alexandrescu}}, \bibinfo {author} {\bibfnamefont {H.}~\bibnamefont
  {Michinel}},\ and\ \bibinfo {author} {\bibfnamefont {V.~M.}\ \bibnamefont
  {P\'erez-Garc\'{\i}a}},\ }\href {https://doi.org/10.1103/PhysRevA.79.013833}
  {\bibfield  {journal} {\bibinfo  {journal} {Phys. Rev. A}\ }\textbf {\bibinfo
  {volume} {79}},\ \bibinfo {pages} {013833} (\bibinfo {year}
  {2009})}\BibitemShut {NoStop}%
\bibitem [{\citenamefont {Paredes}\ \emph {et~al.}(2014)\citenamefont
  {Paredes}, \citenamefont {Feijoo},\ and\ \citenamefont
  {Michinel}}]{FeijooPRL2014}%
  \BibitemOpen
  \bibfield  {author} {\bibinfo {author} {\bibfnamefont {A.}~\bibnamefont
  {Paredes}}, \bibinfo {author} {\bibfnamefont {D.}~\bibnamefont {Feijoo}},\
  and\ \bibinfo {author} {\bibfnamefont {H.}~\bibnamefont {Michinel}},\ }\href
  {https://doi.org/10.1103/PhysRevLett.112.173901} {\bibfield  {journal}
  {\bibinfo  {journal} {Phys. Rev. Lett.}\ }\textbf {\bibinfo {volume} {112}},\
  \bibinfo {pages} {173901} (\bibinfo {year} {2014})}\BibitemShut {NoStop}%
\bibitem [{\citenamefont {Michinel}\ \emph {et~al.}(2006)\citenamefont
  {Michinel}, \citenamefont {Paz-Alonso},\ and\ \citenamefont
  {P\'erez-Garc\'{\i}a}}]{PazAlonsoPRL2006}%
  \BibitemOpen
  \bibfield  {author} {\bibinfo {author} {\bibfnamefont {H.}~\bibnamefont
  {Michinel}}, \bibinfo {author} {\bibfnamefont {M.~J.}\ \bibnamefont
  {Paz-Alonso}},\ and\ \bibinfo {author} {\bibfnamefont {V.~M.}\ \bibnamefont
  {P\'erez-Garc\'{\i}a}},\ }\href
  {https://doi.org/10.1103/PhysRevLett.96.023903} {\bibfield  {journal}
  {\bibinfo  {journal} {Phys. Rev. Lett.}\ }\textbf {\bibinfo {volume} {96}},\
  \bibinfo {pages} {023903} (\bibinfo {year} {2006})}\BibitemShut {NoStop}%
\bibitem [{\citenamefont {Westerberg}\ \emph {et~al.}(2018)\citenamefont
  {Westerberg}, \citenamefont {Wilson}, \citenamefont {Duncan}, \citenamefont
  {Faccio}, \citenamefont {Wright}, \citenamefont {\"Ohberg},\ and\
  \citenamefont {Valiente}}]{westerberg2018self}%
  \BibitemOpen
  \bibfield  {author} {\bibinfo {author} {\bibfnamefont {N.}~\bibnamefont
  {Westerberg}}, \bibinfo {author} {\bibfnamefont {K.~E.}\ \bibnamefont
  {Wilson}}, \bibinfo {author} {\bibfnamefont {C.~W.}\ \bibnamefont {Duncan}},
  \bibinfo {author} {\bibfnamefont {D.}~\bibnamefont {Faccio}}, \bibinfo
  {author} {\bibfnamefont {E.~M.}\ \bibnamefont {Wright}}, \bibinfo {author}
  {\bibfnamefont {P.}~\bibnamefont {\"Ohberg}},\ and\ \bibinfo {author}
  {\bibfnamefont {M.}~\bibnamefont {Valiente}},\ }\href
  {https://doi.org/10.1103/PhysRevA.98.053835} {\bibfield  {journal} {\bibinfo
  {journal} {Phys. Rev. A}\ }\textbf {\bibinfo {volume} {98}},\ \bibinfo
  {pages} {053835} (\bibinfo {year} {2018})}\BibitemShut {NoStop}%
\bibitem [{\citenamefont {Wilson}\ \emph {et~al.}(2018)\citenamefont {Wilson},
  \citenamefont {Westerberg}, \citenamefont {Valiente}, \citenamefont {Duncan},
  \citenamefont {Wright}, \citenamefont {\"Ohberg},\ and\ \citenamefont
  {Faccio}}]{wilson2018observation}%
  \BibitemOpen
  \bibfield  {author} {\bibinfo {author} {\bibfnamefont {K.~E.}\ \bibnamefont
  {Wilson}}, \bibinfo {author} {\bibfnamefont {N.}~\bibnamefont {Westerberg}},
  \bibinfo {author} {\bibfnamefont {M.}~\bibnamefont {Valiente}}, \bibinfo
  {author} {\bibfnamefont {C.~W.}\ \bibnamefont {Duncan}}, \bibinfo {author}
  {\bibfnamefont {E.~M.}\ \bibnamefont {Wright}}, \bibinfo {author}
  {\bibfnamefont {P.}~\bibnamefont {\"Ohberg}},\ and\ \bibinfo {author}
  {\bibfnamefont {D.}~\bibnamefont {Faccio}},\ }\href
  {https://doi.org/10.1103/PhysRevLett.121.133903} {\bibfield  {journal}
  {\bibinfo  {journal} {Phys. Rev. Lett.}\ }\textbf {\bibinfo {volume} {121}},\
  \bibinfo {pages} {133903} (\bibinfo {year} {2018})}\BibitemShut {NoStop}%
\bibitem [{\citenamefont {Vakhitov}\ and\ \citenamefont
  {Kolokolov}(1973)}]{Vakhitov1973}%
  \BibitemOpen
  \bibfield  {author} {\bibinfo {author} {\bibfnamefont {N.~G.}\ \bibnamefont
  {Vakhitov}}\ and\ \bibinfo {author} {\bibfnamefont {A.~A.}\ \bibnamefont
  {Kolokolov}},\ }\href {https://doi.org/10.1007/BF01031343} {\bibfield
  {journal} {\bibinfo  {journal} {Radiophysics and Quantum Electronics}\
  }\textbf {\bibinfo {volume} {16}},\ \bibinfo {pages} {783} (\bibinfo {year}
  {1973})}\BibitemShut {NoStop}%
\bibitem [{\citenamefont {Rasmussen}\ and\ \citenamefont
  {Rypdal}(1986)}]{J.JuulRasmussen_1986}%
  \BibitemOpen
  \bibfield  {author} {\bibinfo {author} {\bibfnamefont {J.~J.}\ \bibnamefont
  {Rasmussen}}\ and\ \bibinfo {author} {\bibfnamefont {K.}~\bibnamefont
  {Rypdal}},\ }\href {https://doi.org/10.1088/0031-8949/33/6/001} {\bibfield
  {journal} {\bibinfo  {journal} {Physica Scripta}\ }\textbf {\bibinfo {volume}
  {33}},\ \bibinfo {pages} {481} (\bibinfo {year} {1986})}\BibitemShut
  {NoStop}%
\bibitem [{\citenamefont {Fibich}(2015)}]{fibich2015nonlinear}%
  \BibitemOpen
  \bibfield  {author} {\bibinfo {author} {\bibfnamefont {G.}~\bibnamefont
  {Fibich}},\ }\href@noop {} {\emph {\bibinfo {title} {The Nonlinear
  Schr{\"o}dinger Equation: Singular Solutions and Optical Collapse}}},\
  Applied Mathematical Sciences\ (\bibinfo  {publisher} {Springer International
  Publishing},\ \bibinfo {year} {2015})\BibitemShut {NoStop}%
\bibitem [{\citenamefont {Haus}(1966)}]{HAUS}%
  \BibitemOpen
  \bibfield  {author} {\bibinfo {author} {\bibfnamefont {H.~A.}\ \bibnamefont
  {Haus}},\ }\href {https://doi.org/10.1063/1.1754519} {\bibfield  {journal}
  {\bibinfo  {journal} {Applied Physics Letters}\ }\textbf {\bibinfo {volume}
  {8}},\ \bibinfo {pages} {128} (\bibinfo {year} {1966})}\BibitemShut {NoStop}%
\bibitem [{\citenamefont {Yankauskas}(1966)}]{Yankauskas1966}%
  \BibitemOpen
  \bibfield  {author} {\bibinfo {author} {\bibfnamefont {Z.~K.}\ \bibnamefont
  {Yankauskas}},\ }\href {https://doi.org/10.1007/BF01038975} {\bibfield
  {journal} {\bibinfo  {journal} {Soviet Radiophysics}\ }\textbf {\bibinfo
  {volume} {9}},\ \bibinfo {pages} {261} (\bibinfo {year} {1966})}\BibitemShut
  {NoStop}%
\bibitem [{\citenamefont {Chiao}\ \emph {et~al.}(1964)\citenamefont {Chiao},
  \citenamefont {Garmire},\ and\ \citenamefont
  {Townes}}]{ChiaoGarmireTownes64}%
  \BibitemOpen
  \bibfield  {author} {\bibinfo {author} {\bibfnamefont {R.~Y.}\ \bibnamefont
  {Chiao}}, \bibinfo {author} {\bibfnamefont {E.}~\bibnamefont {Garmire}},\
  and\ \bibinfo {author} {\bibfnamefont {C.~H.}\ \bibnamefont {Townes}},\
  }\href {https://doi.org/10.1103/PhysRevLett.13.479} {\bibfield  {journal}
  {\bibinfo  {journal} {Phys. Rev. Lett.}\ }\textbf {\bibinfo {volume} {13}},\
  \bibinfo {pages} {479} (\bibinfo {year} {1964})}\BibitemShut {NoStop}%
\bibitem [{\citenamefont {Donley}\ \emph {et~al.}(2001)\citenamefont {Donley},
  \citenamefont {Claussen}, \citenamefont {Cornish}, \citenamefont {Roberts},
  \citenamefont {Cornell},\ and\ \citenamefont {Wieman}}]{Donley2001}%
  \BibitemOpen
  \bibfield  {author} {\bibinfo {author} {\bibfnamefont {E.~A.}\ \bibnamefont
  {Donley}}, \bibinfo {author} {\bibfnamefont {N.~R.}\ \bibnamefont
  {Claussen}}, \bibinfo {author} {\bibfnamefont {S.~L.}\ \bibnamefont
  {Cornish}}, \bibinfo {author} {\bibfnamefont {J.~L.}\ \bibnamefont
  {Roberts}}, \bibinfo {author} {\bibfnamefont {E.~A.}\ \bibnamefont
  {Cornell}},\ and\ \bibinfo {author} {\bibfnamefont {C.~E.}\ \bibnamefont
  {Wieman}},\ }\href {https://doi.org/10.1038/35085500} {\bibfield  {journal}
  {\bibinfo  {journal} {Nature}\ }\textbf {\bibinfo {volume} {412}},\ \bibinfo
  {pages} {295} (\bibinfo {year} {2001})}\BibitemShut {NoStop}%
\bibitem [{\citenamefont {Khaykovich}\ \emph {et~al.}(2002)\citenamefont
  {Khaykovich}, \citenamefont {Schreck}, \citenamefont {Ferrari}, \citenamefont
  {Bourdel}, \citenamefont {Cubizolles}, \citenamefont {Carr}, \citenamefont
  {Castin},\ and\ \citenamefont {Salomon}}]{Salomon}%
  \BibitemOpen
  \bibfield  {author} {\bibinfo {author} {\bibfnamefont {L.}~\bibnamefont
  {Khaykovich}}, \bibinfo {author} {\bibfnamefont {F.}~\bibnamefont {Schreck}},
  \bibinfo {author} {\bibfnamefont {G.}~\bibnamefont {Ferrari}}, \bibinfo
  {author} {\bibfnamefont {T.}~\bibnamefont {Bourdel}}, \bibinfo {author}
  {\bibfnamefont {J.}~\bibnamefont {Cubizolles}}, \bibinfo {author}
  {\bibfnamefont {L.~D.}\ \bibnamefont {Carr}}, \bibinfo {author}
  {\bibfnamefont {Y.}~\bibnamefont {Castin}},\ and\ \bibinfo {author}
  {\bibfnamefont {C.}~\bibnamefont {Salomon}},\ }\href
  {https://doi.org/10.1126/science.1071021} {\bibfield  {journal} {\bibinfo
  {journal} {Science}\ }\textbf {\bibinfo {volume} {296}},\ \bibinfo {pages}
  {1290} (\bibinfo {year} {2002})}\BibitemShut {NoStop}%
\bibitem [{\citenamefont {Strecker}\ \emph {et~al.}(2002)\citenamefont
  {Strecker}, \citenamefont {Partridge}, \citenamefont {Truscott},\ and\
  \citenamefont {Hulet}}]{Strecker2002}%
  \BibitemOpen
  \bibfield  {author} {\bibinfo {author} {\bibfnamefont {K.~E.}\ \bibnamefont
  {Strecker}}, \bibinfo {author} {\bibfnamefont {G.~B.}\ \bibnamefont
  {Partridge}}, \bibinfo {author} {\bibfnamefont {A.~G.}\ \bibnamefont
  {Truscott}},\ and\ \bibinfo {author} {\bibfnamefont {R.~G.}\ \bibnamefont
  {Hulet}},\ }\href {https://doi.org/10.1038/nature747} {\bibfield  {journal}
  {\bibinfo  {journal} {Nature}\ }\textbf {\bibinfo {volume} {417}},\ \bibinfo
  {pages} {150} (\bibinfo {year} {2002})}\BibitemShut {NoStop}%
\bibitem [{\citenamefont {Eigen}\ \emph {et~al.}(2016)\citenamefont {Eigen},
  \citenamefont {Gaunt}, \citenamefont {Suleymanzade}, \citenamefont {Navon},
  \citenamefont {Hadzibabic},\ and\ \citenamefont {Smith}}]{Eigen}%
  \BibitemOpen
  \bibfield  {author} {\bibinfo {author} {\bibfnamefont {C.}~\bibnamefont
  {Eigen}}, \bibinfo {author} {\bibfnamefont {A.~L.}\ \bibnamefont {Gaunt}},
  \bibinfo {author} {\bibfnamefont {A.}~\bibnamefont {Suleymanzade}}, \bibinfo
  {author} {\bibfnamefont {N.}~\bibnamefont {Navon}}, \bibinfo {author}
  {\bibfnamefont {Z.}~\bibnamefont {Hadzibabic}},\ and\ \bibinfo {author}
  {\bibfnamefont {R.~P.}\ \bibnamefont {Smith}},\ }\href
  {https://doi.org/10.1103/PhysRevX.6.041058} {\bibfield  {journal} {\bibinfo
  {journal} {Phys. Rev. X}\ }\textbf {\bibinfo {volume} {6}},\ \bibinfo {pages}
  {041058} (\bibinfo {year} {2016})}\BibitemShut {NoStop}%
\bibitem [{\citenamefont {Weinstein}(1983)}]{weinstein1982nonlinear}%
  \BibitemOpen
  \bibfield  {author} {\bibinfo {author} {\bibfnamefont {M.~I.}\ \bibnamefont
  {Weinstein}},\ }\href {https://doi.org/10.1007/BF01208265} {\bibfield
  {journal} {\bibinfo  {journal} {Communications in Mathematical Physics}\
  }\textbf {\bibinfo {volume} {87}},\ \bibinfo {pages} {567} (\bibinfo {year}
  {1983})}\BibitemShut {NoStop}%
\bibitem [{\citenamefont {Azam}(2021)}]{pierreThesis}%
  \BibitemOpen
  \bibfield  {author} {\bibinfo {author} {\bibfnamefont {P.}~\bibnamefont
  {Azam}},\ }\emph {\bibinfo {title} {Fluides quantiques de lumière avec des
  vapeurs atomiques chaudes}},\ \href {http://www.theses.fr/2021COAZ4059}
  {Ph.D. thesis} (\bibinfo {year} {2021}),\ \bibinfo {note} {thèse de doctorat
  dirigée par Kaiser, Robin Physique Université Côte d'Azur
  2021}\BibitemShut {NoStop}%
\bibitem [{\citenamefont {Moll}\ \emph {et~al.}(2003)\citenamefont {Moll},
  \citenamefont {Gaeta},\ and\ \citenamefont {Fibich}}]{MollGaetaFibich}%
  \BibitemOpen
  \bibfield  {author} {\bibinfo {author} {\bibfnamefont {K.~D.}\ \bibnamefont
  {Moll}}, \bibinfo {author} {\bibfnamefont {A.~L.}\ \bibnamefont {Gaeta}},\
  and\ \bibinfo {author} {\bibfnamefont {G.}~\bibnamefont {Fibich}},\ }\href
  {https://doi.org/10.1103/PhysRevLett.90.203902} {\bibfield  {journal}
  {\bibinfo  {journal} {Phys. Rev. Lett.}\ }\textbf {\bibinfo {volume} {90}},\
  \bibinfo {pages} {203902} (\bibinfo {year} {2003})}\BibitemShut {NoStop}%
\bibitem [{\citenamefont {Bakkali-Hassani}\ \emph {et~al.}(2021)\citenamefont
  {Bakkali-Hassani}, \citenamefont {Maury}, \citenamefont {Zou}, \citenamefont
  {Le~Cerf}, \citenamefont {Saint-Jalm}, \citenamefont {Castilho},
  \citenamefont {Nascimbene}, \citenamefont {Dalibard},\ and\ \citenamefont
  {Beugnon}}]{bakkali2021realization}%
  \BibitemOpen
  \bibfield  {author} {\bibinfo {author} {\bibfnamefont {B.}~\bibnamefont
  {Bakkali-Hassani}}, \bibinfo {author} {\bibfnamefont {C.}~\bibnamefont
  {Maury}}, \bibinfo {author} {\bibfnamefont {Y.-Q.}\ \bibnamefont {Zou}},
  \bibinfo {author} {\bibfnamefont {E.}~\bibnamefont {Le~Cerf}}, \bibinfo
  {author} {\bibfnamefont {R.}~\bibnamefont {Saint-Jalm}}, \bibinfo {author}
  {\bibfnamefont {P.~C.~M.}\ \bibnamefont {Castilho}}, \bibinfo {author}
  {\bibfnamefont {S.}~\bibnamefont {Nascimbene}}, \bibinfo {author}
  {\bibfnamefont {J.}~\bibnamefont {Dalibard}},\ and\ \bibinfo {author}
  {\bibfnamefont {J.}~\bibnamefont {Beugnon}},\ }\href
  {https://doi.org/10.1103/PhysRevLett.127.023603} {\bibfield  {journal}
  {\bibinfo  {journal} {Phys. Rev. Lett.}\ }\textbf {\bibinfo {volume} {127}},\
  \bibinfo {pages} {023603} (\bibinfo {year} {2021})}\BibitemShut {NoStop}%
\bibitem [{\citenamefont {Chen}\ and\ \citenamefont
  {Hung}(2020)}]{chen2021observation}%
  \BibitemOpen
  \bibfield  {author} {\bibinfo {author} {\bibfnamefont {C.-A.}\ \bibnamefont
  {Chen}}\ and\ \bibinfo {author} {\bibfnamefont {C.-L.}\ \bibnamefont
  {Hung}},\ }\href {https://doi.org/10.1103/PhysRevLett.125.250401} {\bibfield
  {journal} {\bibinfo  {journal} {Phys. Rev. Lett.}\ }\textbf {\bibinfo
  {volume} {125}},\ \bibinfo {pages} {250401} (\bibinfo {year}
  {2020})}\BibitemShut {NoStop}%
\bibitem [{\citenamefont {Chen}\ and\ \citenamefont {Hung}(2021)}]{Cheng2021}%
  \BibitemOpen
  \bibfield  {author} {\bibinfo {author} {\bibfnamefont {C.-A.}\ \bibnamefont
  {Chen}}\ and\ \bibinfo {author} {\bibfnamefont {C.-L.}\ \bibnamefont
  {Hung}},\ }\href {https://doi.org/10.1103/PhysRevLett.127.023604} {\bibfield
  {journal} {\bibinfo  {journal} {Phys. Rev. Lett.}\ }\textbf {\bibinfo
  {volume} {127}},\ \bibinfo {pages} {023604} (\bibinfo {year}
  {2021})}\BibitemShut {NoStop}%
\bibitem [{\citenamefont {Berg{\'e}}(1998)}]{BERGE1998259}%
  \BibitemOpen
  \bibfield  {author} {\bibinfo {author} {\bibfnamefont {L.}~\bibnamefont
  {Berg{\'e}}},\ }\href
  {https://doi.org/https://doi.org/10.1016/S0370-1573(97)00092-6} {\bibfield
  {journal} {\bibinfo  {journal} {Physics Reports}\ }\textbf {\bibinfo {volume}
  {303}},\ \bibinfo {pages} {259} (\bibinfo {year} {1998})}\BibitemShut
  {NoStop}%
\bibitem [{\citenamefont {Ferioli}\ \emph {et~al.}(2020)\citenamefont
  {Ferioli}, \citenamefont {Semeghini}, \citenamefont {Terradas-Brians\'o},
  \citenamefont {Masi}, \citenamefont {Fattori},\ and\ \citenamefont
  {Modugno}}]{SemeghiniTerradas}%
  \BibitemOpen
  \bibfield  {author} {\bibinfo {author} {\bibfnamefont {G.}~\bibnamefont
  {Ferioli}}, \bibinfo {author} {\bibfnamefont {G.}~\bibnamefont {Semeghini}},
  \bibinfo {author} {\bibfnamefont {S.}~\bibnamefont {Terradas-Brians\'o}},
  \bibinfo {author} {\bibfnamefont {L.}~\bibnamefont {Masi}}, \bibinfo {author}
  {\bibfnamefont {M.}~\bibnamefont {Fattori}},\ and\ \bibinfo {author}
  {\bibfnamefont {M.}~\bibnamefont {Modugno}},\ }\href
  {https://doi.org/10.1103/PhysRevResearch.2.013269} {\bibfield  {journal}
  {\bibinfo  {journal} {Phys. Rev. Research}\ }\textbf {\bibinfo {volume}
  {2}},\ \bibinfo {pages} {013269} (\bibinfo {year} {2020})}\BibitemShut
  {NoStop}%
\bibitem [{\citenamefont {Otajonov}\ \emph {et~al.}(2020)\citenamefont
  {Otajonov}, \citenamefont {Tsoy},\ and\ \citenamefont
  {Abdullaev}}]{PhysRevE.102.062217}%
  \BibitemOpen
  \bibfield  {author} {\bibinfo {author} {\bibfnamefont {S.~R.}\ \bibnamefont
  {Otajonov}}, \bibinfo {author} {\bibfnamefont {E.~N.}\ \bibnamefont {Tsoy}},\
  and\ \bibinfo {author} {\bibfnamefont {F.~K.}\ \bibnamefont {Abdullaev}},\
  }\href {https://doi.org/10.1103/PhysRevE.102.062217} {\bibfield  {journal}
  {\bibinfo  {journal} {Phys. Rev. E}\ }\textbf {\bibinfo {volume} {102}},\
  \bibinfo {pages} {062217} (\bibinfo {year} {2020})}\BibitemShut {NoStop}%
\bibitem [{\citenamefont {St\"urmer}\ \emph {et~al.}(2021)\citenamefont
  {St\"urmer}, \citenamefont {Tengstrand}, \citenamefont {Sachdeva},\ and\
  \citenamefont {Reimann}}]{Sachdeva}%
  \BibitemOpen
  \bibfield  {author} {\bibinfo {author} {\bibfnamefont {P.}~\bibnamefont
  {St\"urmer}}, \bibinfo {author} {\bibfnamefont {M.~N.}\ \bibnamefont
  {Tengstrand}}, \bibinfo {author} {\bibfnamefont {R.}~\bibnamefont
  {Sachdeva}},\ and\ \bibinfo {author} {\bibfnamefont {S.~M.}\ \bibnamefont
  {Reimann}},\ }\href {https://doi.org/10.1103/PhysRevA.103.053302} {\bibfield
  {journal} {\bibinfo  {journal} {Phys. Rev. A}\ }\textbf {\bibinfo {volume}
  {103}},\ \bibinfo {pages} {053302} (\bibinfo {year} {2021})}\BibitemShut
  {NoStop}%
\bibitem [{\citenamefont {Fort}\ and\ \citenamefont
  {Modugno}(2021)}]{FortModugno}%
  \BibitemOpen
  \bibfield  {author} {\bibinfo {author} {\bibfnamefont {C.}~\bibnamefont
  {Fort}}\ and\ \bibinfo {author} {\bibfnamefont {M.}~\bibnamefont {Modugno}},\
  }\href {https://www.mdpi.com/2076-3417/11/2/866} {\bibfield  {journal}
  {\bibinfo  {journal} {Applied Sciences}\ }\textbf {\bibinfo {volume} {11}}
  (\bibinfo {year} {2021})}\BibitemShut {NoStop}%
\bibitem [{\citenamefont {Labeyrie}\ and\ \citenamefont
  {Bortolozzo}(2011)}]{Labeyrie:11}%
  \BibitemOpen
  \bibfield  {author} {\bibinfo {author} {\bibfnamefont {G.}~\bibnamefont
  {Labeyrie}}\ and\ \bibinfo {author} {\bibfnamefont {U.}~\bibnamefont
  {Bortolozzo}},\ }\href {https://doi.org/10.1364/OL.36.002158} {\bibfield
  {journal} {\bibinfo  {journal} {Opt. Lett.}\ }\textbf {\bibinfo {volume}
  {36}},\ \bibinfo {pages} {2158} (\bibinfo {year} {2011})}\BibitemShut
  {NoStop}%
\bibitem [{\citenamefont {Silberberg}(1990)}]{Silberberg:90}%
  \BibitemOpen
  \bibfield  {author} {\bibinfo {author} {\bibfnamefont {Y.}~\bibnamefont
  {Silberberg}},\ }\href {https://doi.org/10.1364/OL.15.001282} {\bibfield
  {journal} {\bibinfo  {journal} {Opt. Lett.}\ }\textbf {\bibinfo {volume}
  {15}},\ \bibinfo {pages} {1282} (\bibinfo {year} {1990})}\BibitemShut
  {NoStop}%
\bibitem [{\citenamefont {Akhmediev}\ \emph {et~al.}(1992)\citenamefont
  {Akhmediev}, \citenamefont {Korneev},\ and\ \citenamefont
  {Nabiev}}]{akhmediev1992modulation}%
  \BibitemOpen
  \bibfield  {author} {\bibinfo {author} {\bibfnamefont {N.~N.}\ \bibnamefont
  {Akhmediev}}, \bibinfo {author} {\bibfnamefont {V.~I.}\ \bibnamefont
  {Korneev}},\ and\ \bibinfo {author} {\bibfnamefont {R.~F.}\ \bibnamefont
  {Nabiev}},\ }\href {https://doi.org/10.1364/OL.17.000393} {\bibfield
  {journal} {\bibinfo  {journal} {Opt. Lett.}\ }\textbf {\bibinfo {volume}
  {17}},\ \bibinfo {pages} {393} (\bibinfo {year} {1992})}\BibitemShut
  {NoStop}%
\bibitem [{\citenamefont {Defenu}\ \emph {et~al.}(2021)\citenamefont {Defenu},
  \citenamefont {Donner}, \citenamefont {Macrì}, \citenamefont {Pagano},
  \citenamefont {Ruffo},\ and\ \citenamefont
  {Trombettoni}}]{https://doi.org/10.48550/arxiv.2109.01063}%
  \BibitemOpen
  \bibfield  {author} {\bibinfo {author} {\bibfnamefont {N.}~\bibnamefont
  {Defenu}}, \bibinfo {author} {\bibfnamefont {T.}~\bibnamefont {Donner}},
  \bibinfo {author} {\bibfnamefont {T.}~\bibnamefont {Macrì}}, \bibinfo
  {author} {\bibfnamefont {G.}~\bibnamefont {Pagano}}, \bibinfo {author}
  {\bibfnamefont {S.}~\bibnamefont {Ruffo}},\ and\ \bibinfo {author}
  {\bibfnamefont {A.}~\bibnamefont {Trombettoni}},\ }\href@noop {} {\bibinfo
  {title} {Long-range interacting quantum systems}} (\bibinfo {year} {2021}),\
  \Eprint {https://arxiv.org/abs/2109.01063} {arXiv:2109.01063
  [cond-mat.quant-gas]} \BibitemShut {NoStop}%
\bibitem [{\citenamefont {Chomaz}\ \emph {et~al.}(2022)\citenamefont {Chomaz},
  \citenamefont {Ferrier-Barbut}, \citenamefont {Ferlaino}, \citenamefont
  {Laburthe-Tolra}, \citenamefont {Lev},\ and\ \citenamefont
  {Pfau}}]{https://doi.org/10.48550/arxiv.2201.02672}%
  \BibitemOpen
  \bibfield  {author} {\bibinfo {author} {\bibfnamefont {L.}~\bibnamefont
  {Chomaz}}, \bibinfo {author} {\bibfnamefont {I.}~\bibnamefont
  {Ferrier-Barbut}}, \bibinfo {author} {\bibfnamefont {F.}~\bibnamefont
  {Ferlaino}}, \bibinfo {author} {\bibfnamefont {B.}~\bibnamefont
  {Laburthe-Tolra}}, \bibinfo {author} {\bibfnamefont {B.~L.}\ \bibnamefont
  {Lev}},\ and\ \bibinfo {author} {\bibfnamefont {T.}~\bibnamefont {Pfau}},\
  }\href@noop {} {\bibinfo {title} {Dipolar physics: A review of experiments
  with magnetic quantum gases}} (\bibinfo {year} {2022}),\ \Eprint
  {https://arxiv.org/abs/2201.02672} {arXiv:2201.02672 [cond-mat.quant-gas]}
  \BibitemShut {NoStop}%
\bibitem [{\citenamefont {Cinti}\ \emph {et~al.}(2017)\citenamefont {Cinti},
  \citenamefont {Cappellaro}, \citenamefont {Salasnich},\ and\ \citenamefont
  {Macr\`{\i}}}]{PhysRevLett.119.215302}%
  \BibitemOpen
  \bibfield  {author} {\bibinfo {author} {\bibfnamefont {F.}~\bibnamefont
  {Cinti}}, \bibinfo {author} {\bibfnamefont {A.}~\bibnamefont {Cappellaro}},
  \bibinfo {author} {\bibfnamefont {L.}~\bibnamefont {Salasnich}},\ and\
  \bibinfo {author} {\bibfnamefont {T.}~\bibnamefont {Macr\`{\i}}},\ }\href
  {https://doi.org/10.1103/PhysRevLett.119.215302} {\bibfield  {journal}
  {\bibinfo  {journal} {Phys. Rev. Lett.}\ }\textbf {\bibinfo {volume} {119}},\
  \bibinfo {pages} {215302} (\bibinfo {year} {2017})}\BibitemShut {NoStop}%
\bibitem [{\citenamefont {Macr{\`\i}}\ \emph {et~al.}(2014)\citenamefont
  {Macr{\`\i}}, \citenamefont {Saccani},\ and\ \citenamefont
  {Cinti}}]{macri2014ground}%
  \BibitemOpen
  \bibfield  {author} {\bibinfo {author} {\bibfnamefont {T.}~\bibnamefont
  {Macr{\`\i}}}, \bibinfo {author} {\bibfnamefont {S.}~\bibnamefont
  {Saccani}},\ and\ \bibinfo {author} {\bibfnamefont {F.}~\bibnamefont
  {Cinti}},\ }\href {https://doi.org/10.1007/s10909-014-1192-7} {\bibfield
  {journal} {\bibinfo  {journal} {Journal of Low Temperature Physics}\ }\textbf
  {\bibinfo {volume} {177}},\ \bibinfo {pages} {59} (\bibinfo {year}
  {2014})}\BibitemShut {NoStop}%
\bibitem [{\citenamefont {Cinti}\ \emph {et~al.}(2014)\citenamefont {Cinti},
  \citenamefont {Macr{\`i}}, \citenamefont {Lechner}, \citenamefont {Pupillo},\
  and\ \citenamefont {Pohl}}]{Cinti2014}%
  \BibitemOpen
  \bibfield  {author} {\bibinfo {author} {\bibfnamefont {F.}~\bibnamefont
  {Cinti}}, \bibinfo {author} {\bibfnamefont {T.}~\bibnamefont {Macr{\`i}}},
  \bibinfo {author} {\bibfnamefont {W.}~\bibnamefont {Lechner}}, \bibinfo
  {author} {\bibfnamefont {G.}~\bibnamefont {Pupillo}},\ and\ \bibinfo {author}
  {\bibfnamefont {T.}~\bibnamefont {Pohl}},\ }\href
  {https://doi.org/10.1038/ncomms4235} {\bibfield  {journal} {\bibinfo
  {journal} {Nature Communications}\ }\textbf {\bibinfo {volume} {5}},\
  \bibinfo {pages} {3235} (\bibinfo {year} {2014})}\BibitemShut {NoStop}%
\bibitem [{\citenamefont {Macr{\`i}}\ and\ \citenamefont
  {Pohl}(2014)}]{PhysRevA.89.011402}%
  \BibitemOpen
  \bibfield  {author} {\bibinfo {author} {\bibfnamefont {T.}~\bibnamefont
  {Macr{\`i}}}\ and\ \bibinfo {author} {\bibfnamefont {T.}~\bibnamefont
  {Pohl}},\ }\href {https://doi.org/10.1103/PhysRevA.89.011402} {\bibfield
  {journal} {\bibinfo  {journal} {Phys. Rev. A}\ }\textbf {\bibinfo {volume}
  {89}},\ \bibinfo {pages} {011402} (\bibinfo {year} {2014})}\BibitemShut
  {NoStop}%
\bibitem [{\citenamefont {Laghi}\ \emph {et~al.}(2017)\citenamefont {Laghi},
  \citenamefont {Macr{\`i}},\ and\ \citenamefont
  {Trombettoni}}]{PhysRevA.96.043605}%
  \BibitemOpen
  \bibfield  {author} {\bibinfo {author} {\bibfnamefont {D.}~\bibnamefont
  {Laghi}}, \bibinfo {author} {\bibfnamefont {T.}~\bibnamefont {Macr{\`i}}},\
  and\ \bibinfo {author} {\bibfnamefont {A.}~\bibnamefont {Trombettoni}},\
  }\href {https://doi.org/10.1103/PhysRevA.96.043605} {\bibfield  {journal}
  {\bibinfo  {journal} {Phys. Rev. A}\ }\textbf {\bibinfo {volume} {96}},\
  \bibinfo {pages} {043605} (\bibinfo {year} {2017})}\BibitemShut {NoStop}%
\bibitem [{\citenamefont {Bakkali-Hassani}\ \emph {et~al.}(2022)\citenamefont
  {Bakkali-Hassani}, \citenamefont {Maury}, \citenamefont {Stringari},
  \citenamefont {Nascimbene}, \citenamefont {Dalibard},\ and\ \citenamefont
  {Beugnon}}]{jdalibard}%
  \BibitemOpen
  \bibfield  {author} {\bibinfo {author} {\bibfnamefont {B.}~\bibnamefont
  {Bakkali-Hassani}}, \bibinfo {author} {\bibfnamefont {C.}~\bibnamefont
  {Maury}}, \bibinfo {author} {\bibfnamefont {S.}~\bibnamefont {Stringari}},
  \bibinfo {author} {\bibfnamefont {S.}~\bibnamefont {Nascimbene}}, \bibinfo
  {author} {\bibfnamefont {J.}~\bibnamefont {Dalibard}},\ and\ \bibinfo
  {author} {\bibfnamefont {J.}~\bibnamefont {Beugnon}},\ }\href@noop {}
  {\bibinfo {title} {The cross-over from {T}ownes solitons to droplets in a
  2{D} {B}ose mixture}} (\bibinfo {year} {2022}),\ \Eprint
  {https://arxiv.org/abs/2207.06939} {arXiv:2207.06939 [cond-mat.quant-gas]}
  \BibitemShut {NoStop}%
\bibitem [{\citenamefont {Dennis}\ \emph {et~al.}(2013)\citenamefont {Dennis},
  \citenamefont {Hope},\ and\ \citenamefont {Johnsson}}]{XMDS2013}%
  \BibitemOpen
  \bibfield  {author} {\bibinfo {author} {\bibfnamefont {G.~R.}\ \bibnamefont
  {Dennis}}, \bibinfo {author} {\bibfnamefont {J.~J.}\ \bibnamefont {Hope}},\
  and\ \bibinfo {author} {\bibfnamefont {M.~T.}\ \bibnamefont {Johnsson}},\
  }\href {https://doi.org/https://doi.org/10.1016/j.cpc.2012.08.016} {\bibfield
   {journal} {\bibinfo  {journal} {Computer Physics Communications}\ }\textbf
  {\bibinfo {volume} {184}},\ \bibinfo {pages} {201 } (\bibinfo {year}
  {2013})}\BibitemShut {NoStop}%
\end{thebibliography}%

\end{document}